\documentclass{aa}

\usepackage{graphicx}
\usepackage{txfonts}

\usepackage{subfigure}

\usepackage{multirow}

\usepackage{natbib}

\newcommand{\hcfn}{H$^{13}$CO$^+$ }

\newcommand{\cfh}{C$_4$H}

\newcommand{\cctht}{$c$-C$_3$H$_2$ }

\usepackage[breaklinks=true, colorlinks=true, allcolors=blue]{hyperref}

\begin{document}

   \title{The evolution of C$_4$H and $c$-C$_3$H$_2$ in  molecular cores}

   \author{Yijia Liu\inst{1,2},
          Jun Zhi  Wang\inst{1}$^{,\textcolor{blue}{\star}}$,
          Ningyu Tang\inst{2},
         Yang Lu\inst{3},
         Donghui Quan \inst{4},
           Juan Li\inst{5},
             Kai Yang\inst{6,7},
           Shu Liu \inst{8},
          Yuqiang  Li\inst{9},
          Siqi  Zheng\inst{5,10,11},
         and Chao  Ou\inst{1}
                   }

   \institute{Guangxi Key Laboratory for Relativistic Astrophysics, School of Physical Science and Technology, Guangxi University, Nanning 530004,  PR China
         \and 
                      Department of Physics, Anhui Normal University, Wuhu, Anhui 241002, PR China
         \and
             Research Center for Computational Earth and Space Science, Zhejiang Laboratory, Hangzhou 311100, China
             \and
             Department of Physics, Xi'an Jiaotong-Liverpool University, 111 Ren'ai Road, Dushu Lake Science and Education Innovation District, Suzhou 215123, Jiangsu Province, People's Republic of China
          \and 
             State Key Laboratory of Radio Astronomy and Technology, Shanghai Astronomical Observatory, Chinese Academy of Sciences, 80 Nandan Road, Shanghai 200030, China
             \and
             Department of Astronomy, School of Physics and Astronomy, and Shanghai Key Laboratory for Particle Physics and Cosmology, Shanghai Jiao Tong University, Shanghai 200240, People's Republic of China
         \and 
         State Key Laboratory of Dark Matter Physics, Shanghai Jiao Tong University, Shanghai 200240, People's Republic of China
            \and
            National Astronomical Observatories, Chinese Academy of Sciences, Beijing 100101, People’s Republic of China
            \and 
            Korea Astronomy and Space Science Institute, No. 776, Daedeok-daero, Yuseong-gu, Daejeon, Republic of Korea
            \and 
            I. Physikalisches Institut, Universit\"{a}t zu K\"{o}ln, Z\"{u}lpicher Str. 77, 50937 K\"{o}ln, Germany
            \and 
            School of Astronomy and Space Sciences, University of Chinese Academy of Sciences, No. 19A Yuquan Road, Beijing 100049, People's Republic of China
                          }

 \date{Received xx; accepted xxx}
 \authorrunning{Liu et al.}
 
\abstract
   {Linear C$_4$H and cyclic $c$-C$_3$H$_2$, as small unsaturated hydrocarbons, are the key precursors to complex organic molecules and are critical components of the interstellar medium. However, observational constraints on the evolution of these molecules in late-stage massive star-forming regions remain scarce.}
   {We present on-the-fly mapping observations of   C$_4$H 9--8 lines, $c$-C$_3$H$_2$ 2--1, H$^{13}$CO$^+$ 1--0, and  H42$\alpha$  toward a sample of  22  massive star-forming regions using  the IRAM 30m telescope. Our aim is to  further explore the evolution of these carbon-chain molecules by combining observational results obtained in cold cores.}
{We employed  H$^{13}$CO$^+$ 1--0 and H42$\alpha$ as tracers to probe the positions of molecular cloud cores and ionised hydrogen regions (H\,II regions), respectively.
One chemical model in particular,  which includes  gas,  dust grain surface, and  icy mantle phases for C$_4$H and $c$-C$_3$H$_2$ molecules, was used to make comparisons with   observed abundances.}
   {From mapping observations targeting 31 regions across 22 sources, C$_4$H $9$--$8$ ($J=19/2$--$17/2$) and C$_4$H $9$--$8$ ($J=17/2$--$15/2$) were detected in only 17 regions, while H$^{13}$CO$^+$ $1$--$0$ and $c$-C$_3$H$_2$ $2$--$1$ were successfully detected in all 31 regions.
 We find that the emission of C$_4$H 9--8 and $c$-C$_3$H$_2$ 2--1 is concentrated at the edges of H42$\alpha$ emission regions. The C$_4$H/H$^{13}$CO$^+$ and $c$-C$_3$H$_2$/H$^{13}$CO$^+$ relative abundance ratios range from 0.17 to 1.77 (median $\sim$ 0.57) and 1.42 to 6.69 (median $\sim$ 4.19), respectively, with a median C$_4$H/$c$-C$_3$H$_2$ ratio of 0.13. By combining the  observational results of cold cores, we find that  C$_4$H/H$^{13}$CO$^+$ and $c$-C$_3$H$_2$/H$^{13}$CO$^+$ ratios show a strong decreasing trend as molecular cores evolve.}   
  {The decreasing trends in C$_4$H/H$^{13}$CO$^+$ and $c$-C$_3$H$_2$/H$^{13}$CO$^+$ ratios imply that small unsaturated hydrocarbons can be consumed and converted into other organic molecules during the evolution of molecular cores. The spatial concentration of C$_4$H and $c$-C$_3$H$_2$ emission at the edges of H42$\alpha$ regions further supports their role as precursors in the chemical pathways that lead to complex organic molecules in the interstellar medium.}

\maketitle
 
\nolinenumbers

\begingroup
\renewcommand{\thefootnote}{} 
\footnotetext{\textsuperscript{$\star$} Corresponding author; junzhiwang@gxu.edu.cn}
\endgroup
\section{Introduction} \label{sec:intro}

The interstellar medium  harbours a rich diversity of carbon-bearing molecules. Among these, unsaturated hydrocarbon molecules play a pivotal role  in tracing chemical evolution, with  small unsaturated hydrocarbons such as  C$_4$H and $c$-C$_3$H$_2$ 
serving  as  crucial  fundamental building blocks for the formation of more complex organic  molecules. 
The evolution of  unsaturated organic molecules in molecular clouds is critically important, as it represents a crucial stellar feedback mechanism that provides essential constraints for understanding star formation processes \citep{2022MNRAS.511.3832A}.
Regarding C$_4$H and $c$-C$_3$H$_2$, strong emission has been detected  in TMC-1 \citep{1981ApJ...248L.113I,1985ApJ...299L..63T,1989AJ.....97.1403M}, the starless core L1521F  \citep{2008ApJ...685..272H,2011ApJ...728..101T}, and the cold molecular cores \citep{2024ApJ...969...33L}, where their relative abundances are relatively high.
However, studies examining the evolution of carbon-chain molecules in massive star-forming regions are still relatively limited.

C$_4$H, a prototypical linear carbon chain molecule, is predominantly formed via gas-phase processes and exists primarily as a gas-phase species.
Drawing on several  astrochemical networks, C$_4$H can be formed by some routes for the C$_n$H family  and  through  $\mathrm{C} + \mathrm{C_3H_2} \rightarrow \mathrm{C_4H} + \mathrm{H}$ \citep{2014MNRAS.437..930L,2023ApJ...944L..45R}.
 The destruction of C$_4$H is dominated by the reaction with oxygen, $\mathrm{C_4H + O \to CO + C_3H}$ \citep{2014MNRAS.437..930L}, and through  the radiative association  of an electron, a process  that disrupts its molecular structure and forms C$_4$H$^-$ \citep{2008ApJ...685..272H,2016JPhB...49t4003G}.
The cyclic isomer  $c$-C$_3$H$_2$  is observed to be comparatively abundant relative to its linear counterparts $l$-C$_3$H$_2$ within the interstellar medium.
 It forms in the gas phase via dissociative recombination of $\mathrm{C_3H_3^+ + e^- \rightarrow \mathit{c}\mbox{-}\mathrm{C_3H_2 + H}}$) \citep{2001ApJ...552..168F} and isomerisation of $l$-C$_3$H$_2$   ($\mathrm{H + \mathit{l}\mathchar`-\mathrm{C_3H_2} \rightarrow H + \mathit{c}\mathchar`-\mathrm{C_3H_2}}$) \citep{2017MNRAS.470.4075L}.
 The primary destruction pathway of  $c$-C$_3$H$_2$ corresponds to the gas-phase reaction with atomic oxygen (O), HC$_3$O, and other oxygen-bearing fragments,  while its reaction with C atoms  facilitates  carbon-chain growth to form molecules such as C$_4$H \citep{2017MNRAS.470.4075L}.   This reaction exhibits high efficiency at low temperatures.

\cite{2024ApJ...969...33L}  found distinct spatial distributions of  C$_4$H 9--8 and  $c$-C$_3$H$_2$ 2--1  across 19 Galactic cold cores in the early stages of star formation.
Furthermore, it was found that C$_4$H  and $c$-C$_3$H$_2$  may not have a tight chemical link in cold molecular cores and their abundances relative to H$_2$ are approximately $10^{-9}$  \citep{2024ApJ...969...33L}.
The abundance of these two unsaturated hydrocarbon molecules (C$_4$H and $c$-C$_3$H$_2$) is likely to  decrease as the molecular cloud evolves.
 Consistent with this trend, a survey conducted with the Yebes 40-m telescope by \cite{2024A&A...692A..65T} revealed that C$_4$H and $c$-C$_3$H$_2$ were detected in the majority of 11 intermediate-mass cores, with their abundances being lower than those observed in the low-mass protostar L1527.
In a survey of diffuse and marginally translucent clouds toward compact extragalactic millimeter-continuum sources, \cite{2000A&A...358.1069L}  also detected the hydrocarbons  C$_4$H and $c$-C$_3$H$_2$ within these extragalactic sources.
However, current studies on such unsaturated hydrocarbons in late-stage massive star-forming regions  remain scarce. To gain deeper insights into the  evolution of C$_4$H and $c$-C$_3$H$_2$ in star-forming regions at different evolutionary   stages, mapping observations in the massive star-forming regions  are necessary.

In this paper, we present mapping observations  of C$_4$H 9--8 lines,  $c$-C$_3$H$_2$ 2--1, H$^{13}$CO$^+$ 1--0, and  H42$\alpha$  toward a relatively large sample of 22 Galactic late-stage massive star-forming regions with known 6.7 GHz CH$_3$OH masers using the Institut de Radioastronomie Millimétrique (IRAM) 30-meter telescope. 
We investigate the spatial distributions and relative abundances of C$_4$H and $c$-C$_3$H$_2$ in these regions to better understand the evolution of carbon-chain molecules.
  The observations and data reduction are described in  Sect. \ref{sec:obs}, and the  results  in Sect. \ref{sec:results}. A discussion is presented  in Sect. \ref{sec:discussion} and a brief summary  in Sect. \ref{sec:summary}.

\section{Observation and data reduction}
 \label{sec:obs}
 The 22 targets,  selected  from \cite{2014ApJ...783..130R}, are late-stage massive star-forming regions with 6.7\,GHz CH$_3$OH masers, 
for which  accurate  trigonometric parallaxes have been measured. 
  On-the-fly (OTF) mapping observations were carried out   using the IRAM 30-meter telescope, located at  Pico Veleta, Spain, during  July  2019,  October 2019,   November 2019,  December 2020, and January 2021.
The  observational data were acquired  using  the Eight Mixer Receiver (EMIR)  operating  3 mm (E0) band, coupled with   the Fourier Transform Spectrometers (FTS)  backend, which provided  an instantaneous   bandwidth of 8\, GHz. 
The system delivered a spectral resolution of 195 kHz with dual polarisation capability.
The IRAM 30 m telescope has a beam size of $\sim$ 24$^{''}$ at 85\,GHz. A pixel size of  9$''$  was adopted  for re-gridding of the OTF data.
The system temperatures remained stable at approximately 150 K.
The telescope pointing accuracy was maintained through regular calibration every $\sim$ 2 hours using  strong nearby quasi-stellar objects as reference points. 
The focus was calibrated prior to each observation, as well as at sunrise and sunset.

 The main beam brightness temperature  ($T_{\rm mb}$) was derived  by   $T_{\rm mb}$= $T_{\rm A}^{\ast}\cdot F_{\rm eff}/B_{\rm eff}$, where $T_{\rm A}^{\ast}$ represents the antenna temperature.
 For the 3 mm band observations, the  forward efficiency, $F_{\rm eff}$,  and beam efficiency, $B_{\rm eff}$, were 0.95 and  0.81, respectively.  
The following molecular lines within the observed frequency range were analysed in this study:  C$_4$H 9$-8$ (85.634.0044 MHz, $J$=19/2$-17/2$; 85634.0154 MHz, $J$=19/2$-19/2$; 85672.5793 MHz, $J$=17/2$-15/2$; 85672.5815 MHz, $J$=17/2$-17/2$), $c$-C$_3$H$_2$  2$-1 $ (85338.8940 MHz),  H$^{13}$CO$^+$ 1$-0$ (86754.2884 MHz), and  H42$\alpha$ at  85688.4 MHz.
Owing to significant line blending,  C$_4$H $9$--$8$ ($J=19/2$--$17/2$) and C$_4$H $9$--$8$ ($J=19/2$--$19/2$)  are treated as a single feature and  denoted as C$_4$H $9$--$8$ ($J=19/2$--$17/2$) in this work. An analogous approach was applied to the C$_4$H $9$--$8$ ($J=17/2$--$15/2$) and C$_4$H $9$--$8$ ($J=17/2$--$17/2$) transitions.
The parameters for the molecular lines were extracted from the Cologne Database for Molecular Spectroscopy ({\sc CDMS}\footnote{https://cdms.astro.uni-koeln.de/classic}; \citealt{2005JMoSt.742..215M}) and  summarised in  Table  \ref{table:Physical parameters}.

The data reduction was performed using  the  {\sc GILDAS} software package\footnote{http://www.iram.fr/IRAMFR/GILDAS}, specifically employing the {\sc CLASS}  and {\sc GREG} applications.
For each molecular line detected  in the observed  sources, we first determined the velocity range of the spectral emission features.
 Following the removal of a first-order baseline,  these velocity ranges were utilised as a mask and setup window within the  {\sc CLASS}  software. The spatial distribution of each line was obtained  by applying   the print area  function in {\sc CLASS}   to extract   velocity-integrated flux maps  at each spatial pixel.

\begin{table*}[h]
\setlength{\tabcolsep}{0.18in}
\caption{Physical parameters of C$_4$H, $c$-C$_3$H$_2$ and H$^{13}$CO$^+$ lines.}
\label{table:Physical parameters}
\vspace{-1mm}
\centering
\begin{tabular}{cccccccccc}

\hline\hline
Molecular    &Transition &Q$_{37.50}$ & freq  & $E_{up}$  & $g_{u}\ $   & A &\\

                  &           &                 & (MHz)  &   (K)  &   & $(10^{-5}\ \mathrm{s}^{-1})$ &\\

\hline

C$_4$H  &       N=9--8 J=19/2--17/2       &  660.0312&            85634.0044      &       20.561  &       19      &       1.5175           \\
        &       N=9--8 J=19/2--19/2       & 660.0312&             85634.0154      &       20.561         &       21      &       1.5267          \\
        &               N=9--8 J=17/2--15/2       & 660.0312&     85672.5793      &       20.563         &       17      &       1.5078          \\
        &               N=9--8 J=17/2--17/2       &660.0312&      85672.5815      &       20.563         &       19      &       1.5193          \\
$c$-C$_3$H$_2$  &J=2(1,2)--1(0,1)        &               566.8560        &       85338.8940      &       6.445         &       15      &       2.3221          \\
H$^{13}$CO$^+$  &J=1--0  & 18.3516       &               86754.2884      &       4.164   &       3       &       3.8535          \\

 \hline
\end{tabular}
\end{table*}

\section{Results}
 \label{sec:results}

From the mapping observations, multiple emission peaks were detected in several sources. Thus, a total of 31  regions were selected across 22 sources. C$_4$H ($9$--$8$, $J=19/2$--$17/2$) and C$_4$H ($9$--$8$, $J=17/2$--$15/2$) were detected in only 17 regions, while H$^{13}$CO$^+$ ($1$--$0$) and $c$-C$_3$H$_2$ ($2$--$1$) were successfully detected in all 31 regions.
The detailed detection  results of these molecular lines are summarised in Table \ref{table:spectral detection}.
 
\subsection{Spatial distribution of  C$_4$H, $c$-C$_3$H$_2$, and H$^{13}$CO$^+$ lines}
\label{sec:spatial distribution}

Given the relatively weak individual signals of C$_4$H 9$-8$  ($J$=19/2$-17/2$) and C$_4$H 9$-8$ ($J$=17/2$-15/2$), and considering their nearly identical Einstein coefficients and upper energy levels, it is reasonable to combine them into a single total velocity-integrated intensity. 
 This approach provides a more reliable representation of the spatial distribution of C$_4$H compared to analyzing each line independently.
Hereafter,  C$_4$H 9$-8$  is used to represent  the total velocity-integrated fluxes of C$_4$H 9$-8$ ($J$=19/2$-17/2$) and C$_4$H 9$-8$ ($J$=17/2$-15/2$).

To better demonstrate the spatial distribution  of these molecular lines, two  velocity-integrated  intensity maps were drawn for each source.
An example of the G015.03$-$00.67 source is shown in Fig. \ref{fig-1-eg}, the velocity-integrated intensity distributions of C$_4$H $9$--$8$ (red contours) and $c$-C$_3$H$_2$ $2$--$1$ (blue contours and grey scale) are presented in Fig. \ref{fig:map_1}, while the spatial distributions of H$^{13}$CO$^+$ $1$--$0$ (black contours and grey scale) and H42$\alpha$ (magenta contours) are illustrated in Fig. \ref{fig:map_2}. The remaining spatial distribution maps are illustrated in Appendix \ref{Appendix-1} and the parameter settings for the maps of molecular emission lines in each source are summarised in Table \ref{table:step}.

\begin{figure*}
    \centering
    \addtocounter{figure}{0}
      \subfigure[]{
        \label{fig:map_1}
        \includegraphics[width=0.45\textwidth]{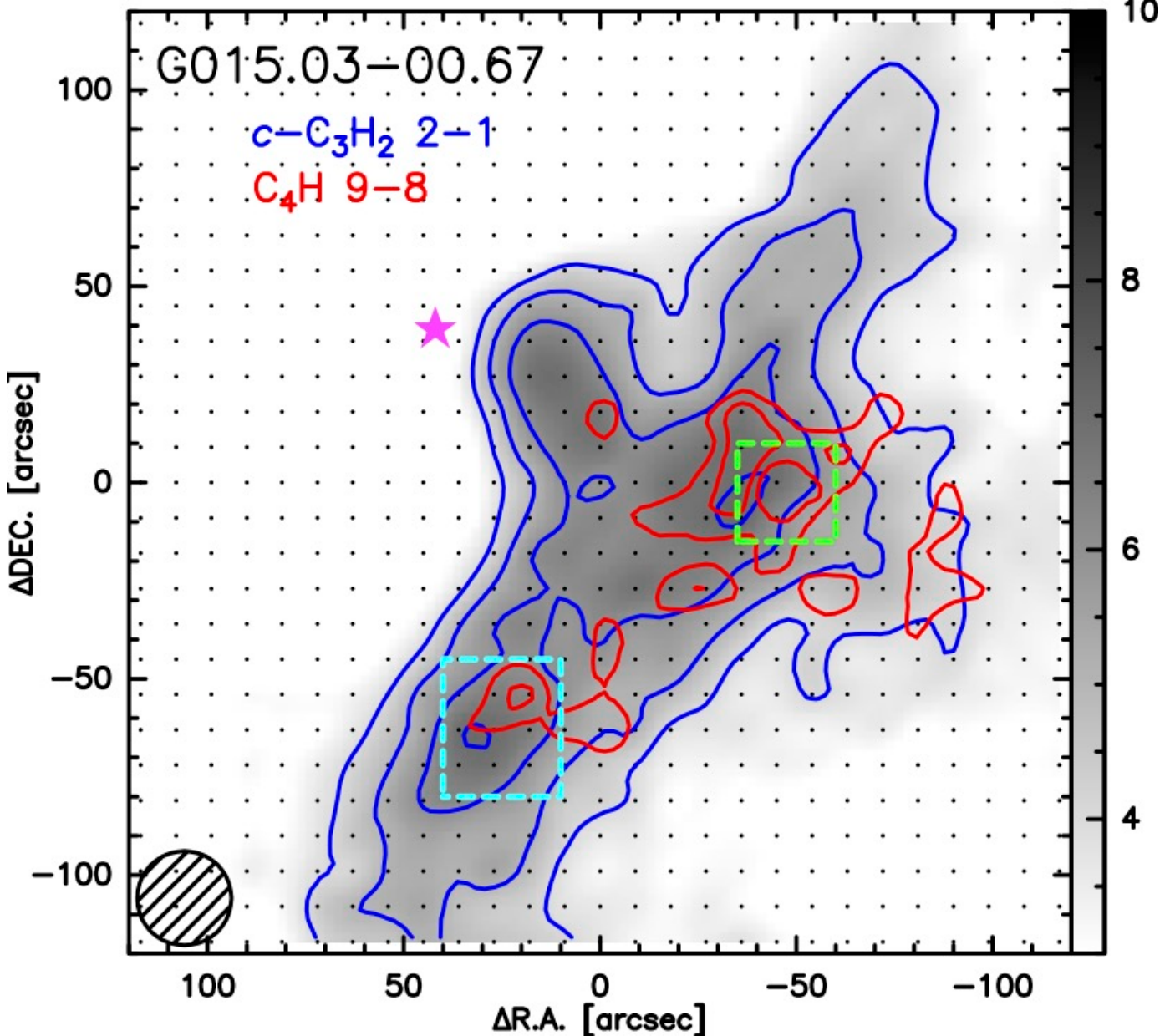}
    }
    \subfigure[]{
        \label{fig:map_2}
        \includegraphics[width=0.35\textwidth]{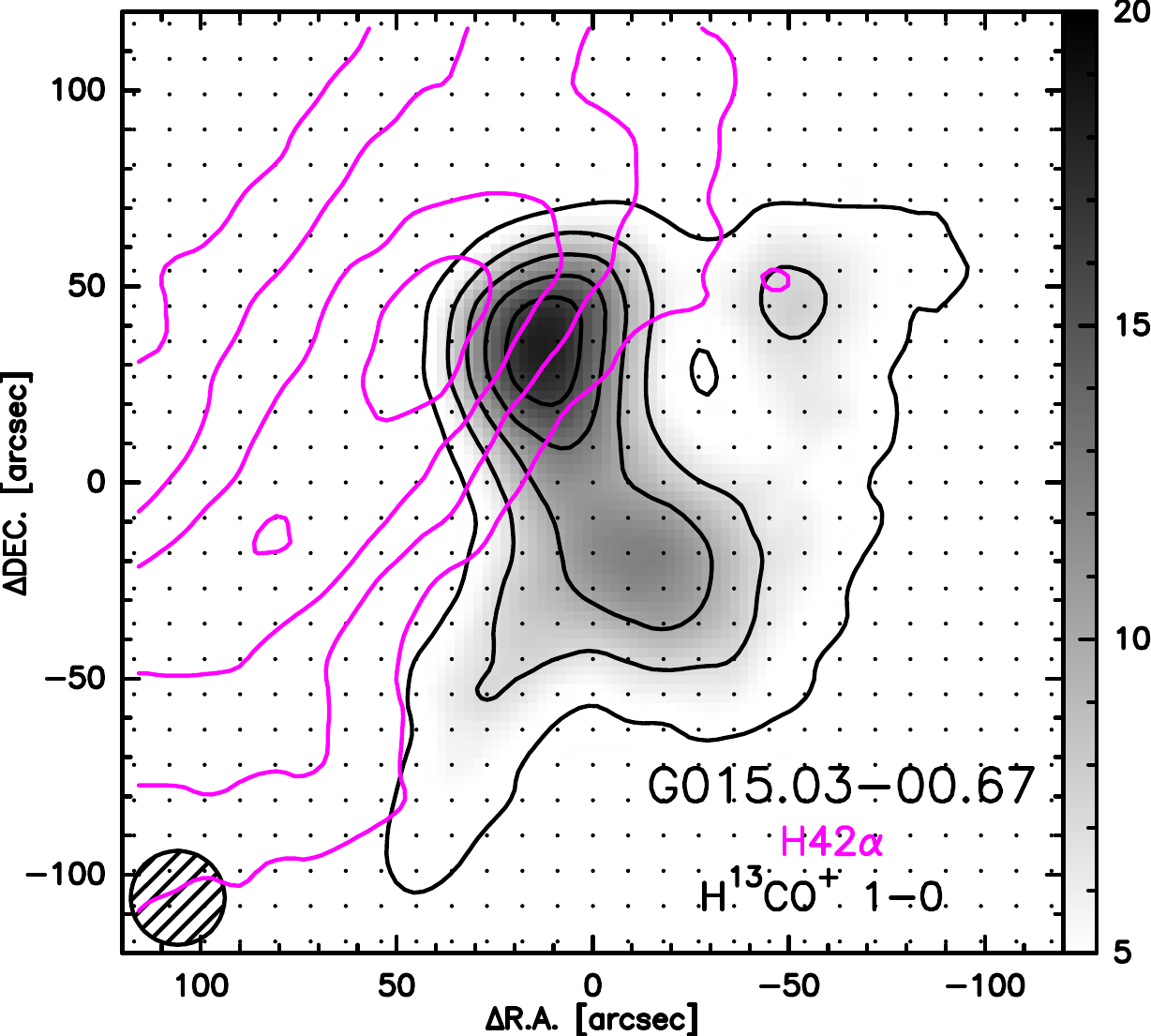}
    }

    \subfigure[]{
        \label{fig:line_1}
        \includegraphics[width=0.35\textwidth]{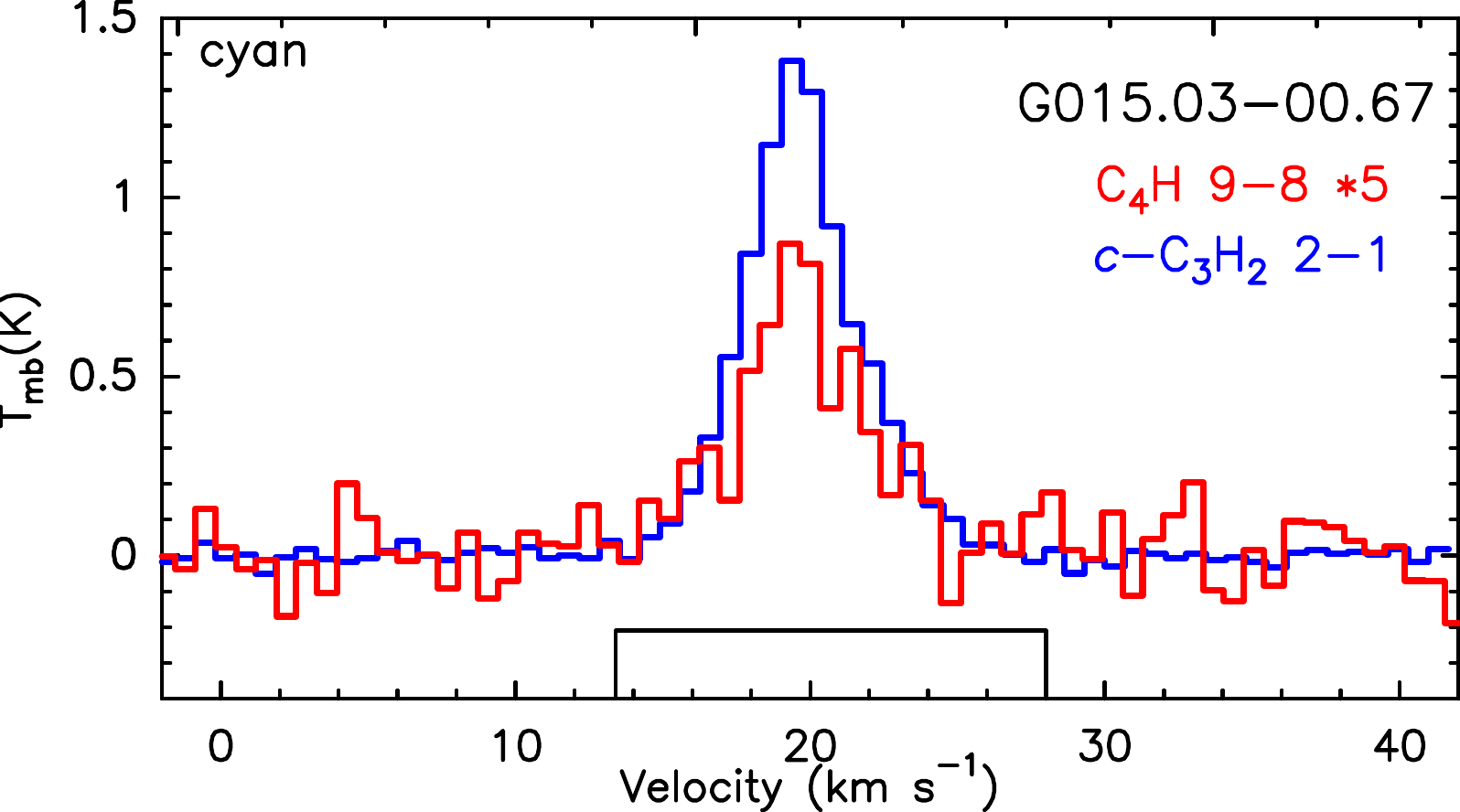}
    }
    \subfigure[]{
        \label{fig:line_2}
        \includegraphics[width=0.35\textwidth]{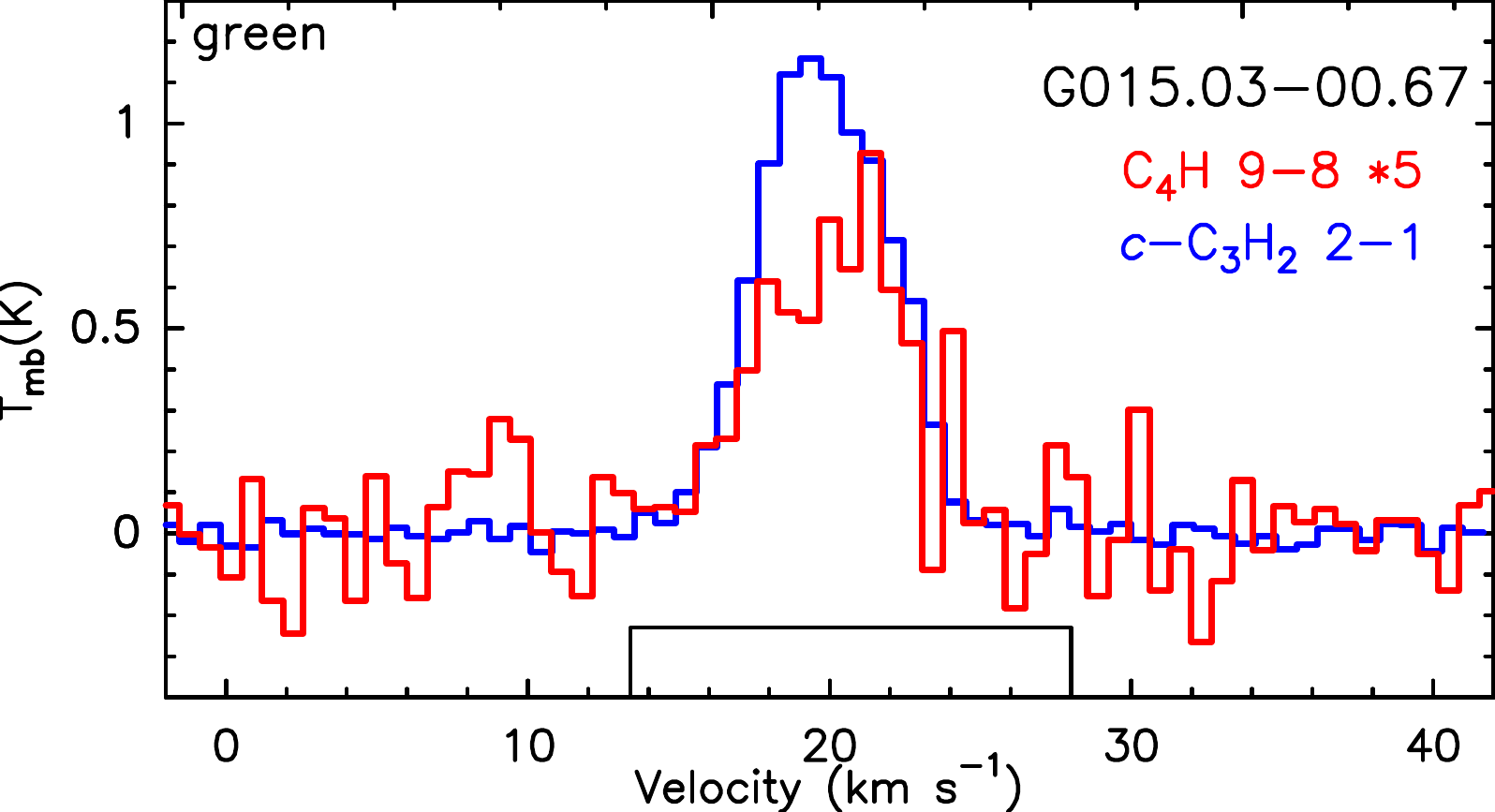}
    }

    \caption{Velocity-integrated  intensity maps and spatial averaged spectra of C$_4$H 9--8, $c$-C$_3$H$_2$ 2--1, H$^{13}$CO$^+$ 1--0, and H42$\alpha$. The source names are presented in the maps and spectra. The grey scale colour at the right is in units of K km s$^{-1}$.  
(a) and (b)  Velocity-integrated intensity maps of G015.03$-$00.67, where panel (a) shows C$_4$H $9$--$8$ (red contours) overlaid on $c$-C$_3$H$_2$ $2$--$1$ (blue contours and grey scale) and panel (b) shows H42$\alpha$ (magenta contours) overlaid on H$^{13}$CO$^+$ $1$--$0$ (black contours and grey scale).
The excitation peak of H42$\alpha$ is marked with a magenta five-pointed star in panel (a),
(c), and (d) Spectra of C$_4$H at 85672.5793 MHz and $c$-C$_3$H$_2$ at 85338.8940 MHz  in the green and cyan  box of G015.03-00.67. The  detailed mapping information of all sources for C$_4$H 9--8, $c$-C$_3$H$_2$ 2--1 and H$^{13}$CO$^+$ 1--0  are listed in Table \ref{table:step}.}
    \label{fig-1-eg}
\end{figure*}

Among the eight sources for which spatial distribution maps of C$_4$H 9$-8$ were obtained, six sources exhibited clear and well-defined spatial structures. However, due to limitations in sensitivity and spatial resolution, the maps of C$_4$H 9$-8$ could not be resolved in the sources G034.39+00.22 and G075.76+00.33. 
In all sources where the $c$-C$_3$H$_2$ 2$-1$ (22 sources), H$^{13}$CO$^+$ 1$-0$ (22 sources), and H42$\alpha$ (9 sources)  were detected, we identified significant and well-resolved spatial distribution structures.
The detailed spatial distribution information of  C$_4$H 9$-8$, $c$-C$_3$H$_2$ 2$-1$, H$^{13}$CO$^+$ 1$-0$, and H42$\alpha$ lines is presented in Table \ref{table:distribution}.

Six  of the eight sources detected in C$_4$H 9$-8$ exhibit spatial distribution structures that differ significantly from those of $c$-C$_3$H$_2$ 2$-1$ and H$^{13}$CO$^+$ 1$-0$. Despite  the fact that the emission of C$_4$H 9$-8$ is relatively weak in both G034.39+00.22 and G075.76+00.33, the emission  peaks are not found to be significantly different from $c$-C$_3$H$_2$ 2$-1$ and H$^{13}$CO$^+$ 1$-0$. Furthermore,
H42$\alpha$ was detected in four of these eight sources, with the spatial structures of three sources exhibiting significant differences, except for G075.76+00.33.

 In 14 of the 22 sources where $c$-C$_3$H$_2$ 2$-1$ was detected (G005.88$-$00.39, G011.91$-$00.61, G015.03$-$00.67, G023.43$-$00.18, G031.28$+$00.66, G034.39$+$00.22, G035.19$-$00.74, G035.19$-$01.74, G037.43$+$01.51, G075.76$+$00.33, G081.19$+$01.51, G081.75$+$00.59, G081.87$+$00.78, and G121.29$+$00.65), the spatial distribution of $c$-C$_3$H$_2$ 2$-1$ was not found to different from  that of H$^{13}$CO$^+$ 1$-0$.
 In contrast, the remaining eight sources (G012.80$-$00.20, G043.16$+$00.01, G049.48$-$00.36, G049.48$-$00.38, G109.87$+$02.11, G111.54$+$00.77, G133.94$+$01.06, and G192.60$-$00.04) exhibited significantly different spatial structures between $c$-C$_3$H$_2$ 2$-1$ and H$^{13}$CO$^+$ 1$-0$.
 In the sources G012.80$-$00.20, G043.16$+$00.01, G049.48$-$00.36, G049.48$-$00.38, and G109.87$+$02.11, the spatial distribution differences between $c$-C$_3$H$_2$ 2$-$1 and H$^{13}$CO$^+$ 1$-$0 are primarily characterised by the significantly weaker emission of $c$-C$_3$H$_2$ 2$-$1 at the peak excitation regions of H$^{13}$CO$^+$ 1$-$0. 
 This observed discrepancy suggests that $c$-C$_3$H$_2$ 2$-$1 may have been depleted or consumed during the process of stellar evolution in these regions, while all nine sources in which H42$\alpha$ was detected with $c$-C$_3$H$_2$ 2$-$1  showed a clearly different spatial distribution. The individual sources providing useful spatial information are annotated below.

{\itshape G015.03-00.67}: The mapping size of C$_4$H 9$-8$, $c$-C$_3$H$_2$ 2$-1$, H$^{13}$CO$^+$ 1$-0$, and H42$\alpha$ is $240'' \times 240''$.
Strong  emissions of C$_4$H 9$-8$, $c$-C$_3$H$_2$ 2$-1$, H$^{13}$CO$^+$ 1$-0$, and H42$\alpha$   were  detected,  each exhibiting a distinct and well-resolved  spatial structure.
 The emission peaks of $c$-C$_3$H$_2$ 2$-1$ and H$^{13}$CO$^+$ 1$-0$  locate at approximately (10$''$, 40$''$) of the map, with no significant spatial differences.
Both transitions display an extended  spatial distribution,   covering a region of about   $180'' \times 240''$,  and they are primarily concentrated along  the edges of the  H42$\alpha$  emission.
In contrast,  the C$_4$H 9$-8$ emission is fragmented and lies primarily along the eastern edge of $c$-C$_3$H$_2$ 2$-1$ region.
Meanwhile, the H42$\alpha$ emission forms a band-like structure predominantly along the south-eastern part of the source G015.03-00.67. Its emission peak lies near  (45$''$, 40$''$), slightly offset from those of $c$-C$_3$H$_2$ 2$-1$ and H$^{13}$CO$^+$ 1$-0$.

{\itshape G023.43-00.18}:
The OTF mode was used to cover $120'' \times 120''$  for C$_4$H 9$-8$, $c$-C$_3$H$_2$ 2$-1$,  H$^{13}$CO$^+$ 1$-0$, and  H42$\alpha$. 
All these lines,  except for  H42$\alpha$, were  successfully detected.      
C$_4$H 9$-8$ and $c$-C$_3$H$_2$ 2$-1$ exhibit   different  spatial distributions. The C$_4$H 9$-8$ emission  is primarily concentrated at south of the $c$-C$_3$H$_2$ 2$-1$, with a slight  offset between their peak positions, approximately centered at (-10$''$, 10$''$) and (0$''$, -10$''$), respectively. In contrast, no significant spatial differences were observed between the distributions of  $c$-C$_3$H$_2$ 2$-1$, and H$^{13}$CO$^+$ 1$-0$.

{\itshape G049.48-00.36}:
A $120'' \times 120''$  map of C$_4$H 9$-8$, $c$-C$_3$H$_2$ 2$-1$,  H$^{13}$CO$^+$ 1$-0$, and  H42$\alpha$ was obtained. Although all lines were  detected,  C$_4$H 9$-8$ and $c$-C$_3$H$_2$ 2$-1$ exhibit markedly different spatial distributions, with partial overlap observed in the south-western region of $c$-C$_3$H$_2$ 2$-1$ emission. 
$c$-C$_3$H$_2$ 2$-1$ emission extends across nearly the entire source, displaying  two  prominent peaks at  approximately  (40$''$, -20$''$) and (-40$''$, -10$''$). In comparison, C$_4$H 9$-8$ is primarily distributed  in the southern and south-western parts of G049.48-00.36, with its peak  around (-35$''$, -15$''$), closely coincides with one of the peak positions of $c$-C$_3$H$_2$ 2$-1$.
Meanwhile, the H42$\alpha$ emission is localised to the south-eastern region of G049.48-00.36, exhibiting two prominent excitation peaks at  (0$''$, 0$''$) and (30$''$, -30$''$).

{\itshape G049.48-00.38}: The mapping size of C$_4$H 9$-8$, $c$-C$_3$H$_2$ 2$-1$, H$^{13}$CO$^+$ 1$-0$, and H42$\alpha$ is $120'' \times 120''$.
The source is located adjacent to G049.48-00.36.
All  the lines were detected, though with varying characteristics.
The  weak C$_4$H 9$-8$ emission presents spatially unresolved due to sensitivity and spatial resolution limitations, with only a marginal detection in the south-eastern region of G049.48-00.38. 
In contrast, the strong $c$-C$_3$H$_2$ 2$-1$,  H$^{13}$CO$^+$ 1$-0$, and H42$\alpha$ emission were clearly observed, extending from the north-west to the south-east across the source.

{\itshape G081.75+00.59}:
In this study, we divided the source G081.75+00.59 into two parts, designated as G081.75+00.59-1 and G081.75+00.59-2, with a detailed analysis of each provided separately. 
For both parts, the observational sizes of C$_4$H 9$-8$, $c$-C$_3$H$_2$ 2$-1$, H$^{13}$CO$^+$ 1$-0$, and H42$\alpha$ are $120'' \times 120''$ and $80'' \times 80''$, respectively.
In this source, all the molecular lines were detected, except  H42$\alpha$. They exhibit extended filamentary structures spanning more than 120$''$, revealing distinct and  resolved spatial distributions.

{\itshape G192.60-00.04}: The mapping size of C$_4$H 9$-8$, $c$-C$_3$H$_2$ 2$-1$, H$^{13}$CO$^+$ 1$-0$, and H42$\alpha$ is $120'' \times 240''$.
 All targeted lines have been detected,  revealing distinct spatial distributions, except for H42$\alpha$. 
 H$^{13}$CO$^+$ 1$-0$ emission extends from the north-west to the south-east over a region exceeding 200$''$, with its peak located at (-15$''$, 40$''$). In contrast, the spatial distribution of C$_4$H 9$-8$ consists of two distinct components: one component is slightly offset from the emission center of $c$-C$_3$H$_2$ 2$-1$, peaking at (-10$''$, 30$''$), while the other lies  at (-30$''$, -60$''$) near the south-western edge of $c$-C$_3$H$_2$ 2$-1$ emission.

 \subsection{Column densities and relative abundances  } 
\label{sec:column densities and relative abundance } 

Based on the spatial distributions of C$_4$H 9$-8$, $c$-C$_3$H$_2$ 2$-1$, and H$^{13}$CO$^+$ 1$-0$  in  Sect. \ref{sec:spatial distribution}, we selected the regions with the strongest emission from each source. 
Since multiple  emission peaks were detected in several sources, a total of 31 distinct regions were selected from 22 sources  to calculate the column densities and relative abundances of the three molecules.
The regions  for extracting the spatially averaged spectra are marked with  green dashed  boxes. For maps with multiple selected regions, the second region is highlighted with a cyan box for clear distinction. 
The velocity-integrated intensities of these three lines were derived  from the spatially averaged spectra   via  single-component Gaussian fitting.
G049.48--00.36  and its adjacent counterpart G049.48--00.38 are both complex regions and major massive protocluster candidates in W51 \citep{W51}. Due to  the absorption of molecular lines toward  free-free continuum emission,   we selected regions away from H42$\alpha$ when calculating abundances.

Next, $c$-C$_3$H$_2$ 2$-1$ and H$^{13}$CO$^+$ 1$-0$  emissions  were detected in all 31 selected regions. 
Among these, C$_4$H 9$-8$ transitions (including $J$=19/2$-17/2$ and $J$=17/2$-15/2$) were  detected in 17  regions, specifically: G015.03-00.67 (green and cyan), G023.43-00.18 (green), G034.39+00.22 (green), G049.48--00.36 (green and cyan), G049.48--00.38 (green and cyan), G075.76+00.33 (green), G081.75+00.59-1 (green and cyan), G081.75+00.59-2 (green and cyan), G081.87+00.78 (green), G133.94+01.06 (green), and G192.60-00.04 (green and cyan). 
The spectroscopic properties of these detections are summarised in  Table \ref{table:Observed date}, which includes source names, molecular line names, velocity-integrated intensities ($\int$$T_{\rm mb}\rm dv$), full-width at half-maximum (FWHM), and peak temperatures ($T_{\rm peak}$) for all 31 regions.
The  velocity-integrated intensities of C$_4$H 9$-8$ transitions $J$=19/2$-17/2$ and $J$=17/2$-15/2$, exhibit a range of  0.11 $\pm$ 0.03 K km\,s$^{-1}$ in G081.87+00.78 (green) to 1.25 $\pm$ 0.12 in G049.48--00.38 (cyan) and 0.13 $\pm$ 0.04 K km\,s$^{-1}$ in G081.87+00.78 (green) to 1.33 $\pm$ 0.10 K km\,s$^{-1}$ in G049.48--00.38 (cyan), respectively.
By contrast, the $c$-C$_3$H$_2$ 2$-1$  were  detected in G015.03-00.67 (green) with the strongest  velocity-integrated fluxes of 6.42 $\pm$ 0.06 K km\,s$^{-1}$, while the weakest emissions were found to be 0.98 $\pm$ 0.06 K km\,s$^{-1}$ in G188.94+00.88 (green).

Under the assumptions of local thermodynamic equilibrium (LTE) and optical thinness, the column density of these  molecules can be calculated using the following formula,
\begin{equation}
\label{lias:2}
N_{\rm tot}=\frac{8\pi k\nu^2}{hc^3A_{ul}}\frac{Q(T_{\rm ex})}{\rm g_u}e^{E_u/kT_{\rm ex}}\int T_{\rm mb}\rm dv  (\rm cm^{-2})
.\end{equation} 
In the above equation, $k$ represents the Boltzmann constant, $\nu$ is the frequency of the molecular emission line, $h$ is the Planck constant, $c$ is the speed of light, $A_{\rm ul}$ is the Einstein emission coefficient, $\rm g_u$ is the upper-level degeneracy, and $E_u$ is the upper-level energy. 
The values for $\nu$, $\rm g_u$, $A_{\rm ul}$, and $E_u$ corresponding to the C$_4$H 9$-8$ ($J$=19/2$-17/2$), C$_4$H 9$-8$ ($J$=19/2$-19/2$), C$_4$H 9$-8$ ($J$=17/2$-15/2$), C$_4$H 9$-8$ ($J$=17/2$-17/2$), $c$-C$_3$H$_2$ 2$-1$, and H$^{13}$CO$^+$ 1$-0$ are shown in Table \ref{table:Physical parameters}, which are taken from the CDMS database. 
The partition function $Q (T_{\rm ex})$ predominantly depends on the excitation temperature $T_{\rm ex}$. In this study a value of  37.5 K was adopted for $T_{\rm ex}$, as all the sources were selected from hot cores.

 The small difference in frequency between C$_4$H 9$-8$ ($J$=19/2$-17/2$) and C$_4$H 9$-8$ ($J$=19/2$-19/2$), as well as between C$_4$H 9$-8$ ($J$=17/2$-15/2$) and C$_4$H 9$-8$ ($J$=17/2$-17/2$) results in a blending of the lines, where only two emission features are observable across these four transitions. 
 Therefore, based on Eq. \ref{lias:2}, and considering the two blended lines with their respective values for $A_{ul_1}$, $g_{u_1}$ and $A_{ul_2}$, $g_{u_2}$, the following equation can be derived to calculate the column density of C$_4$H,
 \begin{equation}
\label{lias:3}
        N_{\rm tot}=\frac{8\pi k\nu^2}{hc^3}\frac{Q(T_{\rm ex})}{A_{ul_1}\rm g_{u_1}+A_{ul_2}\rm g_{u_2}}e^{E_u/kT_{\rm ex}}\int T_{\rm mb}\rm dv (\rm cm^{-2})
.\end{equation} 
Since the difference in $E_u$ between the two blended molecular lines is also very small, we use the $E_u$ values of C$_4$H 9$-8$ ($J$=19/2$-17/2$) and C$_4$H 9$-8$ ($J$=17/2$-15/2$) in Eq. \ref{lias:3}.
For sources with non-detection of   C$_4$H 9$-8$ emission, 3$\sigma$  upper limits for the integrated intensity, $\int$$T_{\rm mb}\rm dv$,  were calculated using the formula $3  \times rms\sqrt{\delta {v}\cdot\Delta {v}}$, where $\delta v$ is the channel separation in velocity,  $\Delta v$ is the  velocity range for integration in km s$^{-1}$, and $rms$  is the root mean square value per channel of the spectrum.

 The column densities of C$_4$H, $c$-C$_3$H$_2$, and H$^{13}$CO$^+$ are displayed in columns 2-4 of Table \ref{table_Colunm density}. The range of column densities for C$_4$H spans from 0.33 $\times$ $10^{13}$ cm$^{-2}$ in G081.87+00.78 (green) to 3.67 $\times$ $10^{13}$ cm$^{-2}$ in G049.48-00.38 (cyan), $c$-C$_3$H$_2$ range from 2.68 $\times$ $10^{13}$ cm$^{-2}$ in G188.94+00.88 (green) to 17.57 $\times$ $10^{13}$ cm$^{-2}$ in G015.03-00.67 (green), and  H$^{13}$CO$^+$   from 0.48 $\times$ $10^{13}$ cm$^{-2}$ in G192.60-00.04 (cyan) to 4.92 $\times$ $10^{13}$ cm$^{-2}$ in G049.48-00.36 (green).

 The relative abundances of the three molecules, which can be derived from the averaged column densities for each core, are more important than the averaged densities for the scientific analysis in this study.
Table \ref{table_Colunm density} exhibits the abundance ratios [C$_4$H/H$^{13}$CO$^+$], [$c$-C$_3$H$_2$/H$^{13}$CO$^+$], and [C$_4$H/$c$-C$_3$H$_2$] in these 26 regions.
The abundance ratio of [C$_4$H/H$^{13}$CO$^+$] ranges from 0.17 $\pm$ 0.04 in G081.87+00.78 (green) to 1.77 $\pm$ 0.11 in G049.48-00.38 (cyan), with a median value of 0.57, while it ranges from 1.42 $\pm$ 0.05  in G081.87+00.78 (green) to  6.69  $\pm$ 0.34 in G192.60-00.04 (cyan) with a median value of 4.19 for [$c$-C$_3$H$_2$/H$^{13}$CO$^+$]. For the ratio [C$_4$H/$c$-C$_3$H$_2$], the values range from 0.07 $\pm$ 0.01 in G081.75+00.59 (green) to 0.29 $\pm$ 0.03 in G133.94+01.06 (green), with a median value of 0.13. By combining data from 25 regions selected from 19 cold cores in a  previous work \citep{2024ApJ...969...33L}  with current   data from  26 regions of 22 hot cores, we obtained the relation  between the abundances of C$_4$H and  $c$-C$_3$H$_2$ normalised by H$^{13}$CO$^+$. It is shown in  Fig. \ref{abundance-1}, with a Pearson  correlation coefficient of 0.83, a  slope of 3.41,  and a least-squares fitting.

\section{Discussion} 
\label{sec:discussion}
 
  \subsection{Relative abundances of C$_4$H  and \textit{c}-C$_3$H$_2$ from cold and hot cores.}
 
\begin{figure}[!htbp] 
\centering
\includegraphics[width=\linewidth]{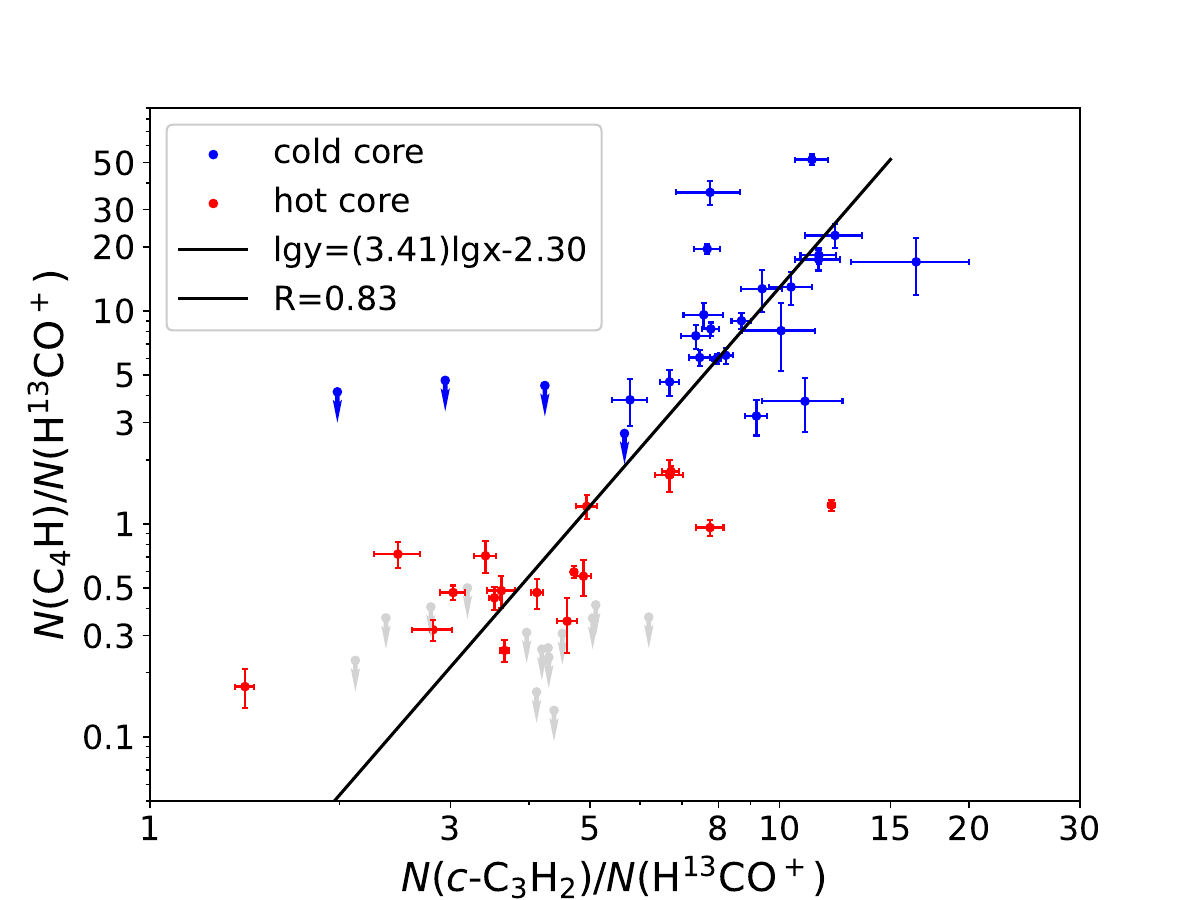} 
\caption{Relation between  $c$-C$_3$H$_2$/H$^{13}$CO$^+$ and  C$_4$H/H$^{13}$CO$^+$ abundance ratios. Data are from 31 regions in 22 hot cores (red points) and 22 regions in 19 cold cores (blue points).}
\label{abundance-1}      
\end{figure}

 As  shown  in Fig. \ref{abundance-1},  the abundance ratios of C$_4$H and $c$-C$_3$H$_2$ relative to  H$^{13}$CO$^+$  decreases from cold cores \citep{2024ApJ...969...33L}  to hot cores, suggesting a downward trend with advancing evolutionary stage of star formation.
  A  least-squares fitting procedure results in the relation: $lg$(C$_4$H/H$^{13}$CO$^+$) = 3.41$lg$($c$-C$_3$H$_2$/H$^{13}$CO$^+$) - 3.20, with a correlation coefficient of 0.83 (excluding C$_4$H upper limits).  
 Among these  31 hot regions, column densities of both $c$-C$_3$H$_2$ and H$^{13}$CO$^+$ were successfully obtained, while  C$_4$H was acquired for 18 regions.
 Specifically, C$_4$H/H$^{13}$CO$^+$ ratio ranges from 0.17 to 1.77 (median $\sim$ 0.57), and  $c$-C$_3$H$_2$/H$^{13}$CO$^+$ ratio spans 1.42 to 6.69 (median $\sim$ 4.19) in hot cores, whereas cold cores exhibit significantly higher ratios of  C$_4$H/H$^{13}$CO$^+$ (3--50) and $c$-C$_3$H$_2$/H$^{13}$CO$^+$ ratios (2--18) \citep{2024ApJ...969...33L}. 
 This observational trend may be closely related to the low temperatures (10--20 K)  that are characteristic of  cold molecular  cores. 
Given that these cloud cores typically form during the early phases of interstellar material accretion, their chemical evolution has not yet been significantly disrupted by star-forming activity \citep{72,73}.
Under such low-temperature conditions, the formation pathways of unsaturated hydrocarbons (C$_4$H   and $c$-C$_3$H$_2$) gain a competitive advantage. 
The low-temperature environment favors simpler reaction mechanisms, particularly because reduced molecular collision rates allow such species to accumulate and sustain relatively high abundances within  cold molecular cores \citep{2017MNRAS.470.4075L}.
In contrast,  massive star-forming regions exhibit higher temperatures (typically 30--100 K), where star formation activity is more mature. 
Under these conditions, intense radiation and physical conditions significantly alter the chemical environment of the molecular clouds. 
The high-temperature conditions may be  expected to accelerate the consumption or conversion of these originally stable small molecules into more complex organic species, leading to reduced abundances of C$_4$H and $c$-C$_3$H$_2$  \citep{2009ARA&A..47..427H,76}, which is also consistent with our observational results.
 
Additionally, it has been established that C$_4$H forms efficiently   via gas-phase ion--molecule reactions, a process particularly effective at low temperatures\ during the early stages of star formation \citep{3}. \cite{7} investigated three sources at different evolutionary stages and found that both C$_4$H  and  $c$-C$_3$H$_2$  exhibit  higher abundances   in the earlier phases.
It was concluded that C$_4$H is also formed through the reaction of C$_2$H$_2$ with C$_2$, a process which was found to be more efficient at lower temperatures \citep{b}.
During the late stages of star formation, thermal processing leads to the desorption of molecules such as H$_2$O, CH$_3$OH, and O$_2$ from dust-grain ice mantles into the gas phase, where C$_4$H is rapidly consumed through reactions with these species \citep{3}.
The derived spatial distribution shows that the emission of C$_4$H 9$-8$ and $c$-C$_3$H$_2$ 2$-1$  is relatively weak across most regions near the excitation peak of H42$\alpha$. In sources such as G015.03-00.67, G049.48-00.36, G049.48-00.38, G075.76+00.33, and G035.19-01.73, the emission is primarily concentrated at the edge of H42$\alpha$. 
Additionally, in regions such as G012.80-00.20 and G043.16+00.01, C$_4$H 9$-8$ was not detected where the emission of  H$^{13}$CO$^+$ 1$-0$  was strong. These positions exhibited $c$-C$_3$H$_2$/H$^{13}$CO$^+$ abundance ratios of 2.17 $\pm$ 0.14 and 2.05 $\pm$ 0.25, respectively. 
This finding  suggests that unsaturated molecules such as C$_4$H   and $c$-C$_3$H$_2$ can be  destroyed or processed into more complex organic molecules during the evolution of molecular cloud cores, leading to a corresponding decrease in their abundances.

  \subsection{Chemical models}

We employed the NAUTILUS three-phase chemical model \citep{2016MNRAS.459.3756R,2024A&A...689A..63W}, which includes gas-phase, grain-surface, and grain-mantle chemistry, to simulate the chemical evolution of \textit{c}-C$_3$H$_2$ and C$_4$H in both cold-core and hot-core environments. For the cold-core model, typical static physical conditions are adopted. The total hydrogen nuclei density is set to n$_H$ = 10$^4$ cm$^{-3}$, the temperature is fixed at T = 10 K, and the cosmic-ray ionisation rate is taken to be $\zeta$ = 1.3$\times 10^{-17}$ s$^{-1}$. The hot-core physical model is divided into two successive stages: a free-fall collapse stage followed by a warm-up stage. During the collapse stage, the temperature remains constant at T = 10 K, while the density increases from 3 $\times$10$^3$ to 3 $\times$10$^6$  cm$^{-3}$. In the warm-up stage, the density is kept constant, and the temperature gradually increases up to 100 K.

\begin{figure}[!htbp] 
\centering
\includegraphics[width=\linewidth]{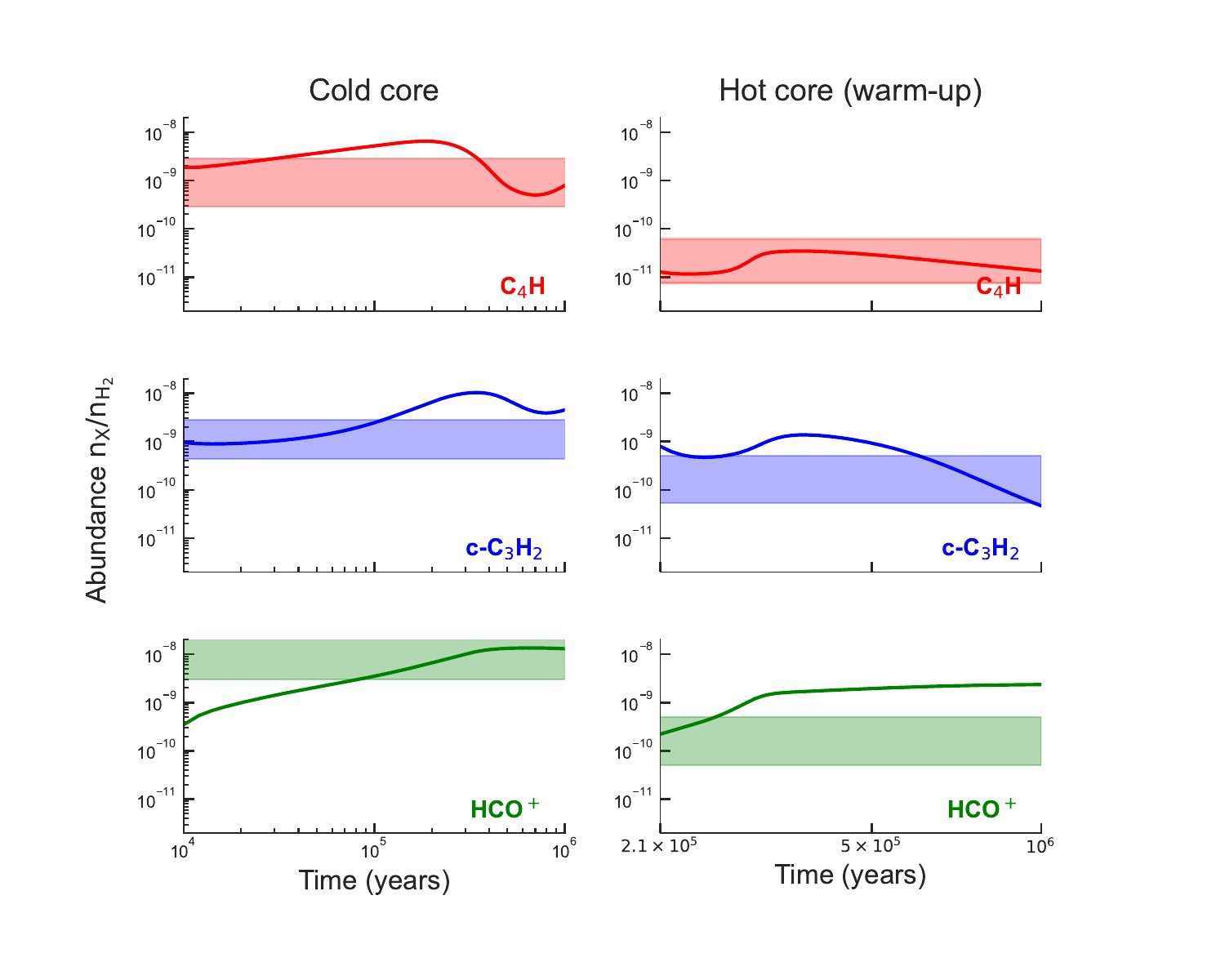} 
\caption{Temporal evolution of \textit{c}-C$_3$H$_2$, C$_4$H, and  HCO$^+$ abundances predicted by models in cold core and hot core (warm-up). } 
 \label{model_predicted_abundance}       
 \end{figure}

 \begin{figure*}
 \centering 

\subfigure[]{ \label{fig:c3h2_pro_a} 
 \includegraphics[width=0.32\textwidth]{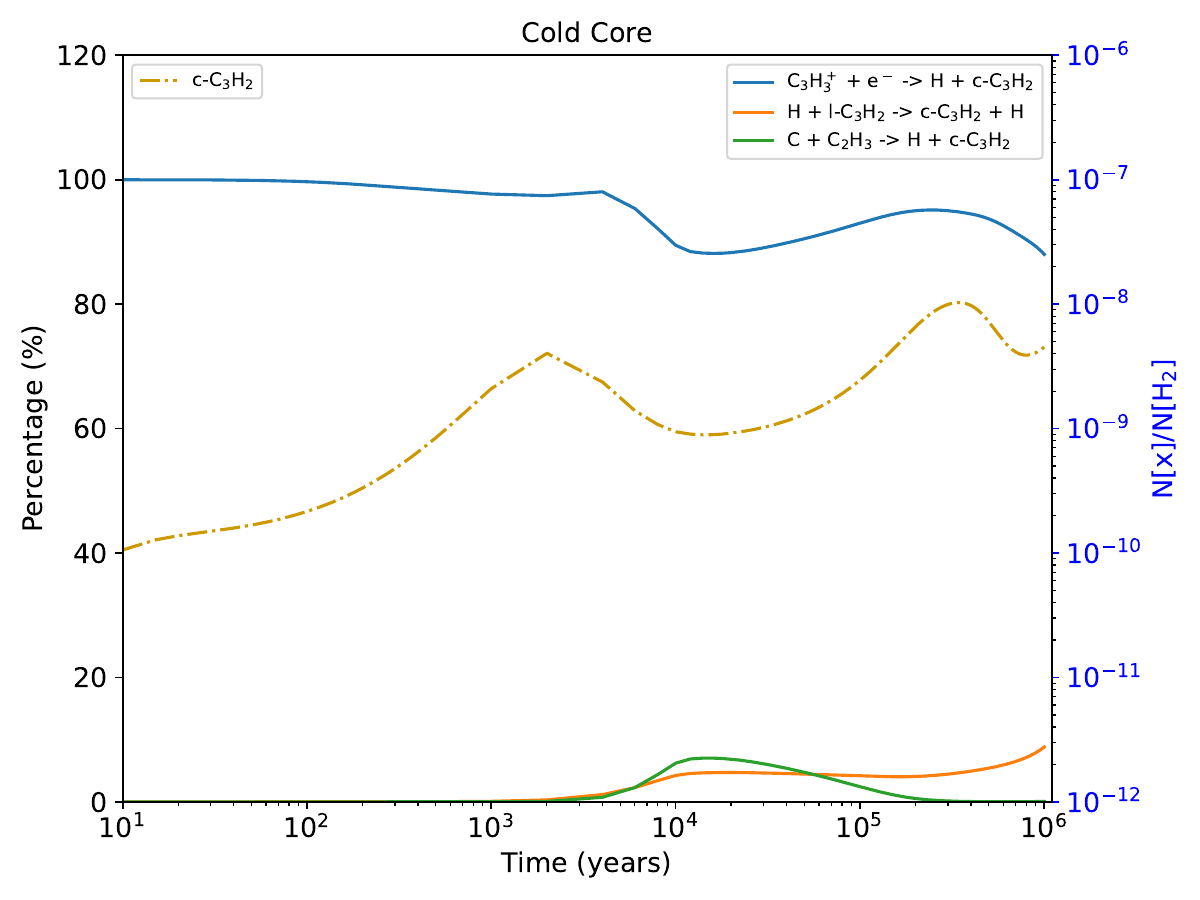} 
} 
\subfigure[]{ \label{fig:c3h2_pro_b} 
\includegraphics[width=0.32\textwidth]{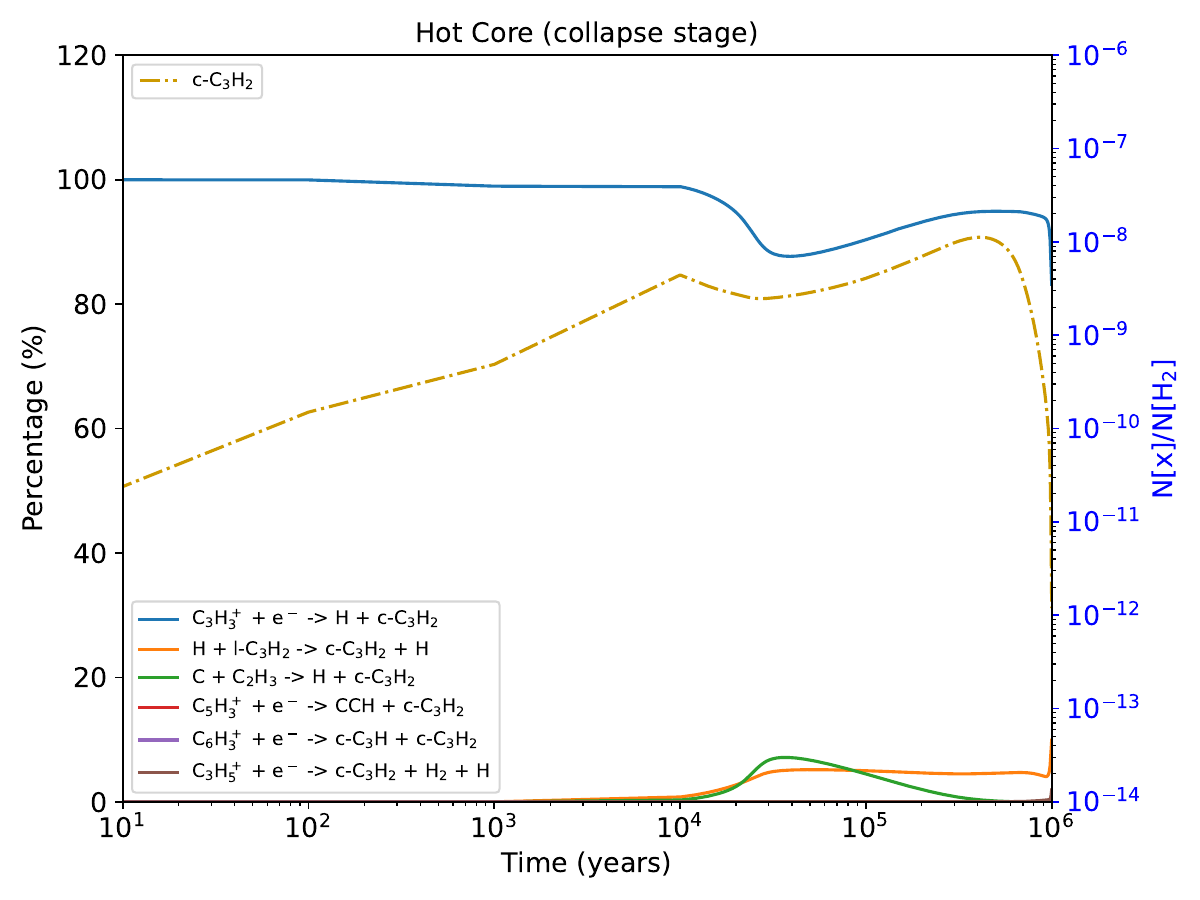}
} 
\subfigure[]{ \label{fig:c3h2_pro_c} 
\includegraphics[width=0.32\textwidth]{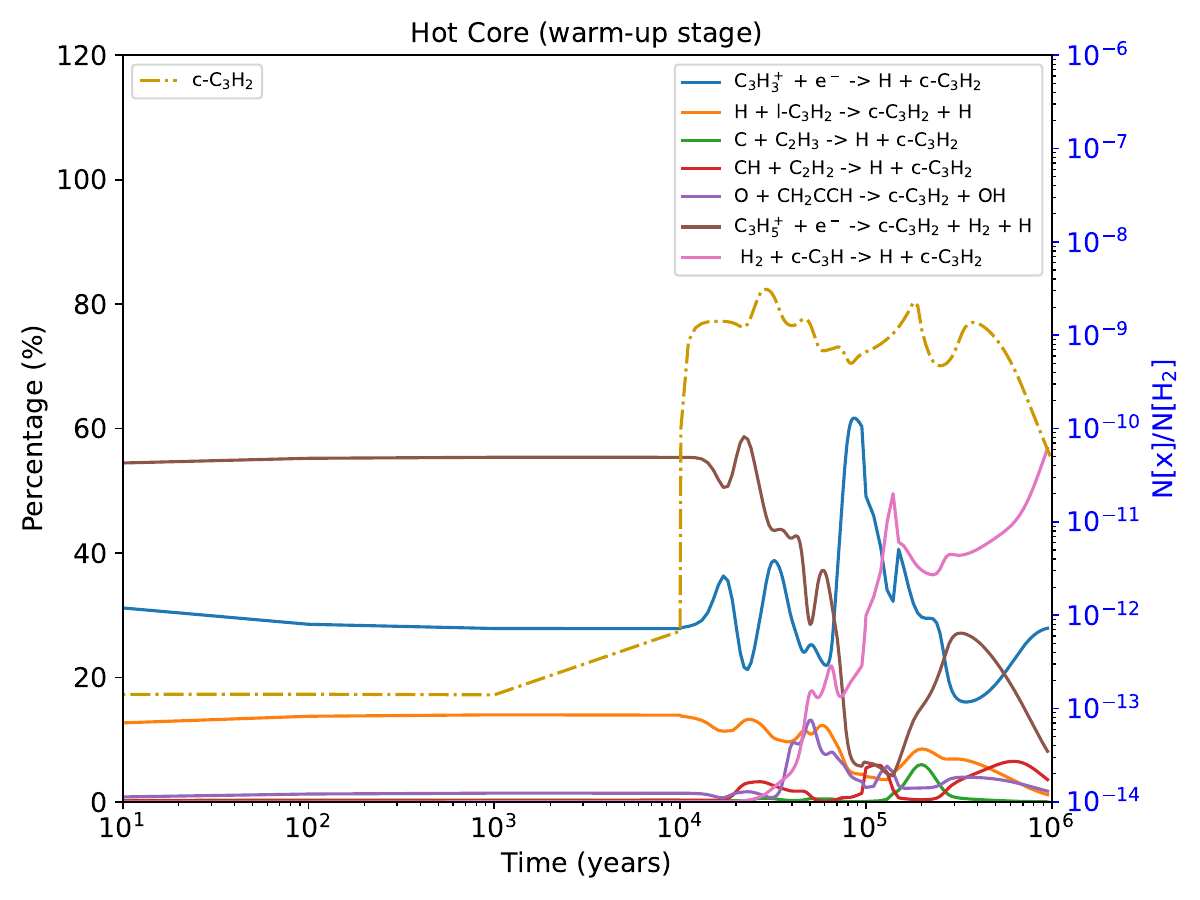}
}

\subfigure[]{ \label{fig:c3h2_des_d} 
\includegraphics[width=0.32\textwidth]{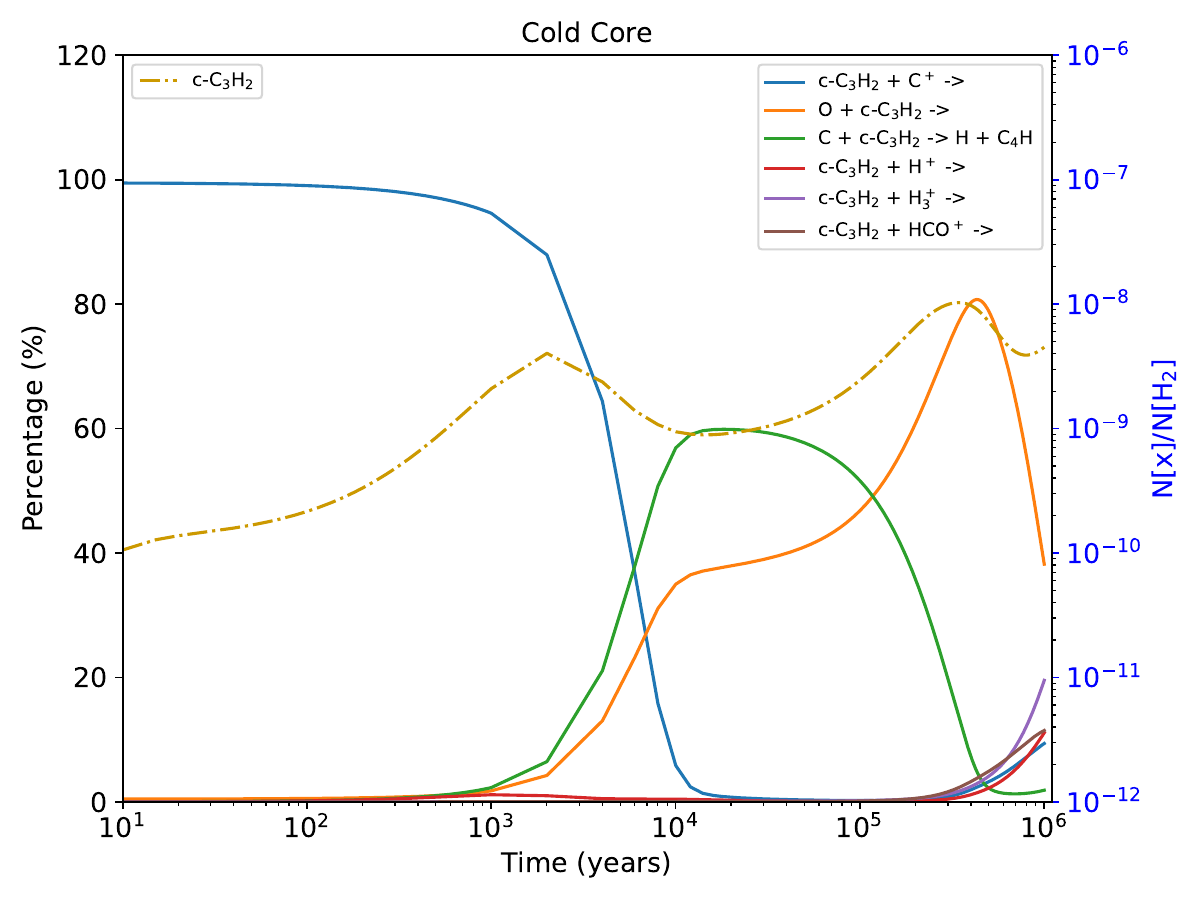}
}
\subfigure[]{ \label{fig:c3h2_des_e} 
\includegraphics[width=0.32\textwidth]{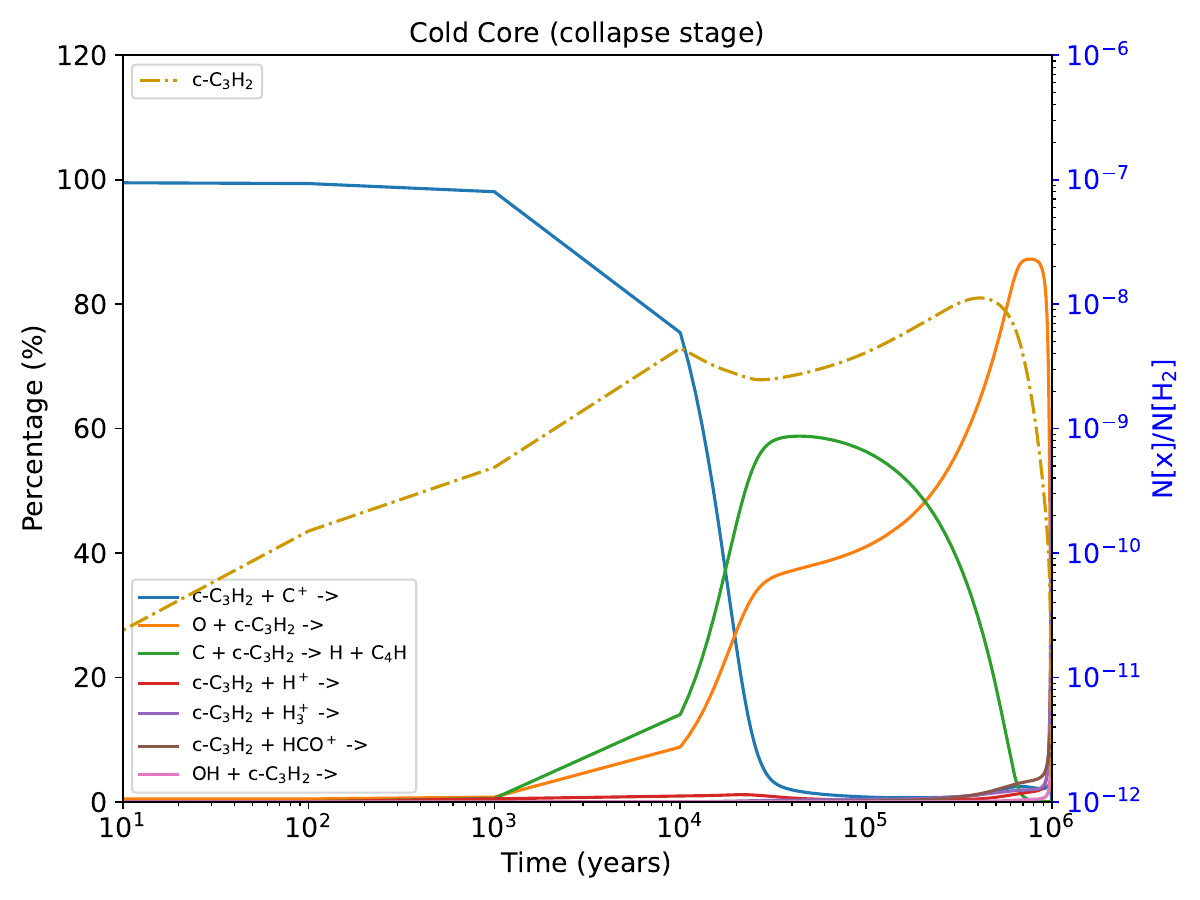}
} 
\subfigure[]{ \label{fig:c3h2_des_f} 
\includegraphics[width=0.32\textwidth]{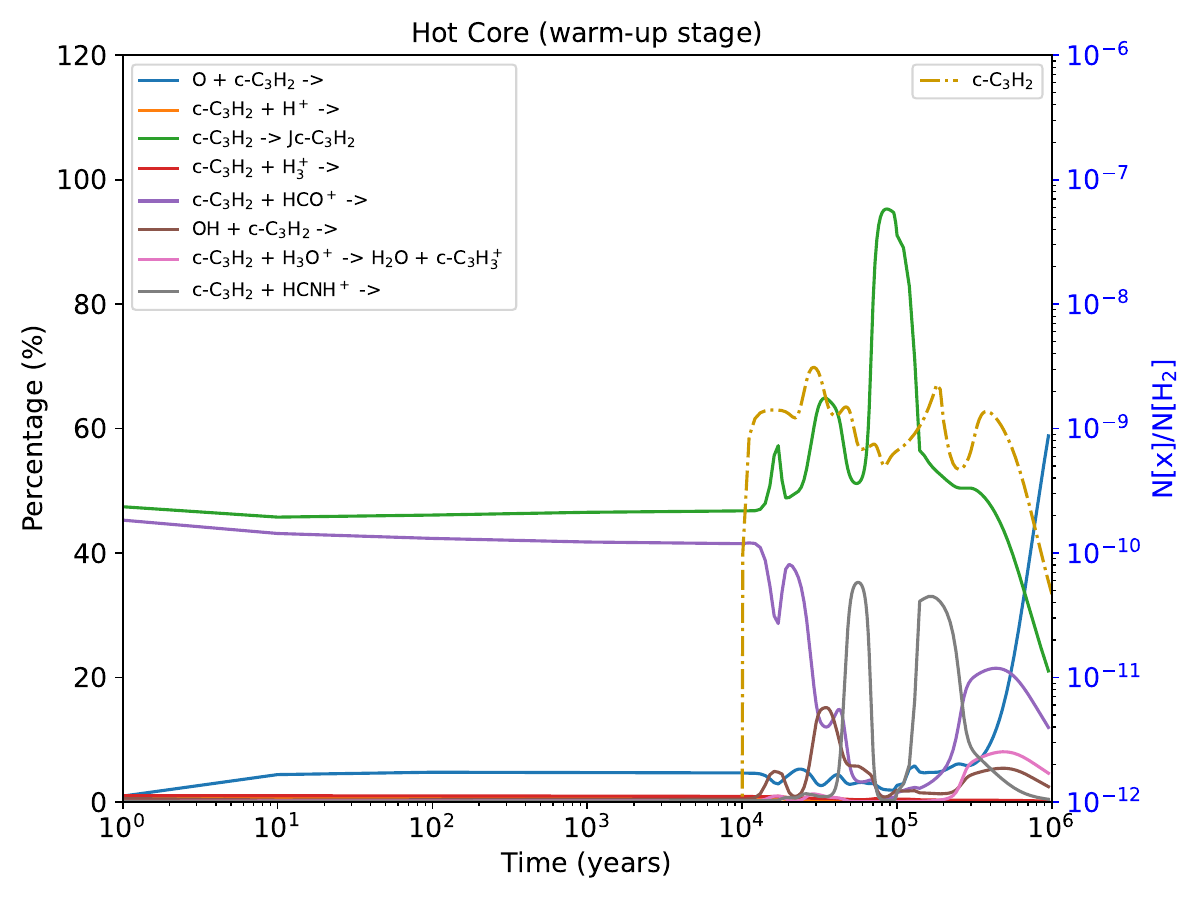}
}

\caption{Net percentage contributions of the formation and destruction pathways of \textit{c}-C$_3$H$_2$ in cold and hot cores. 
(a) Net percentage contributions of the main formation pathways of \textit{c}-C$_3$H$_2$ in cold molecular cloud cores.
(b) Net percentage contributions of the main formation pathways of \textit{c}-C$_3$H$_2$ during the hot-core collapse stage.
(c) Net percentage contributions of the main formation pathways of \textit{c}-C$_3$H$_2$ during the hot-core warm-up stage. 
(d) Net percentage contributions of the main destruction pathways of \textit{c}-C$_3$H$_2$ in cold molecular cloud cores. 
(e) Net percentage contributions of the main destruction pathways of \textit{c}-C$_3$H$_2$ during the hot-core collapse stage. 
(f) Net percentage contributions of the main destruction pathways of \textit{c}-C$_3$H$_2$ during the hot-core warm-up stage.}
\label{fig:c3h2_pathway}
\end{figure*}
\begin{figure*}
 \centering 

\subfigure[]{ \label{fig:c4h_pro_a} 
 \includegraphics[width=0.32\textwidth]{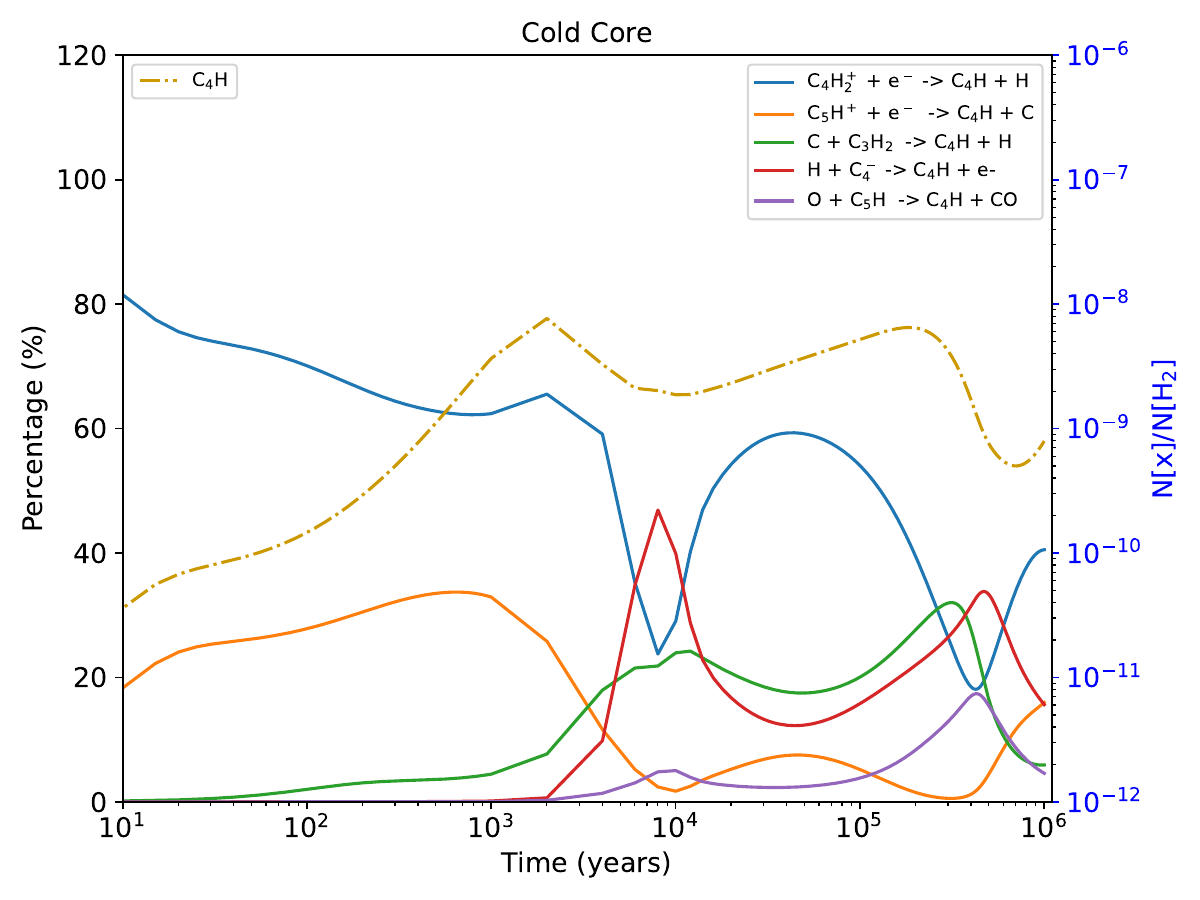} 
} 
\subfigure[]{ \label{fig:c4h_pro_b} 
\includegraphics[width=0.32\textwidth]{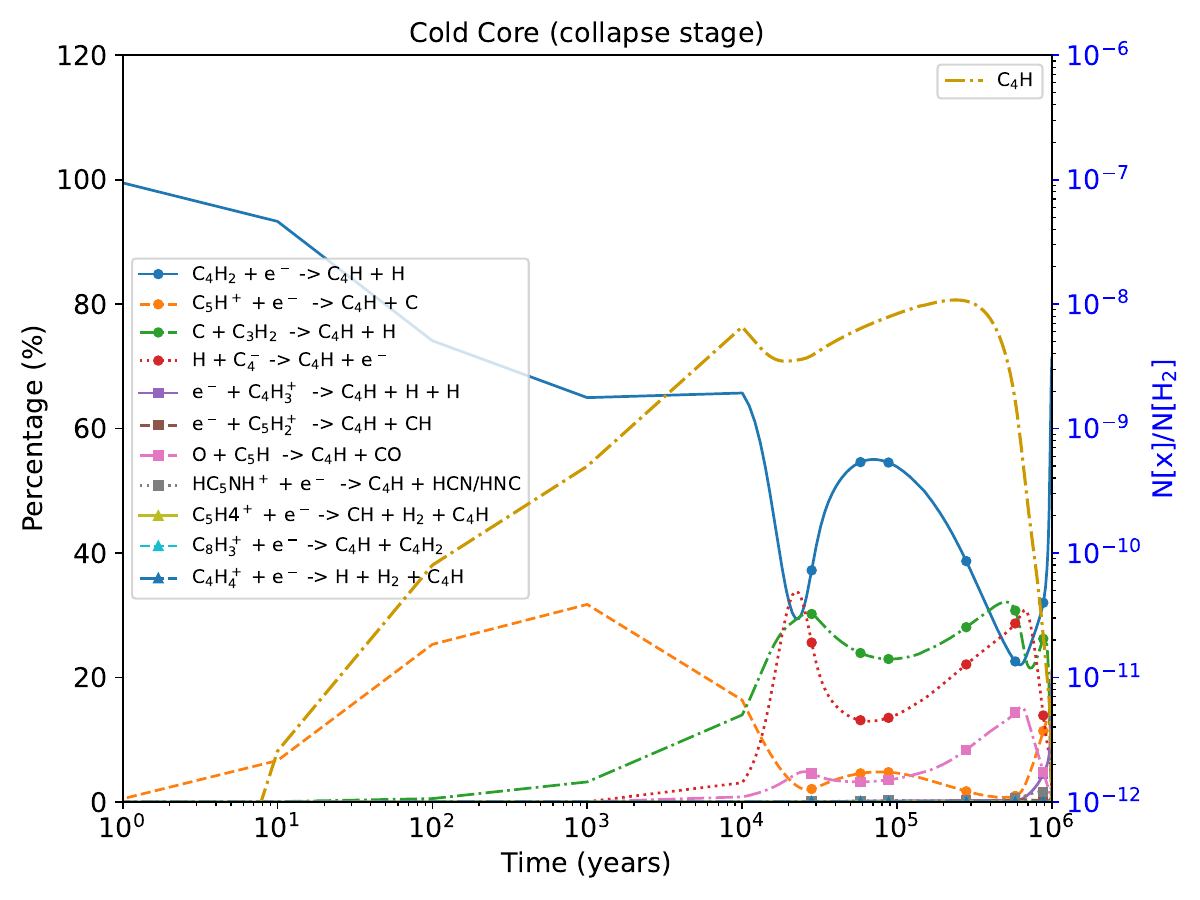}
} 
\subfigure[]{ \label{fig:c4h_pro_c} 
\includegraphics[width=0.32\textwidth]{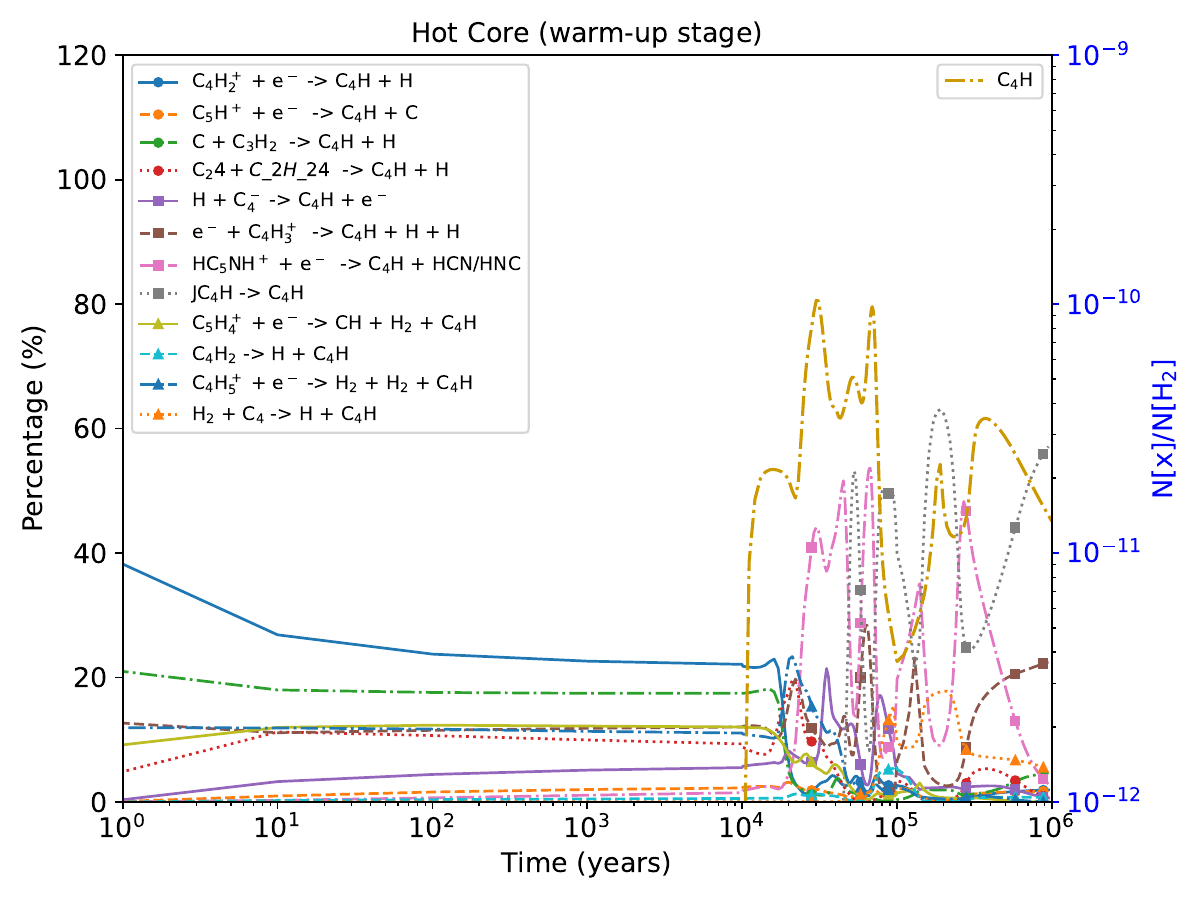}
}

\subfigure[]{ \label{fig:c4h_des_d} 
\includegraphics[width=0.32\textwidth]{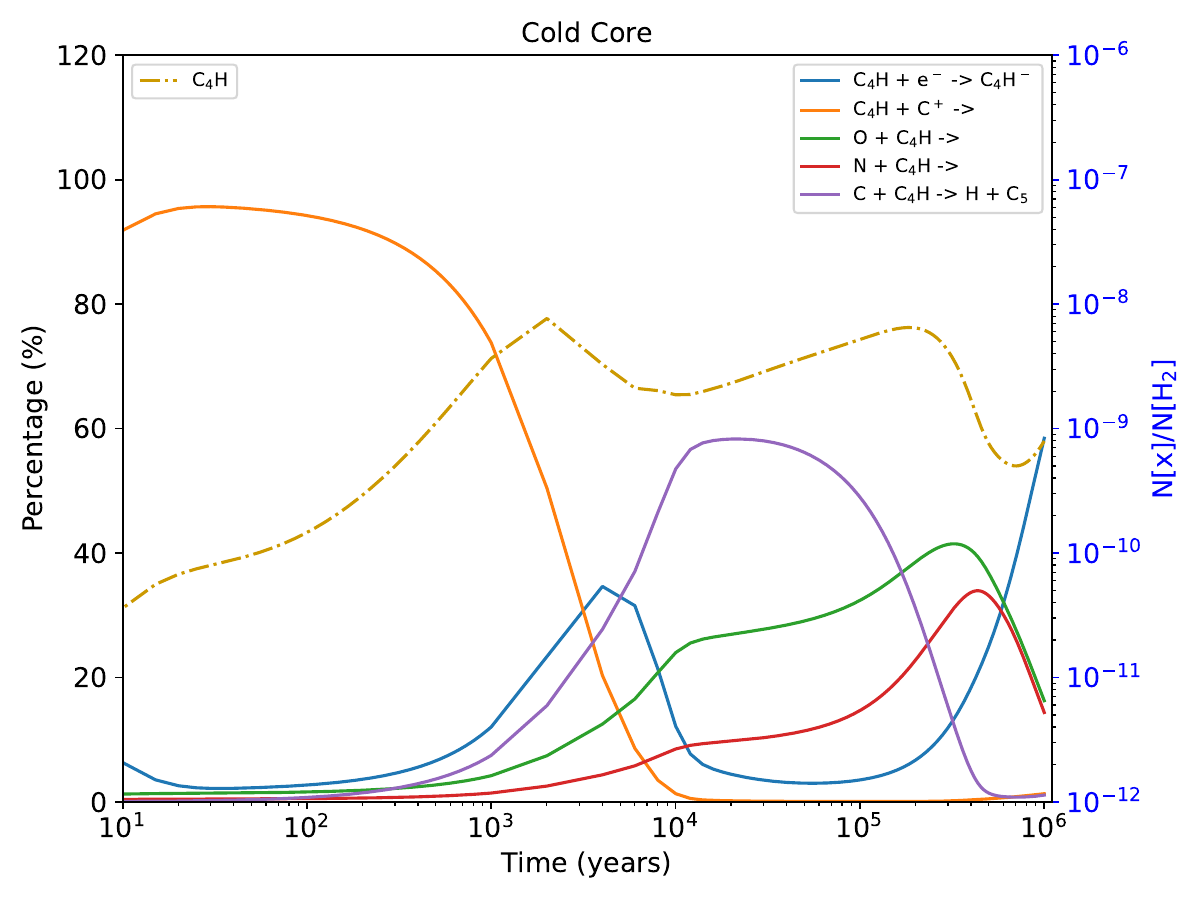}
}
\subfigure[]{ \label{fig:c4h_des_e} 
\includegraphics[width=0.32\textwidth]{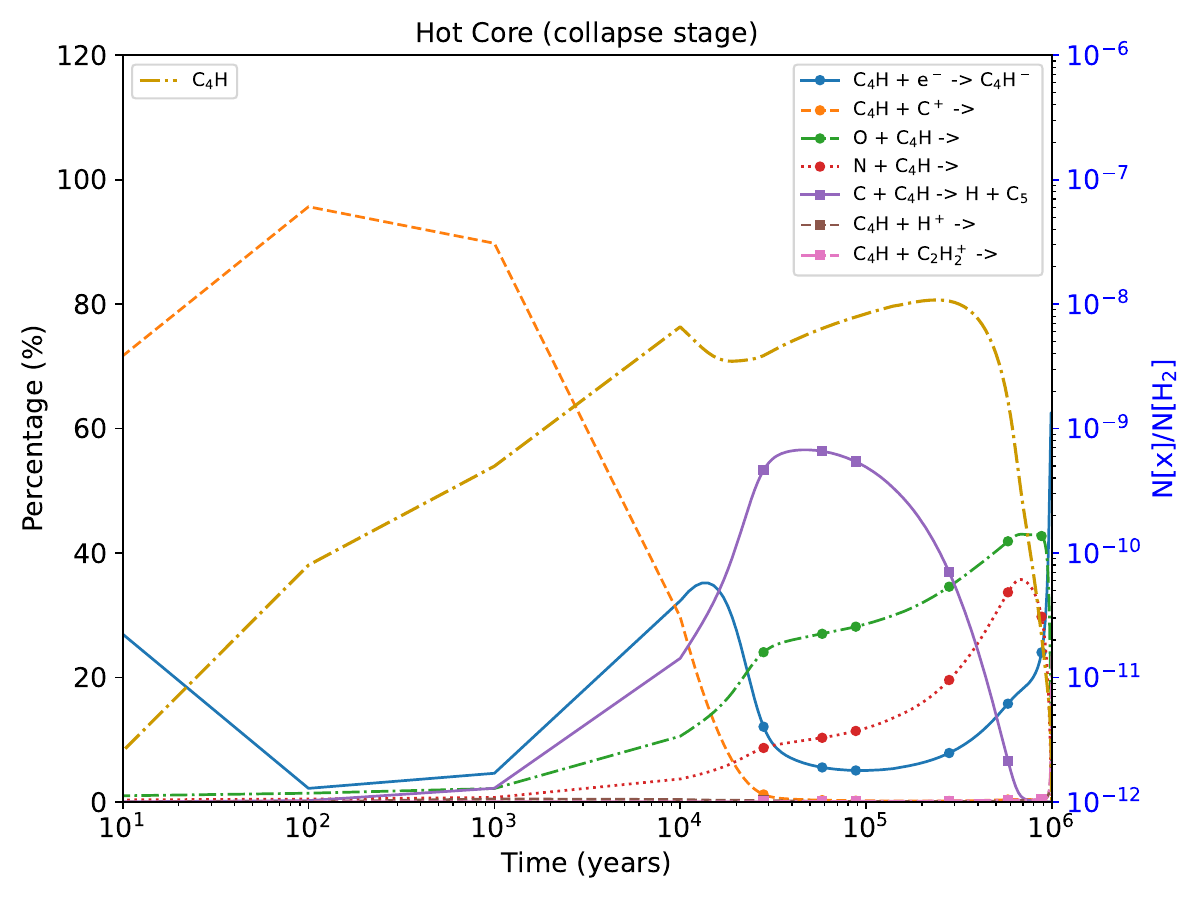}
} 
\subfigure[]{ \label{fig:c4h_des_f} 
\includegraphics[width=0.32\textwidth]{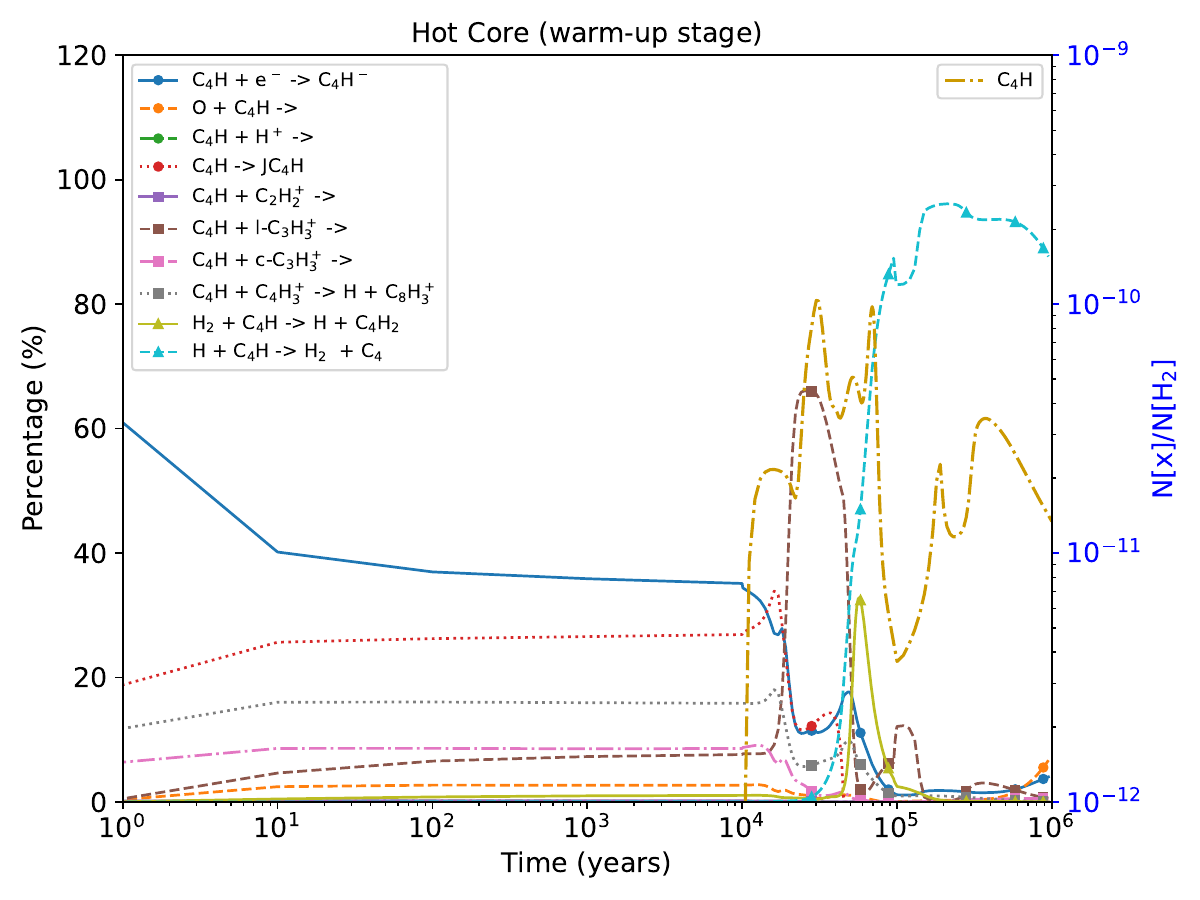}
}

\caption{Net percentage contributions of the formation and destruction pathways of C$_4$H in cold and hot cores. 
(a) Net percentage contributions of the main formation pathways of C$_4$H in cold molecular cloud cores. 
(b) Net percentage contributions of the main formation pathways of C$_4$H during the hot-core collapse stage. 
(c) Net percentage contributions of the main formation pathways of C$_4$H during the hot-core warm-up stage. 
(d) Net percentage contributions of the main destruction pathways of C$_4$H in cold molecular cloud cores. 
(e) Net percentage contributions of the main destruction pathways of C$_4$H during the hot-core collapse stage. 
(f) Net percentage contributions of the main destruction pathways of C$_4$H during the hot-core warm-up stage.}
\label{fig:c4h_pathway}
\end{figure*}

The optically thin H$^{13}$CO$^+$  1--0 isotopologue line was employed to normalise the relative abundances of  c-C$_3$H$_2$ and C$_4$H relative to  H$^{13}$CO$^+$ from their observed intensities.
Figure \ref{model_predicted_abundance} shows the abundances of \textit{c}-C$_3$H$_2$,  C$_4$H,    and HCO$^+$, as the major isotopic molecule of H$^{13}$CO$^+$,  predicted by the models.
The shaded regions indicate the observational abundance ranges of the corresponding species. 
Observations show that the abundance of C$_4$H     and HCO$^+$  in cold cores is approximately two orders of magnitude higher than that in hot cores, while the abundance of \textit{c}-C$_3$H$_2$ in cold cores is about one order of magnitude higher than that in hot cores.
Figures~\ref{fig:c3h2_pathway} and~\ref{fig:c4h_pathway} show the net percentage contributions of the formation and destruction pathways of \textit{c}-C$_3$H$_2$ and C$_4$H, respectively, in cold and hot cores.
Because the chemistry of these species involves a large number of reactions (e.g. more than 100 reactions for C$_4$H), only reaction pathways with contributions greater than 5\% are shown in the figures.
Reaction pathways for which the products are not explicitly labeled correspond to reactions with multiple product branching channels.
Each panel shows the net percentage contributions of the formation or destruction pathways on the left-hand y-axis, while the abundances of \textit{c}-C$_3$H$_2$ or C$_4$H are indicated on the right-hand y-axis.

According to the model results, in cold cores, \textit{c}-C$_3$H$_2$ is mainly formed via the dissociative recombination of C$_3$H$_3^+$ with $e^-$, while C$_4$H is primarily produced through the dissociative recombination of C$_4$H$_2^+$ with $e^-$. 
During the collapse stage of the hot-core model, the formation pathways of \textit{c}-C$_3$H$_2$ and C$_4$H are nearly identical to those in cold cores. 
However, as the density increases, both species are almost completely depleted toward the end of the collapse phase, with abundances falling below $10^{-12}$.
In the subsequent warm-up stage, \textit{c}-C$_3$H$_2$ and C$_4$H are gradually reformed. 
Their formation pathways become more complex and differ significantly from those in cold cores. 
Meanwhile, the increasing temperature enhances their destruction efficiencies, which ultimately leads to lower abundances of both species in hot cores compared to cold cores.
During the warm-up stage, \textit{c}-C$_3$H$_2$ is mainly formed through the dissociative recombination of C$_3$H$_5^+$ with $e^-$, with an additional contribution from the dissociative recombination of C$_3$H$_3^+$. 
For C$_4$H, the dominant formation pathways include reactions involving HC$_5$NH$^+$ with $e^-$, as well as thermal desorption from grain surfaces as the temperature increases.

In both cold-core and hot-core environments, the destruction of \textit{c}-C$_3$H$_2$ and C$_4$H is dominated by neutral--neutral reactions. 
In cold cores, however, the low temperatures significantly reduce the efficiency of neutral--neutral destruction pathways. 
As a result, the destruction rates are lower than the formation rates driven by dissociative recombination, allowing the abundances of \textit{c}-C$_3$H$_2$ and C$_4$H to accumulate and reach higher levels than in hot cores.

In hot cores, the increasing gas density during the collapse stage and the rising temperature during the warm-up stage both enhance neutral--neutral destruction reactions. 
Consequently, the destruction of \textit{c}-C$_3$H$_2$ and C$_4$H becomes more efficient, leading to lower abundances of these species compared to those in cold cores.

 \section{Summary and conclusions}
  \label{sec:summary}

We observed C$_4$H  9--8 lines, \textit{c}-C$_3$H$_2$ 2--1, H$^{13}$CO$^+$  1--0, and  H42$\alpha$ toward 22 late-stage massive star-forming regions with the IRAM 30m telescope. Column densities and abundance ratios were derived under local thermodynamic equilibrium, combined with data from 19 cold cores to explore evolutionary trends.

C$_4$H 9--8 lines were detected in 17 regions, specifically: G015.03--00.67 (green and cyan), G023.43--00.18 (green), G034.39+00.22 (green), G049.48--00.36 (green and cyan), G049.48--00.38 (green and cyan), G075.76+00.33 (green), G081.75+00.59-1 (green and cyan), G081.75+00.59-2 (green and cyan), G081.87+00.78 (green), G133.94+01.06 (green), and G192.60--00.04 (green and cyan), out of 31 regions toward 22 sources, while \textit{c}-C$_3$H$_2$  2--1and  H$^{13}$CO$^+$ 1--0 were detected in all regions.
 Both carbon-chain molecules are spatially concentrated at the edges of H42$\alpha$ emission regions, avoiding harsh H\,II region core environments. The abundance ratios in late-stage regions are:  C$_4$H/H$^{13}$CO$^+$ (0.170-1.77, median $\sim 0.57$),  $c$-C$_3$H$_2$/H$^{13}$CO$^+$  (1.42 - 6.69, median $\sim 4.19$), and C$_4$H/$c$-C$_3$H$_2$  (0.07- 0.29, median $\sim 0.13$), all significantly lower than in cold cores.

NAUTILUS three-phase model simulations confirm this evolutionary trend, predicting   higher abundances in cold cores by one to two orders of magnitude than in hot cores. Formation mechanisms shift from dissociative recombination (cold cores) to gas-phase reactions and grain desorption (late-stage regions), with destruction enhanced by neutral-neutral reactions in high-temperature and high-density environments.

\begin{acknowledgements}

This work is supported by National Key R$\&$D Program of China under grant 2023YFA1608204, the National Natural Science Foundation of China grant 12550003 and the Guangxi Talent Programme (Highland of Innovation Talents). This work is based on observations carried out under project number 042-19, with the IRAM 30m telescope. 
N. Y.-T. acknowledges support  from the National Natural Science Foundation of China (grant Nos. 12473023), the University Annual Scientific Research Plan of Anhui Province (Nos. 2023AH030052).
Y. L. acknowledges support from  the Leading Innovation and Entrepreneurship Team of Zhejiang Province of China (grant No. 2023R01008) and the Key R$\&$D Program of Zhejiang, China (grant No. 2024SSYS0012).
D.H.-Q.  acknowledges support from the  National Key R$\&$D program of China grant (2025YFE0108200) and  the National Natural Science Foundation of China (No. 12373026).
K.Y. acknowledges supports from the National Natural Science Foundation of China under Grant Number 12503031, the Postdoctoral Fellowship Program of CPSF under Grant Number GZC20252099, the Shanghai Post-doctoral Excellence Program (No. 2024379), the Natural Science Foundation of Shanghai (No. 25ZR1402267) and the Yangyang Development Fund.
IRAM is supported by INSU/CNRS (France), MPG (Germany) and IGN (Spain).

\end{acknowledgements}

 \onecolumn
\begin{appendix}
\section{The spatial distribution maps and spectral lines of all sources}

\begin{figure}[h]
\begin{center}
   
 \includegraphics[width=0.25\columnwidth]{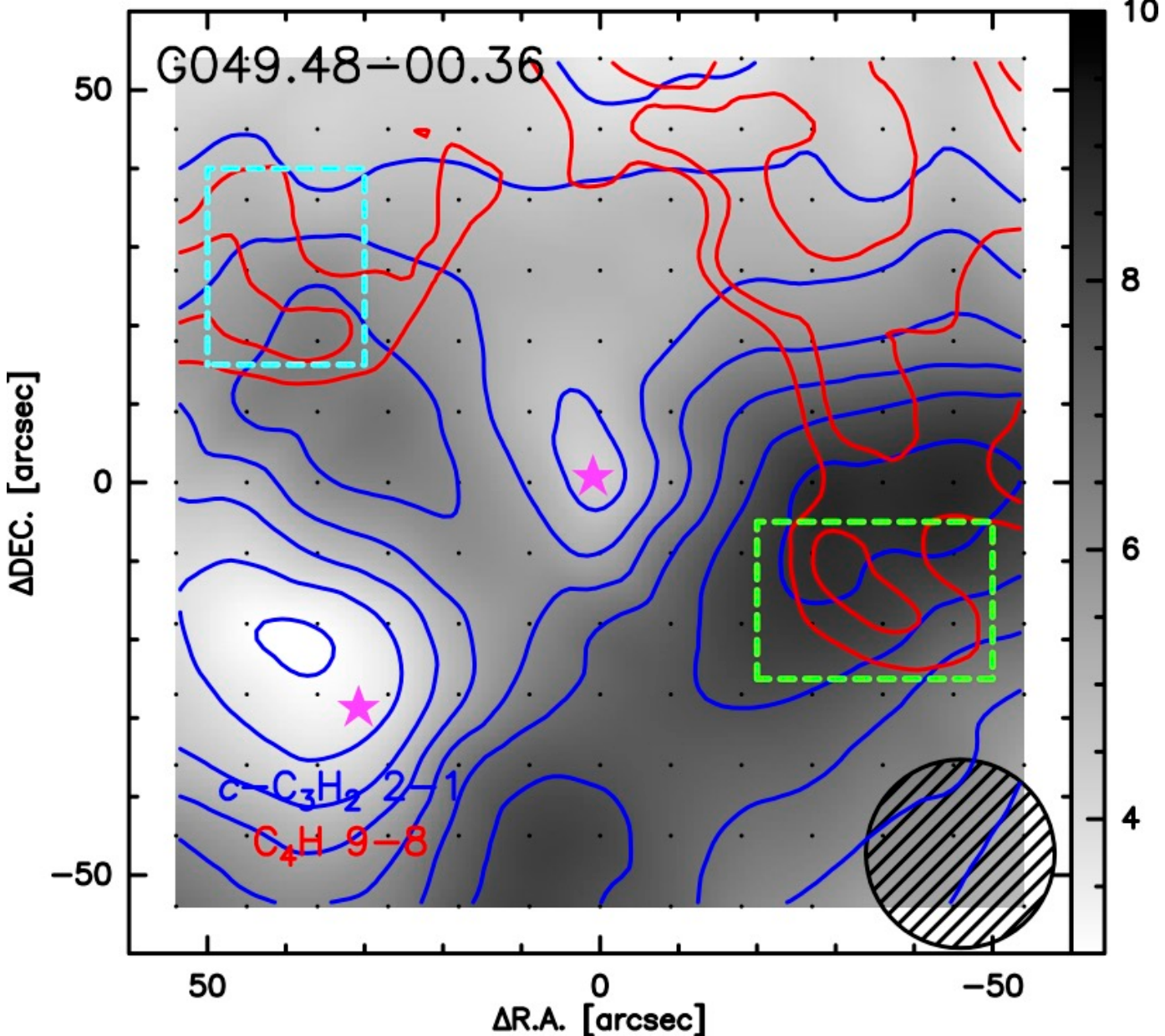}
\includegraphics[width=0.20\columnwidth]{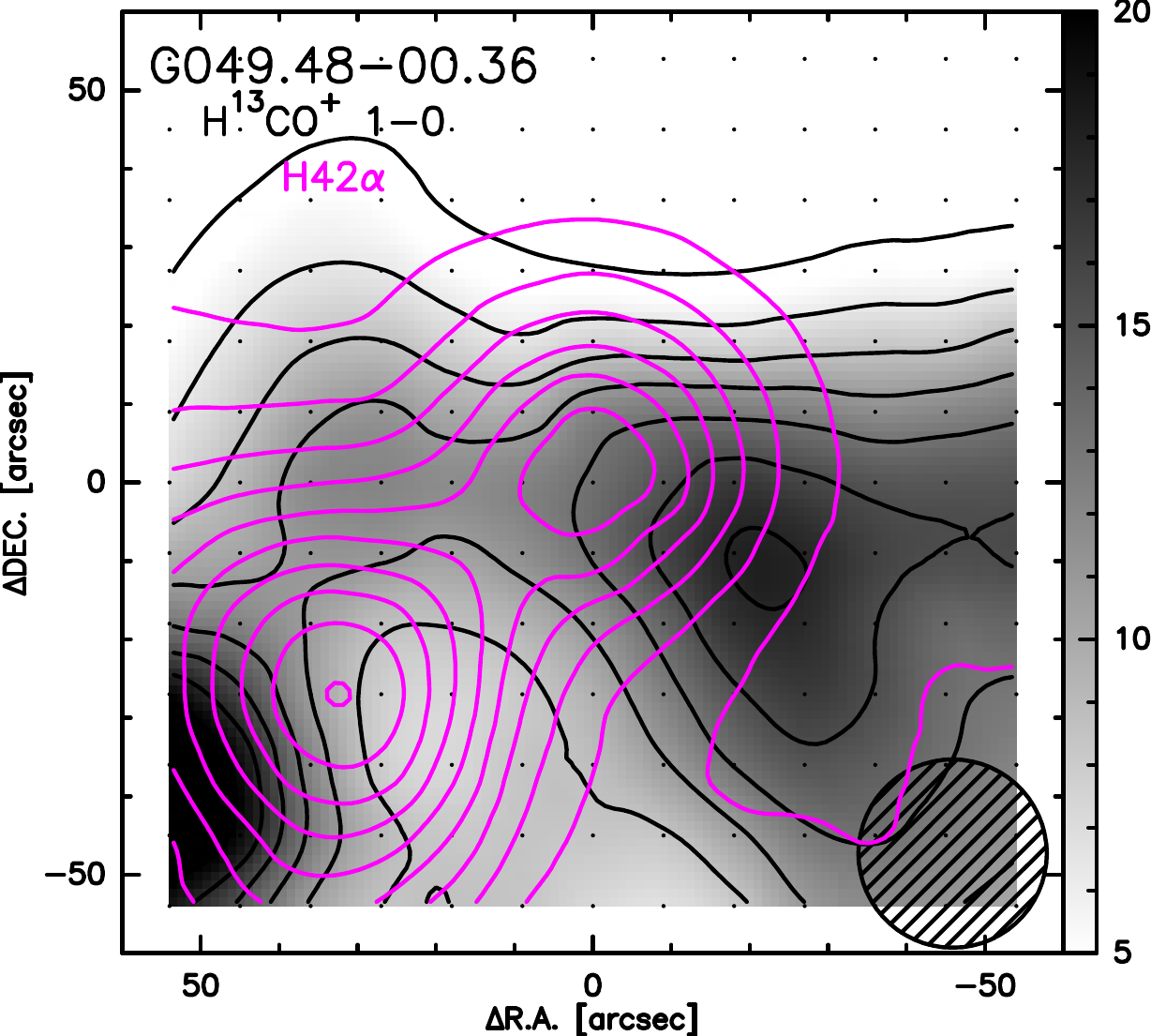}
\includegraphics[width=0.25\columnwidth]{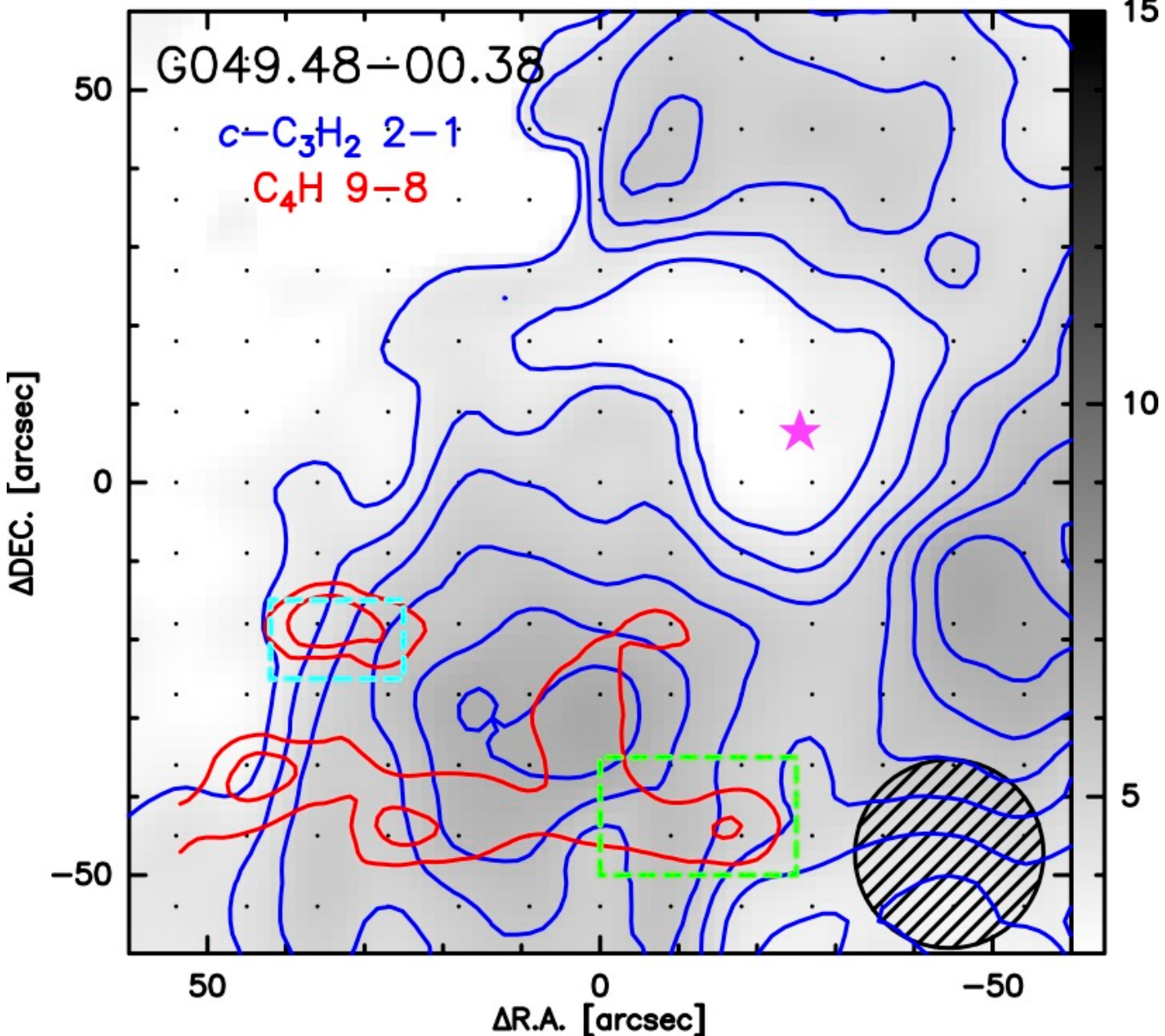}
\includegraphics[width=0.20\columnwidth]{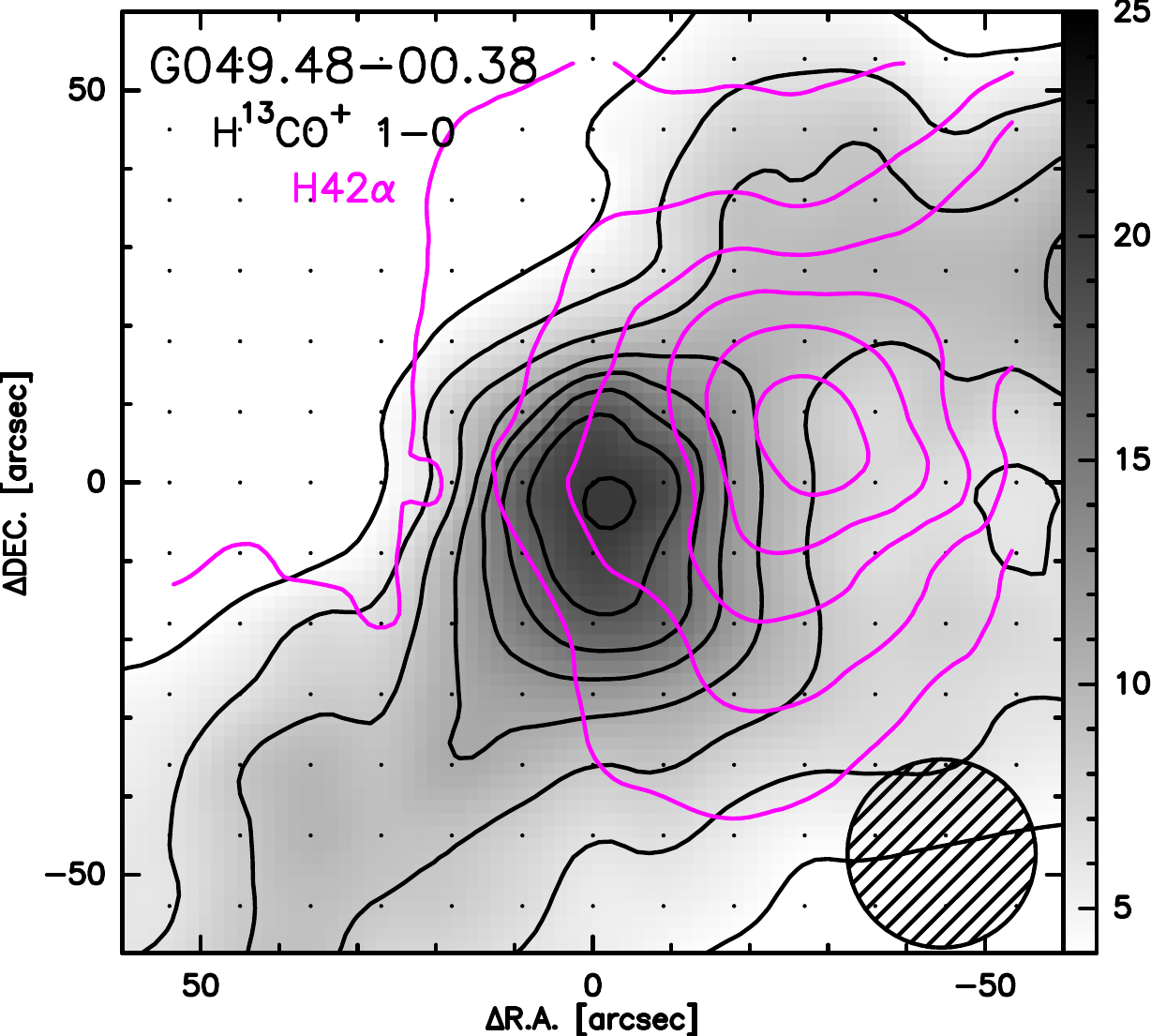}

\includegraphics[width=0.22\columnwidth]{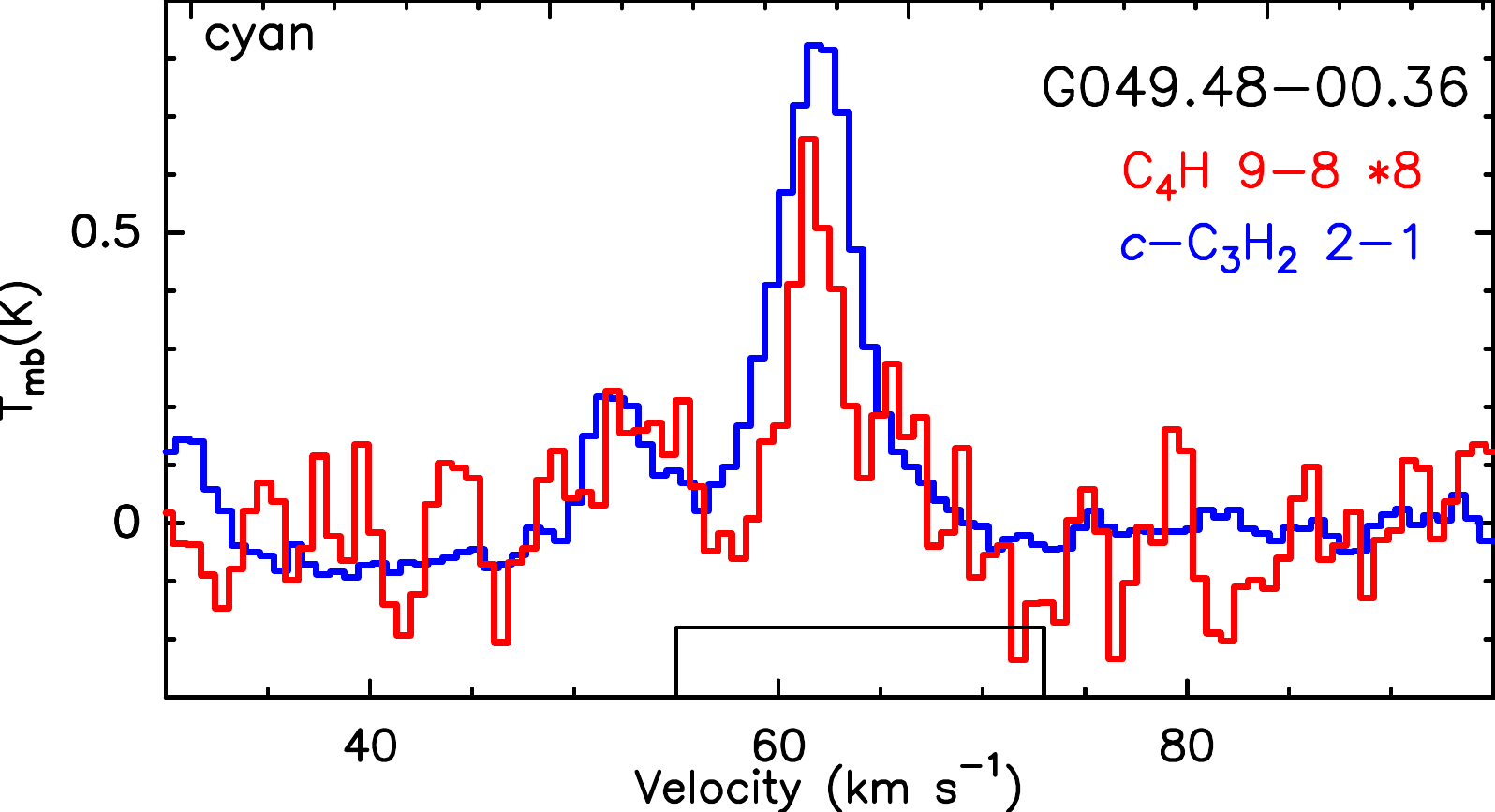}
\includegraphics[width=0.22\columnwidth]{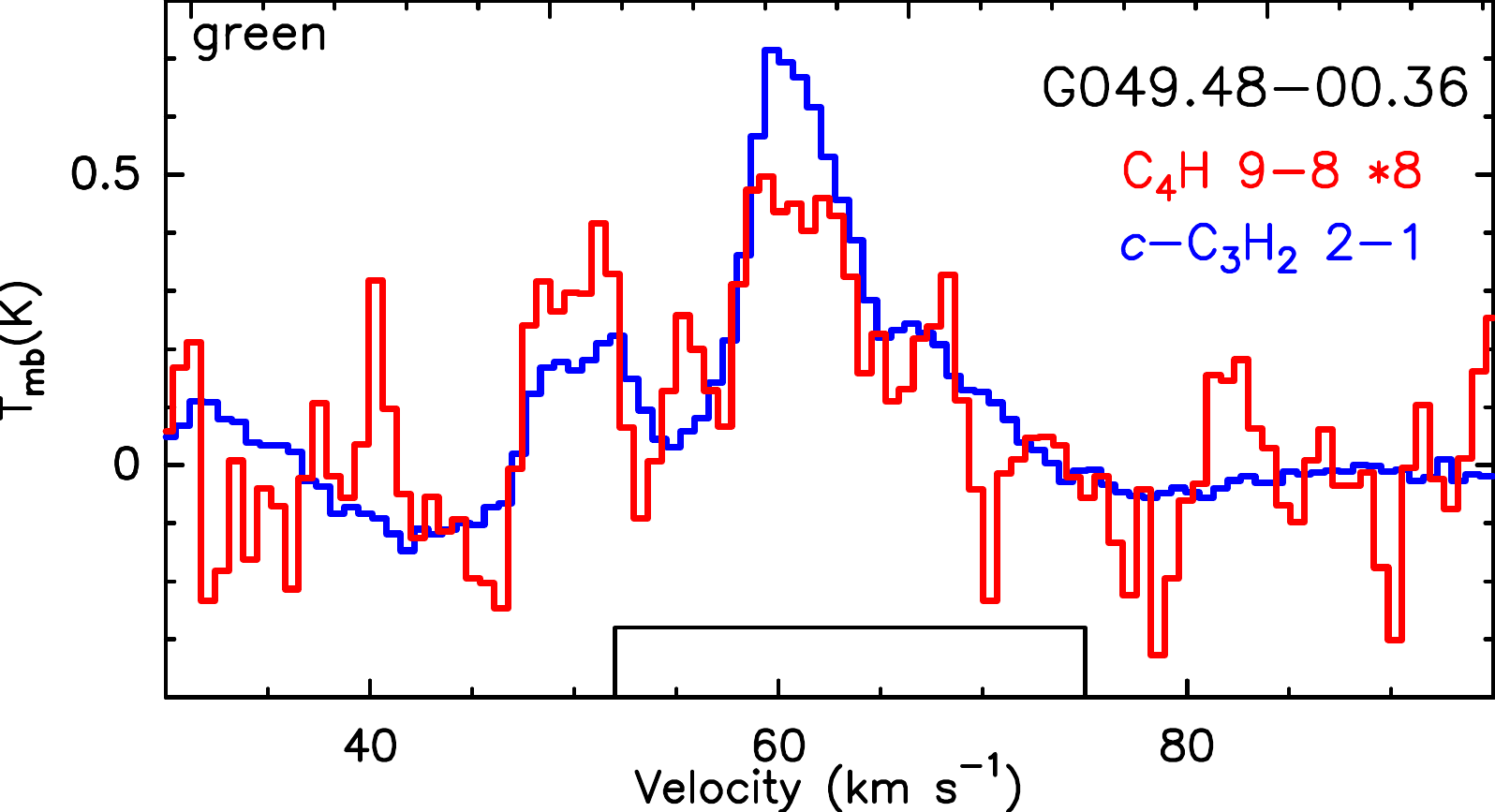}
\includegraphics[width=0.22\columnwidth]{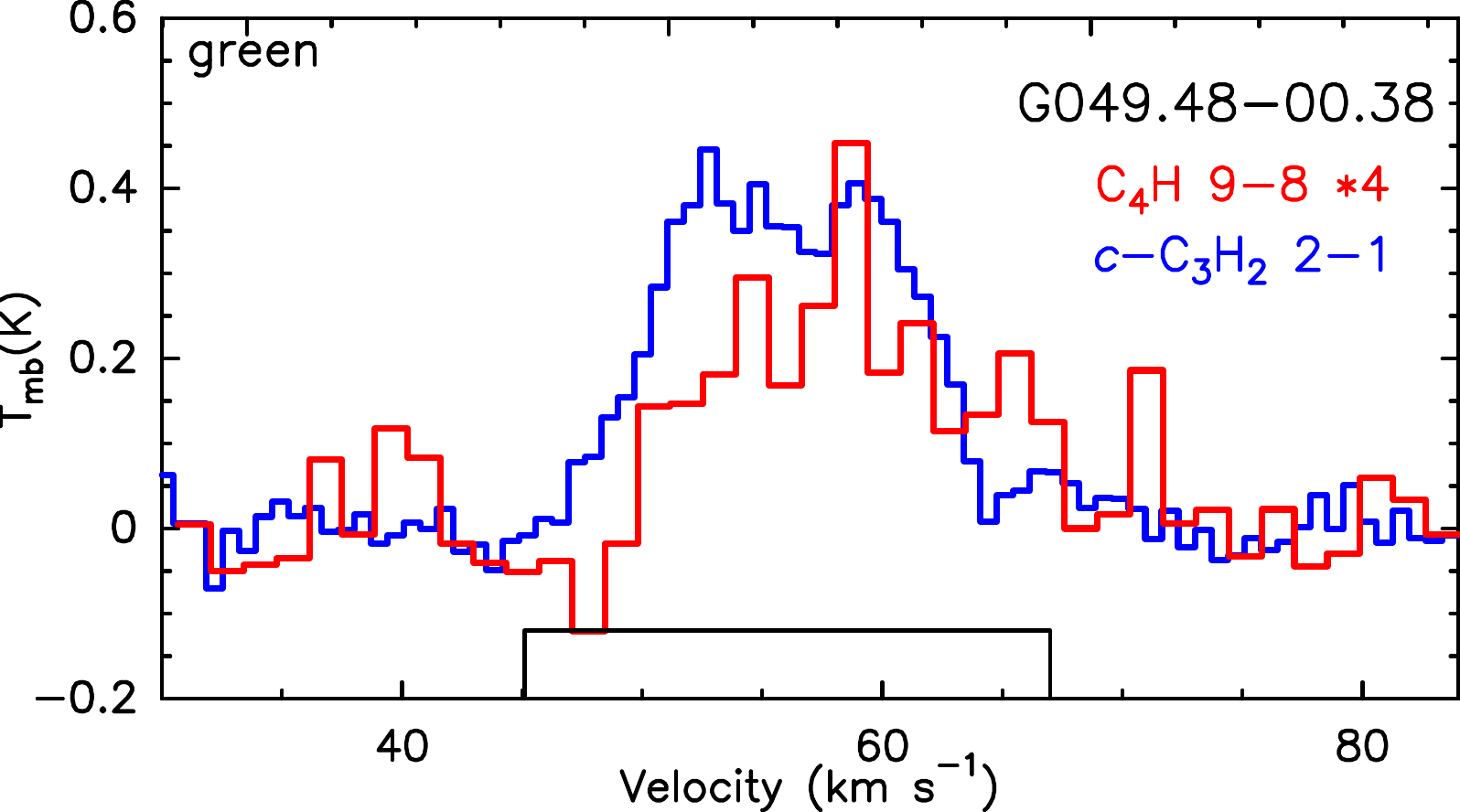}
\includegraphics[width=0.22\columnwidth]{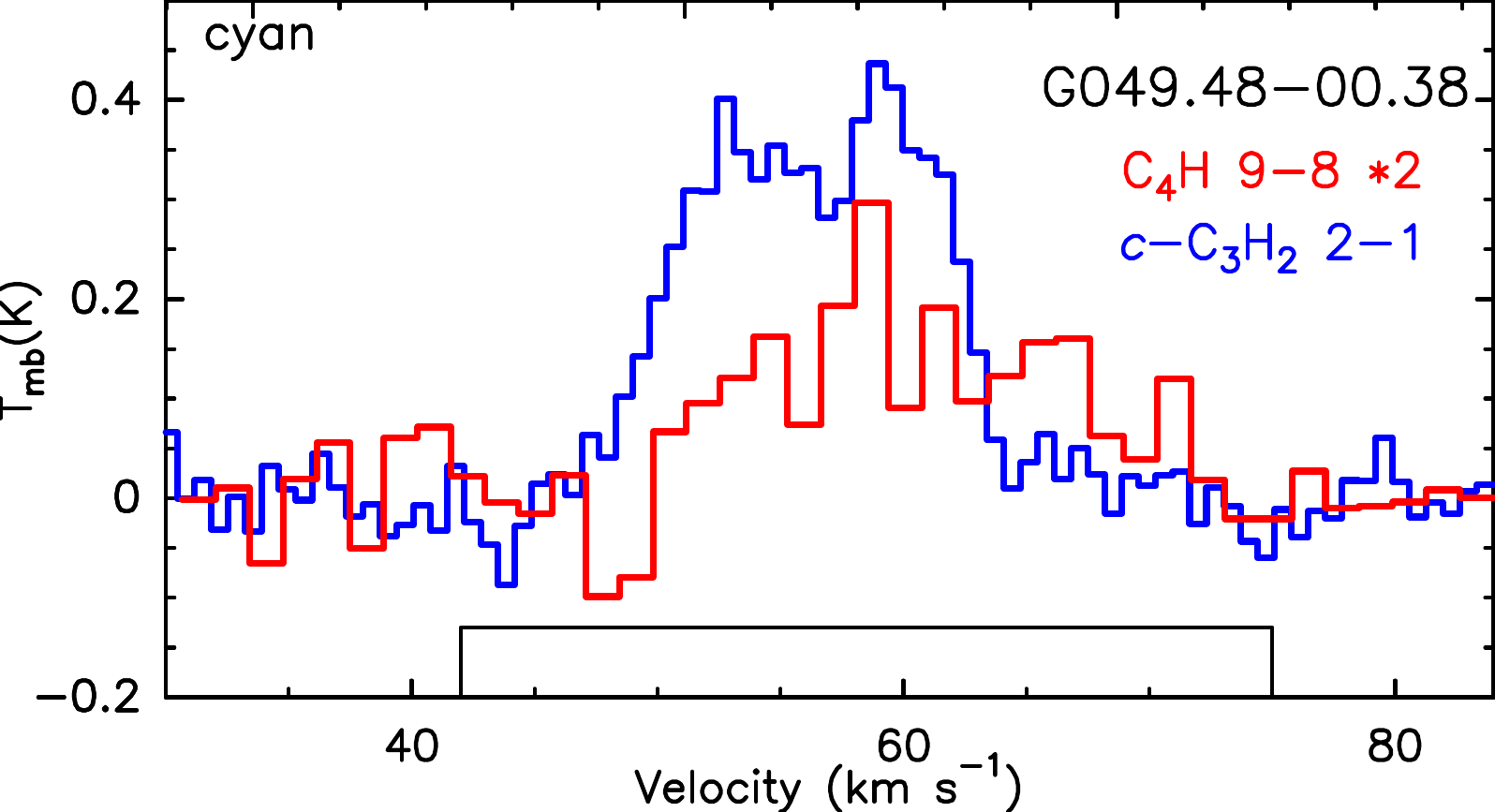}

 \includegraphics[width=0.25\columnwidth]{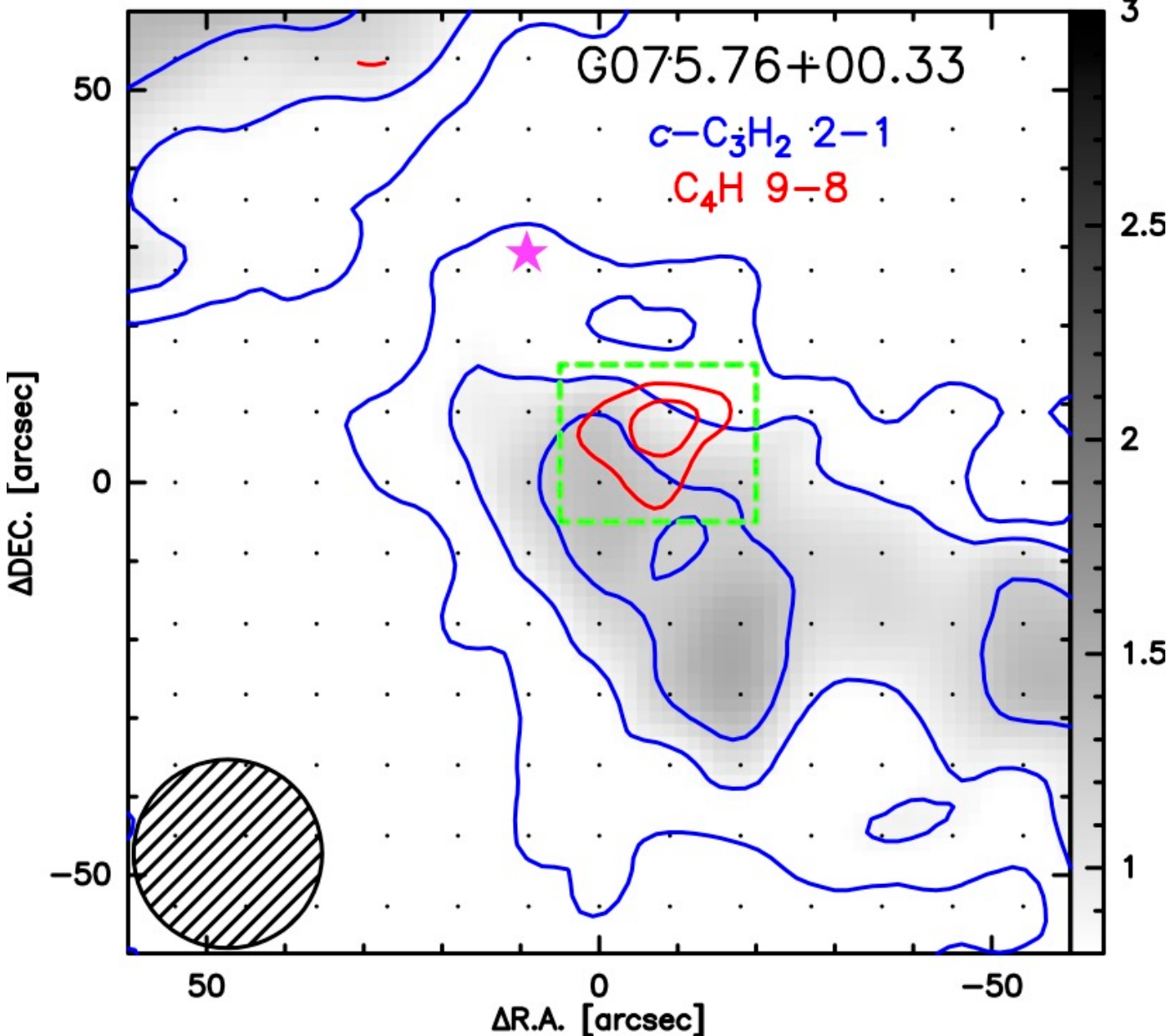}
\includegraphics[width=0.20\columnwidth]{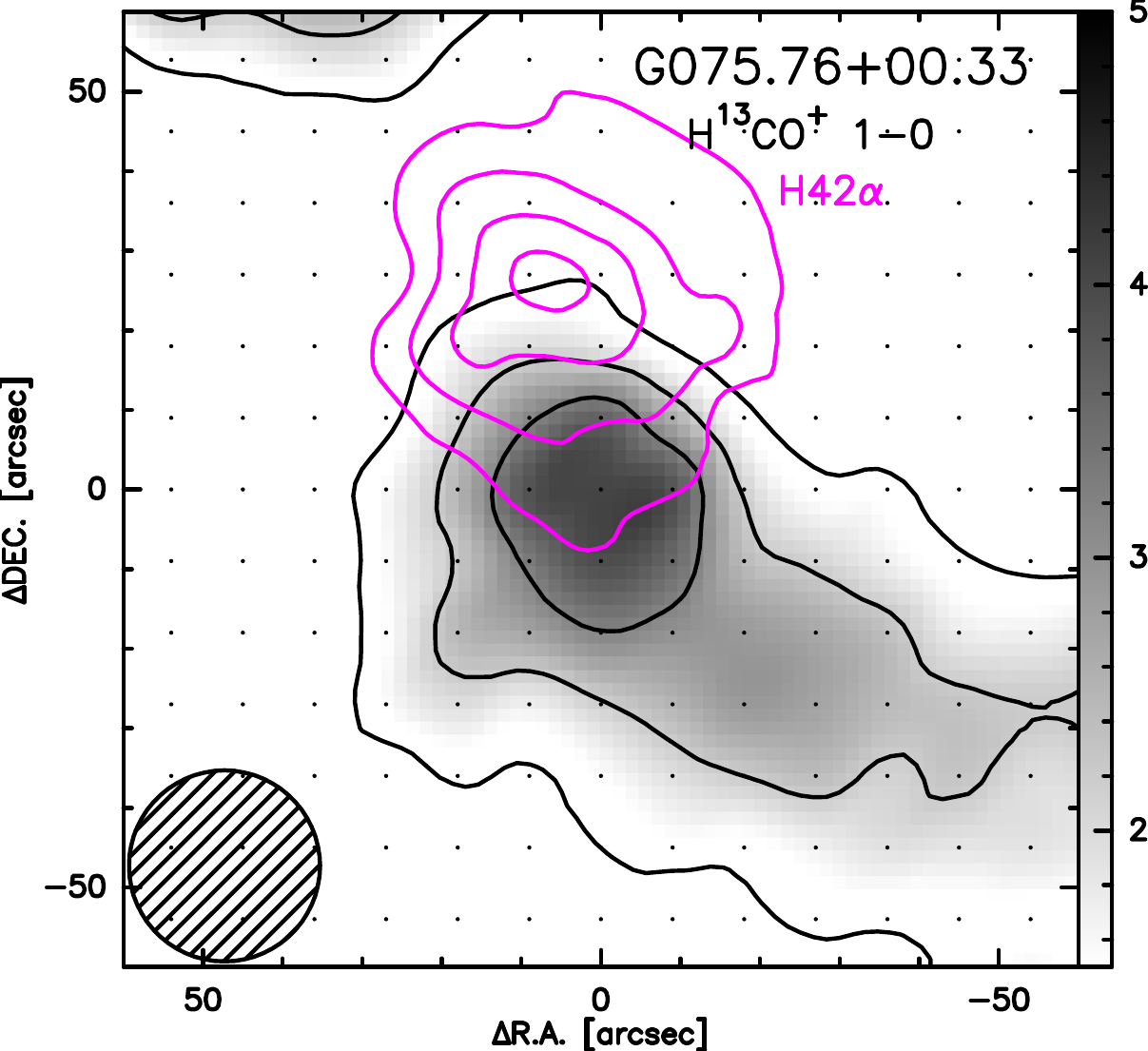}
\includegraphics[width=0.22\columnwidth]{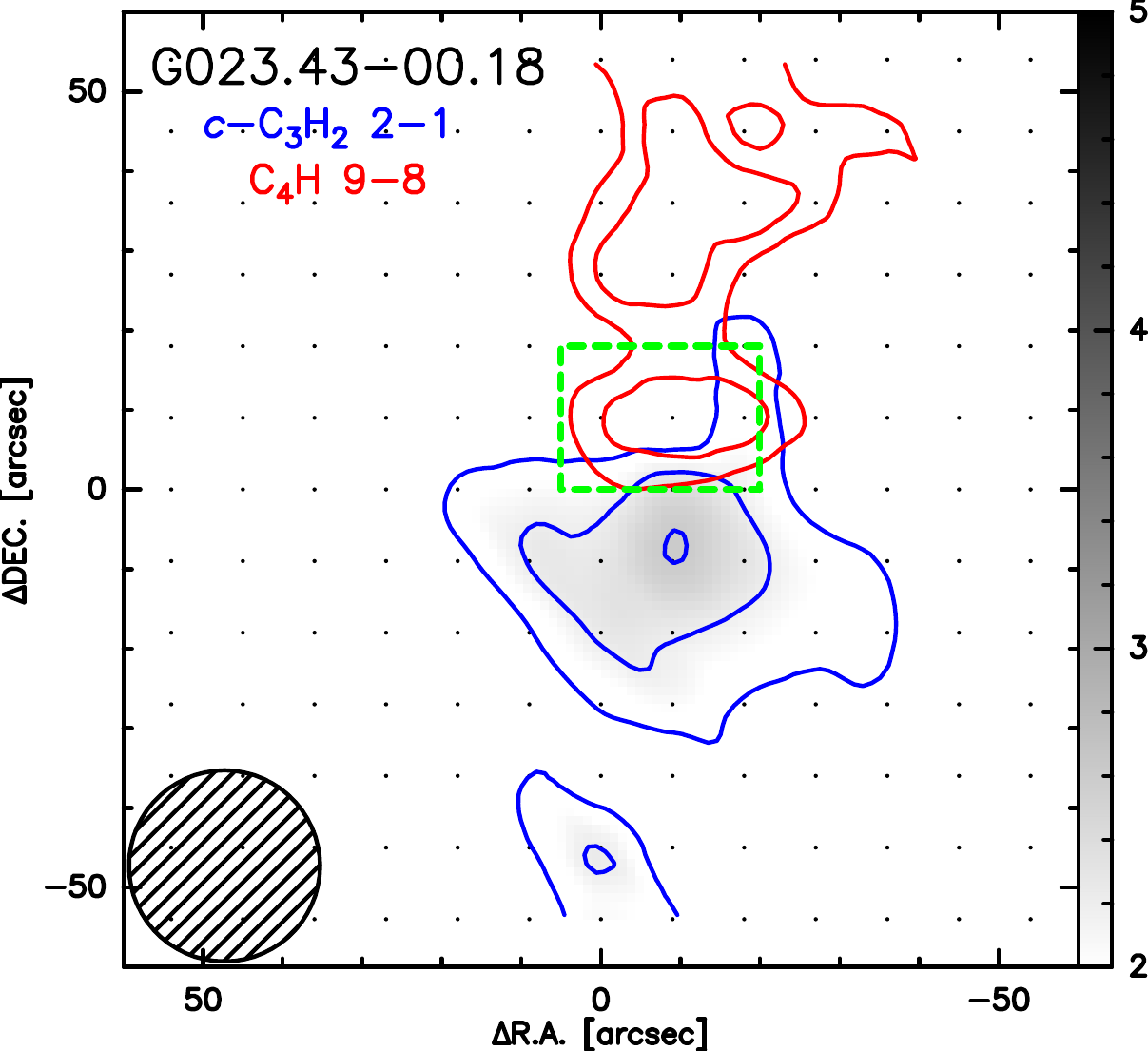}
\includegraphics[width=0.22\columnwidth]{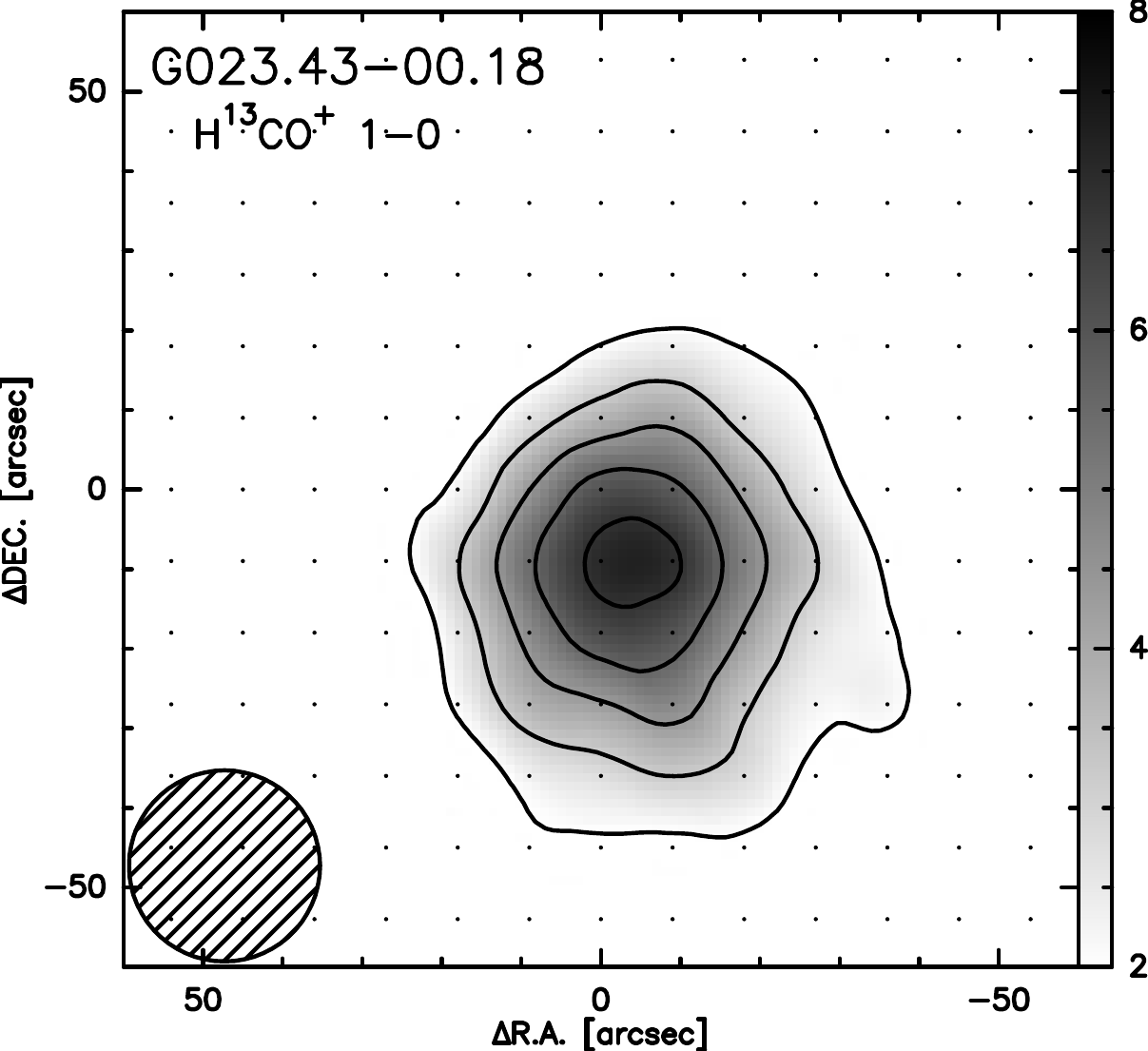}
 
\includegraphics[width=0.22\columnwidth]{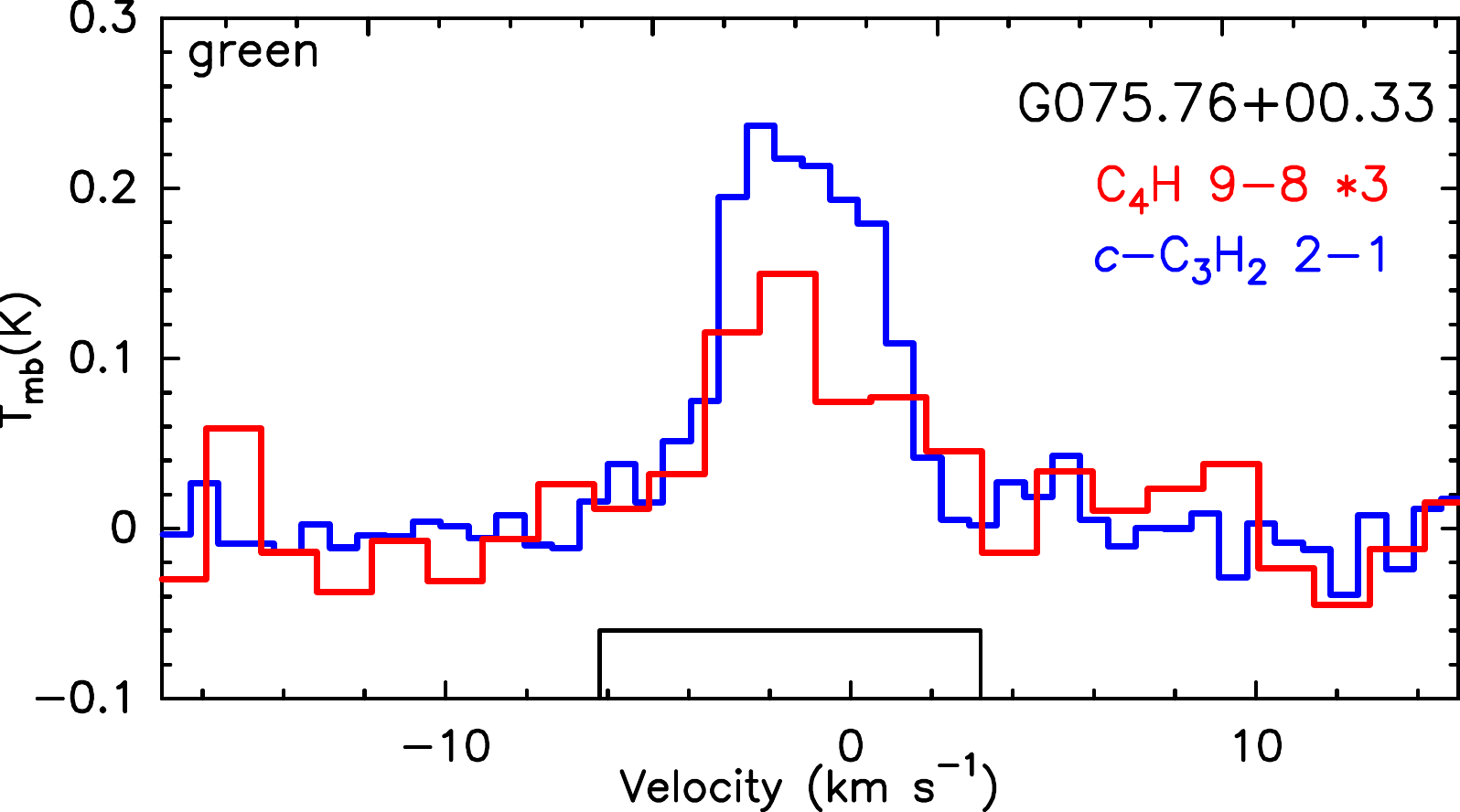}
\includegraphics[width=0.22\columnwidth]{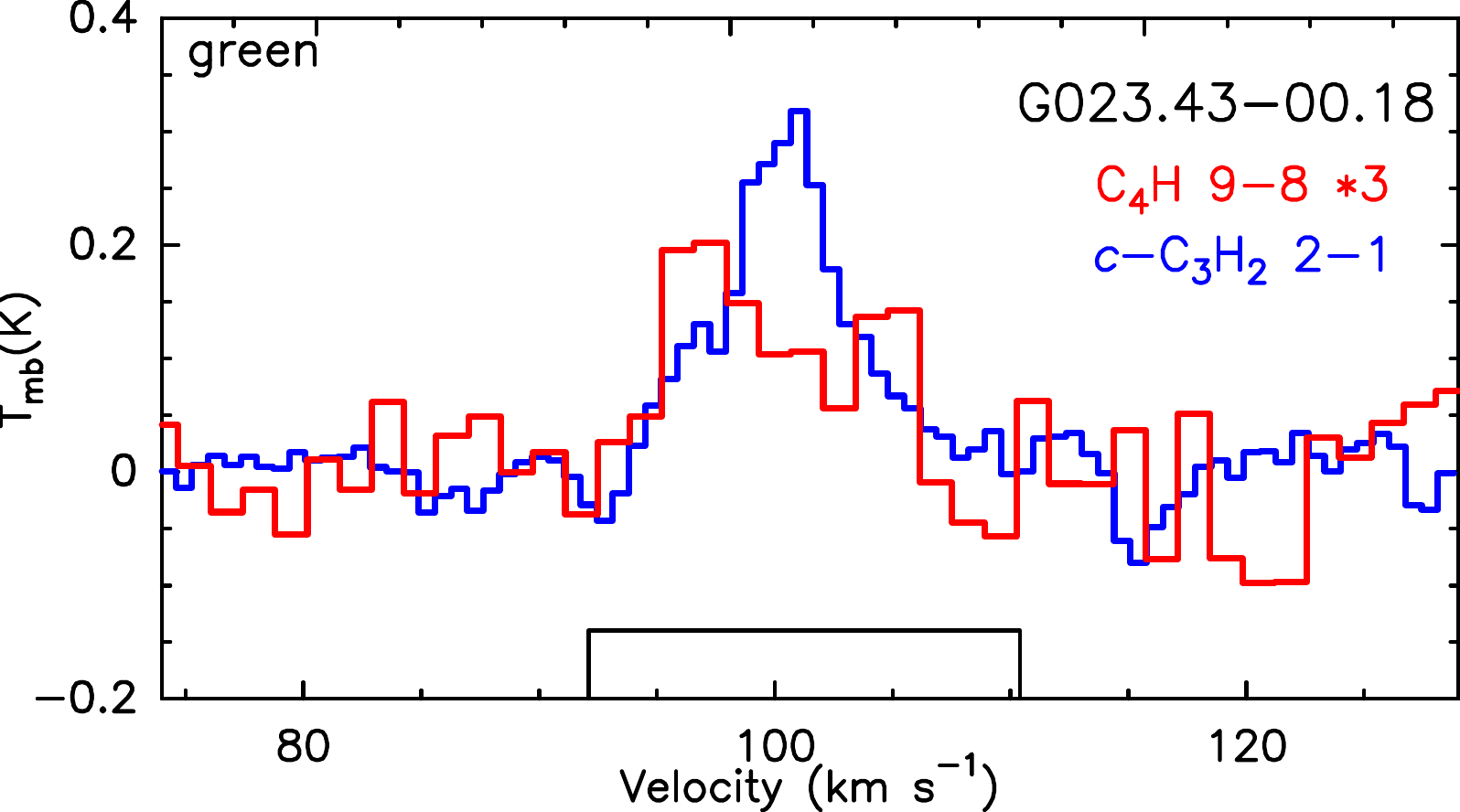}
\includegraphics[width=0.22\columnwidth]{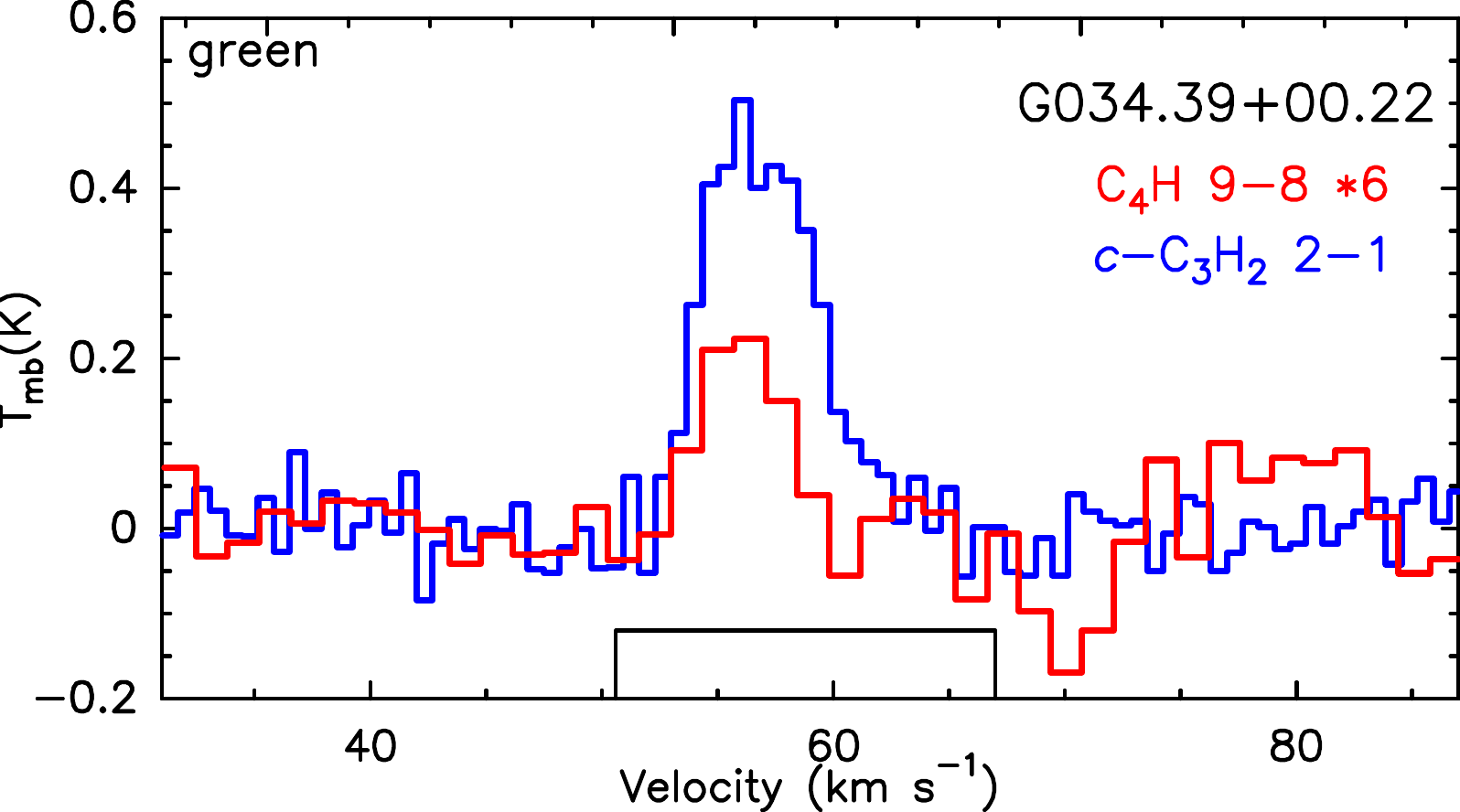}
  \includegraphics[width=0.22\columnwidth]{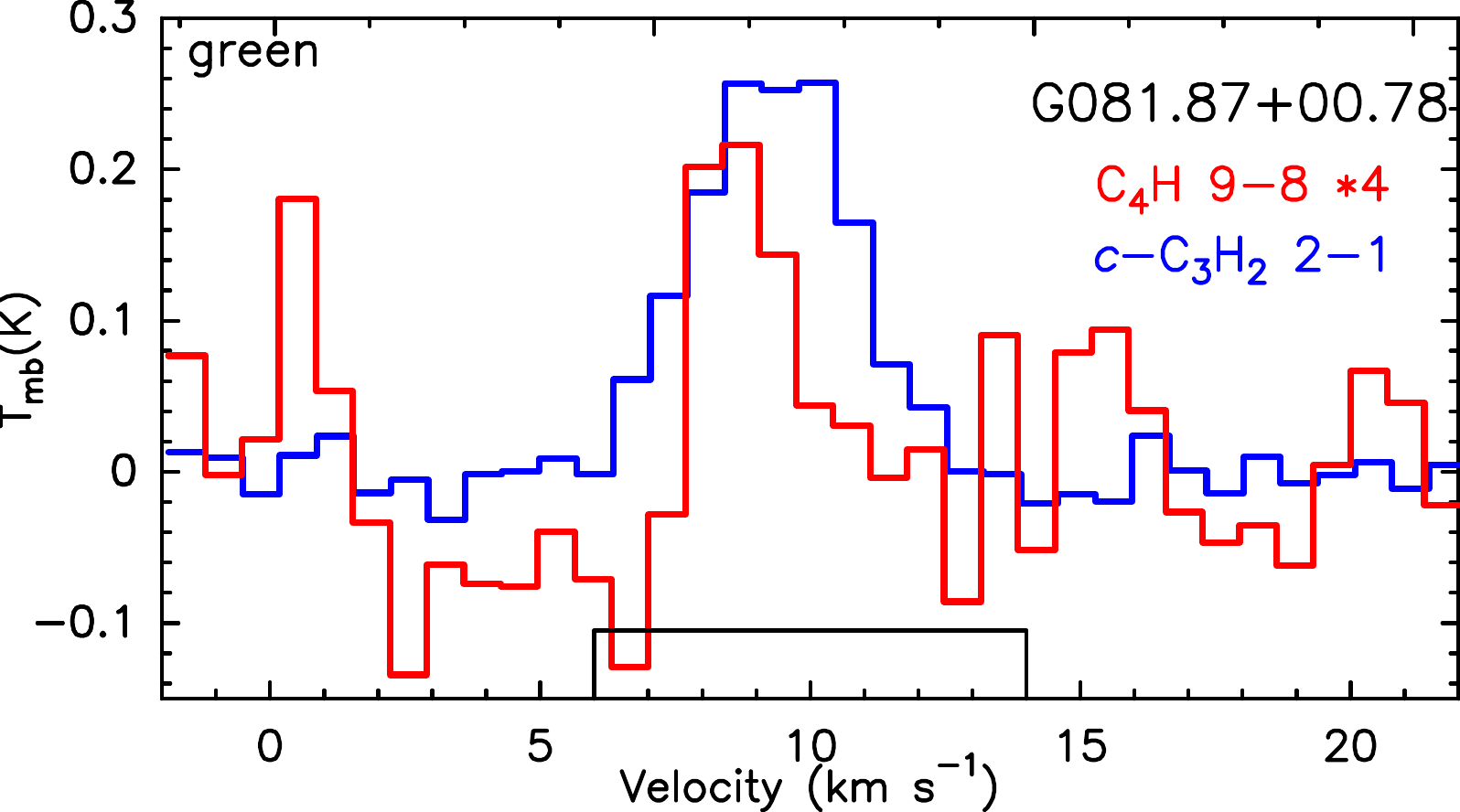}

 \includegraphics[width=0.22\columnwidth]{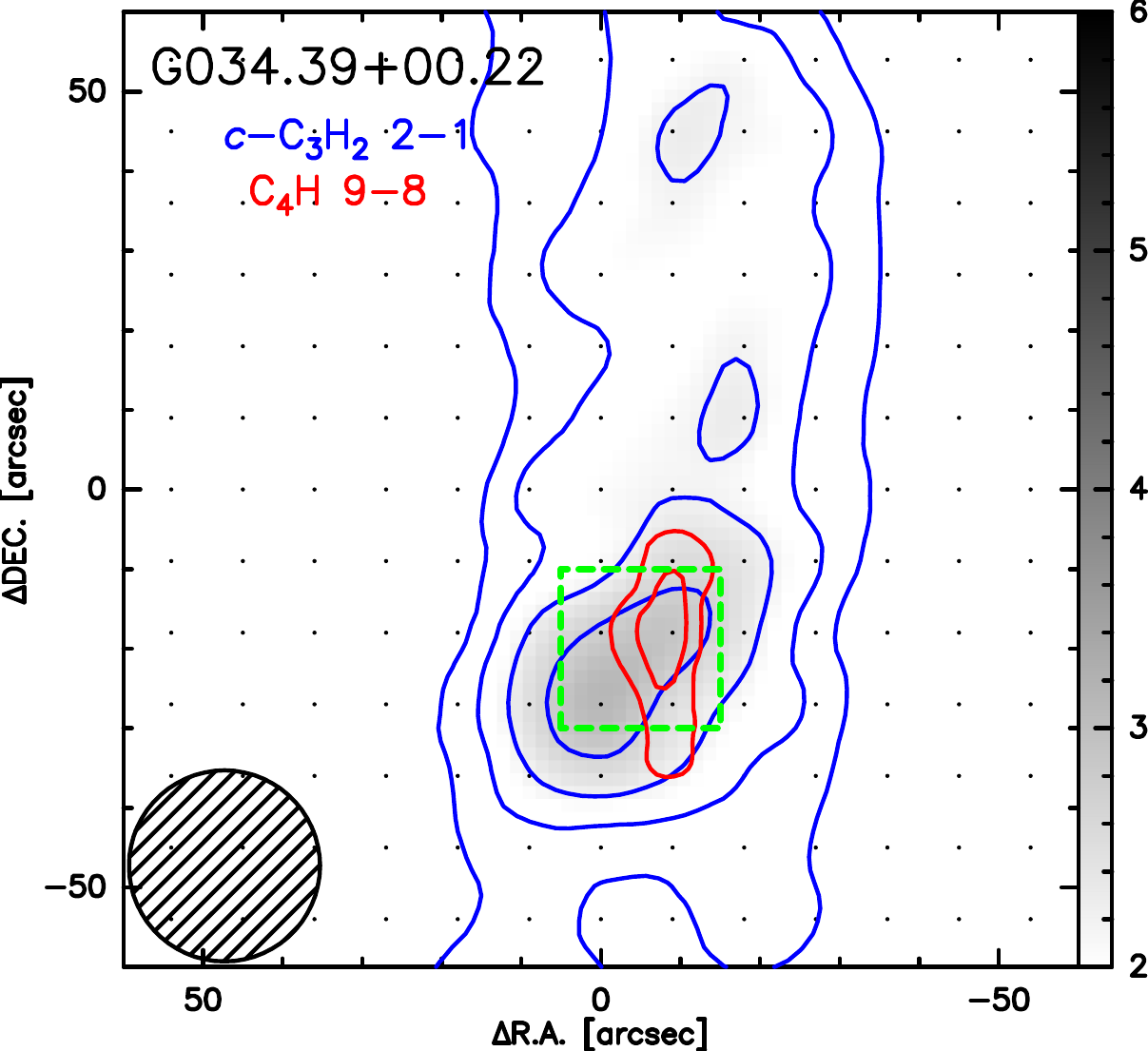}
 \includegraphics[width=0.22\columnwidth]{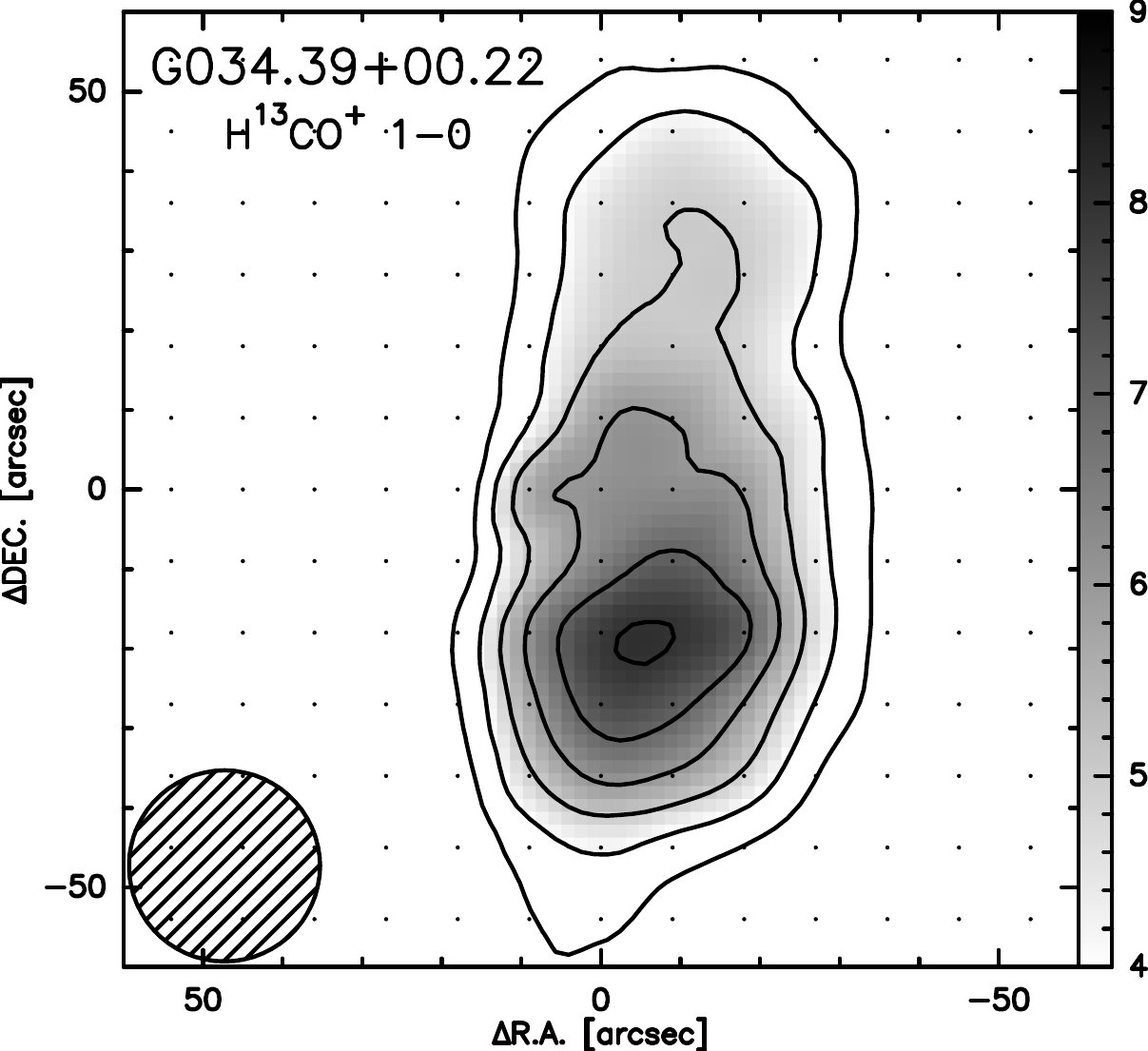}
  \includegraphics[width=0.22\columnwidth]{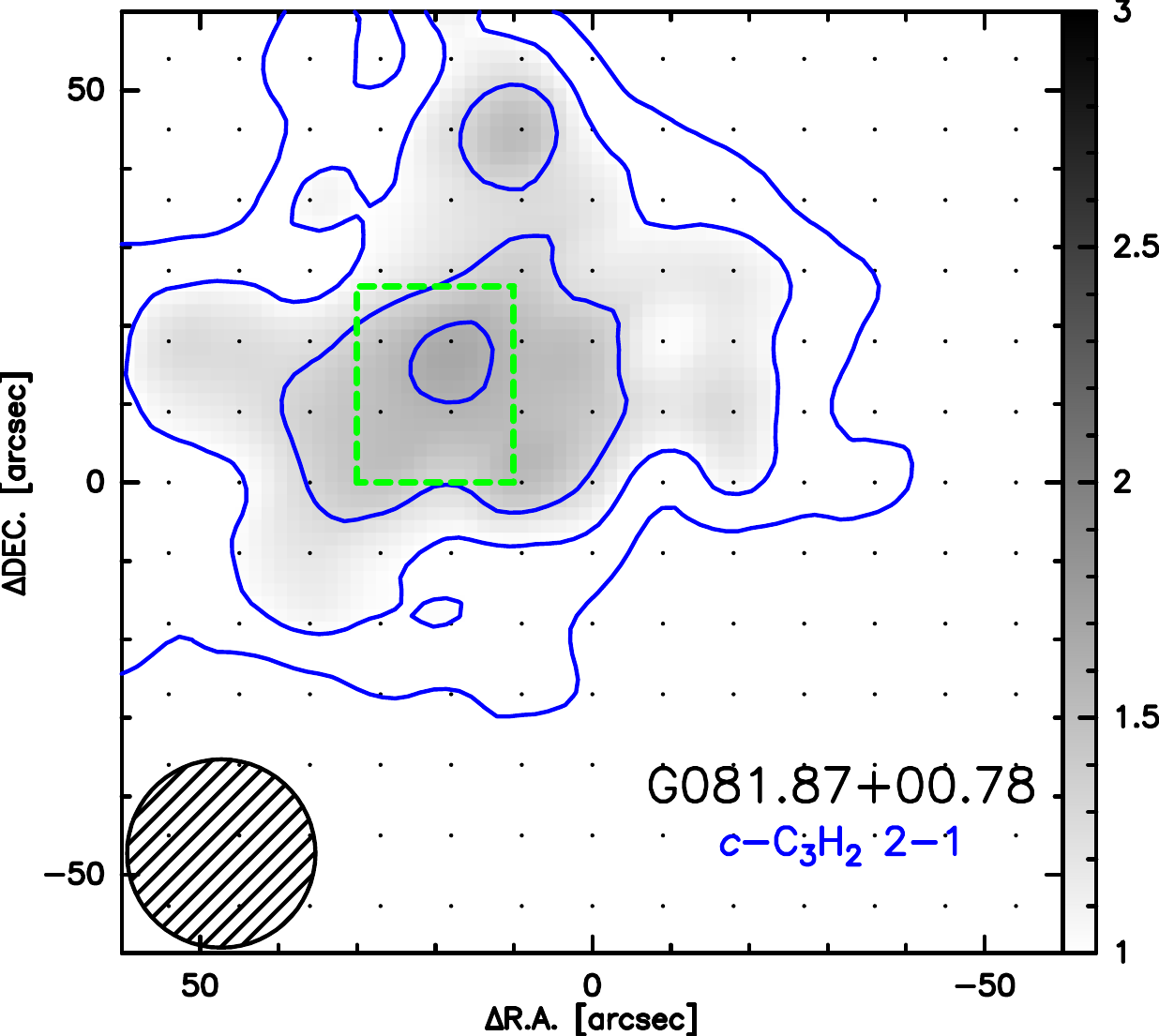}
  \includegraphics[width=0.22\columnwidth]{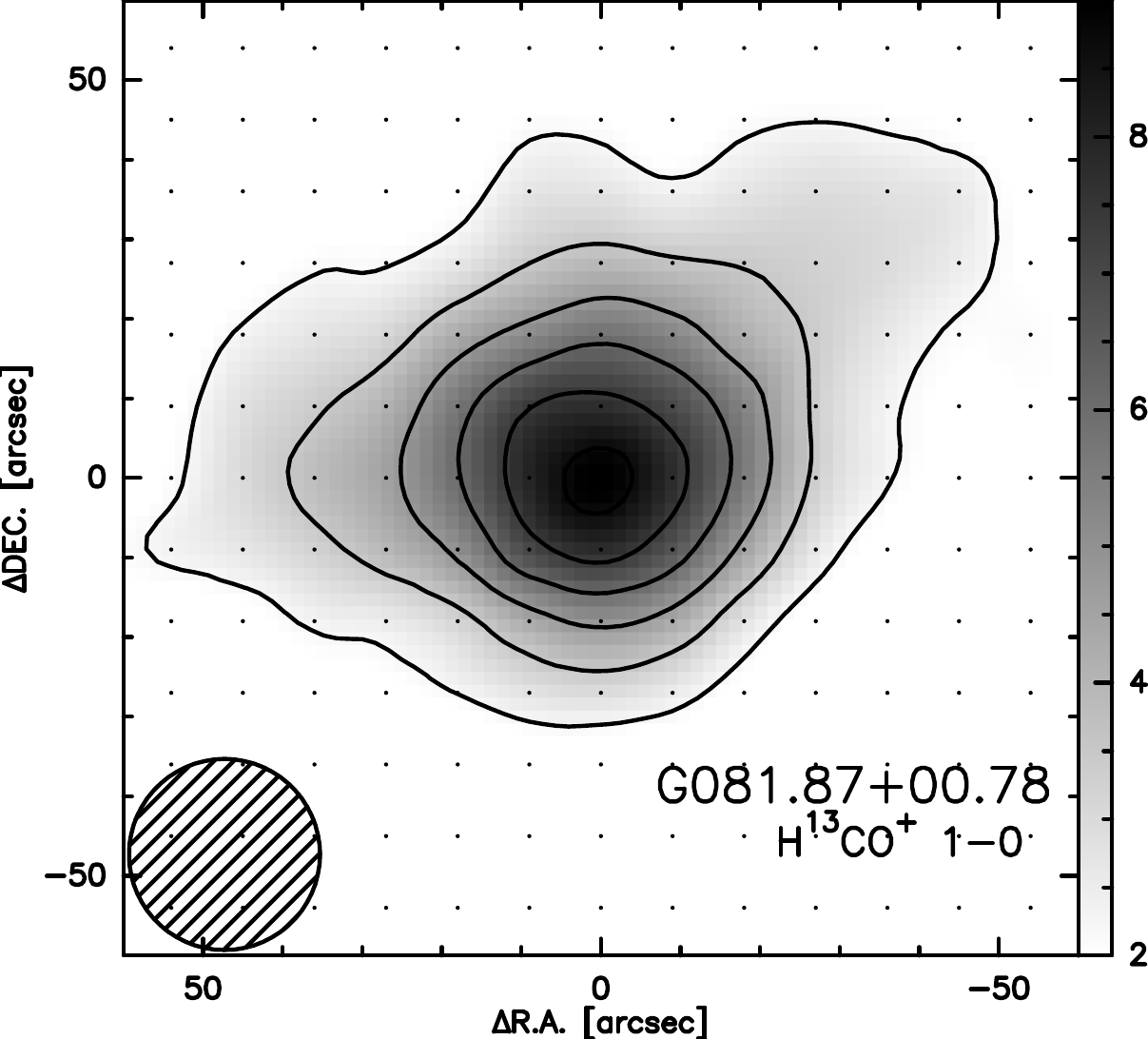}

\caption{ Same as Fig. 1 but for more sources.}
 
\label{Appendix-1}
\end{center}
\end{figure}

 \clearpage
 
\begin{figure}[h]
\begin{center}
\addtocounter{figure}{-1}
\centering

\centering 
\includegraphics[width=0.22\columnwidth]{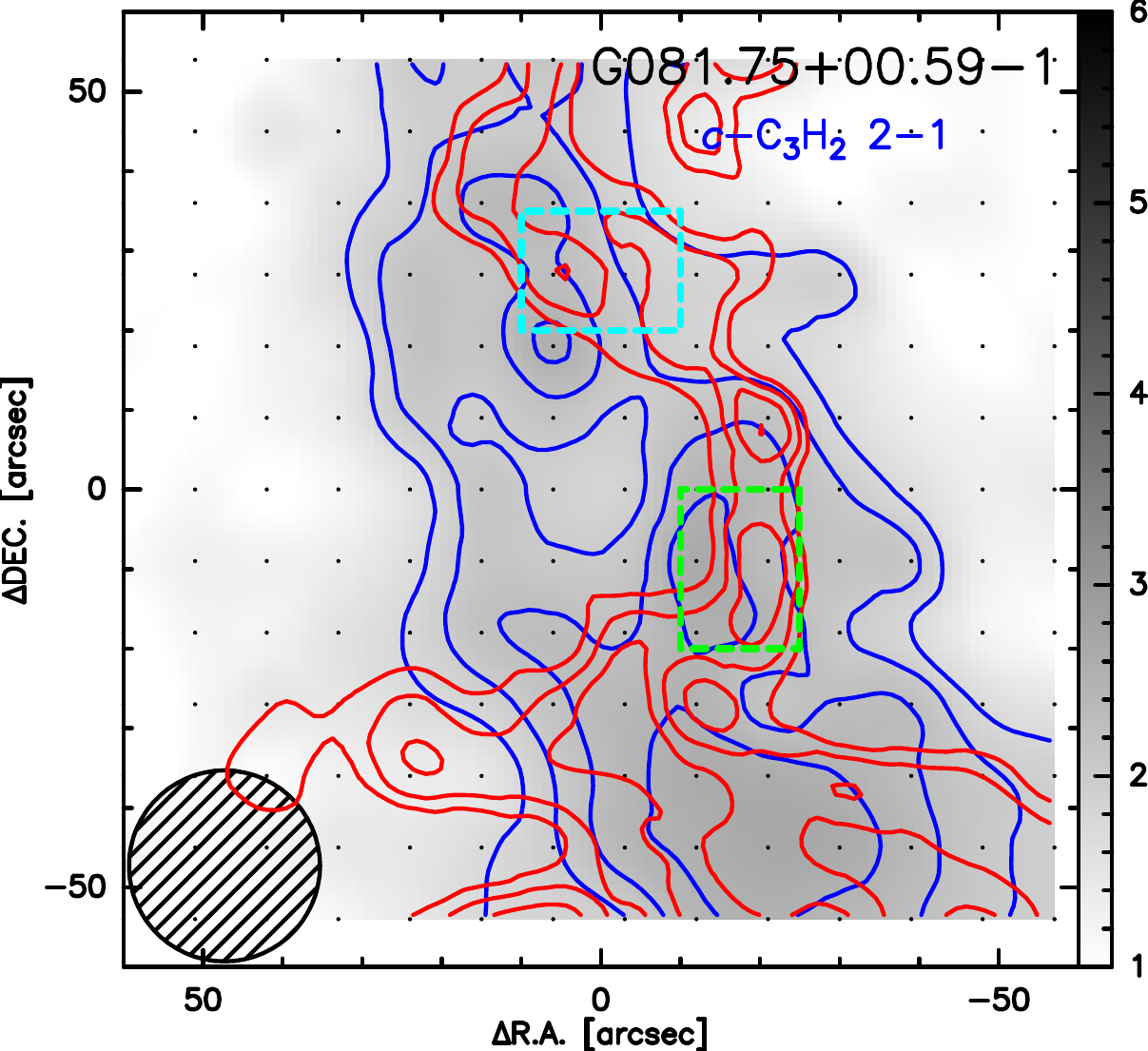}
  \includegraphics[width=0.22\columnwidth]{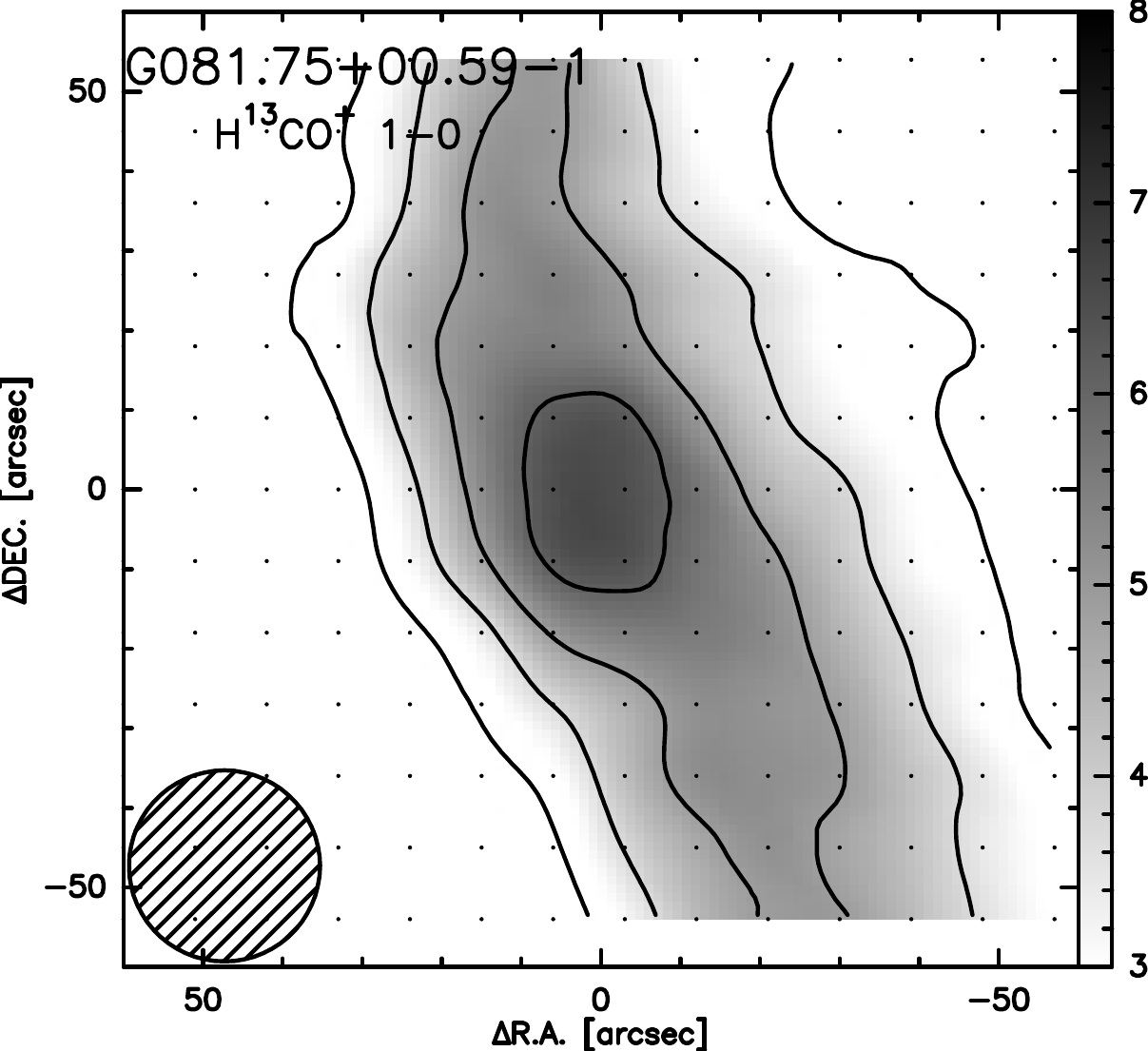}
 \includegraphics[width=0.22\columnwidth]{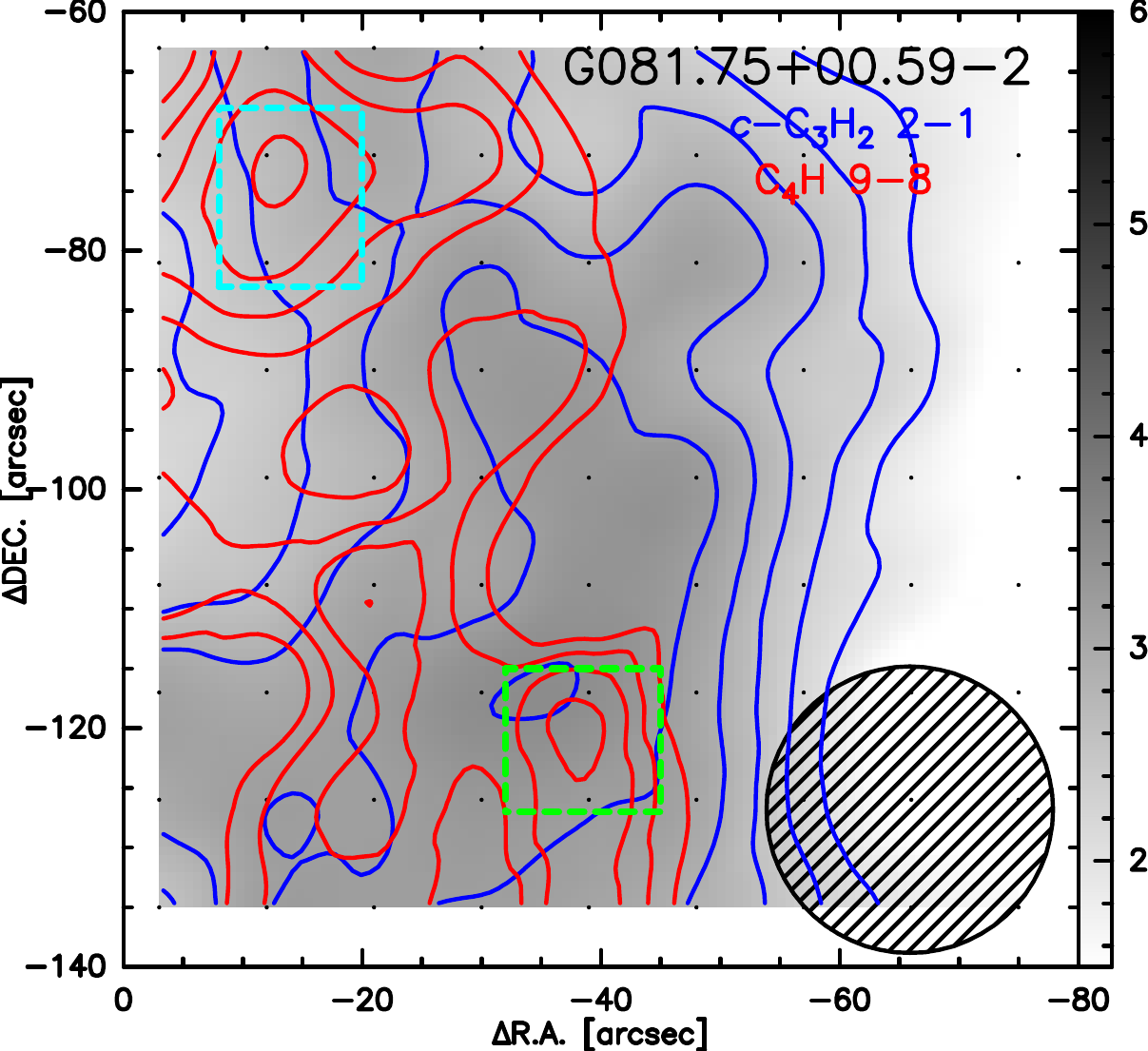}
 \includegraphics[width=0.22\columnwidth]{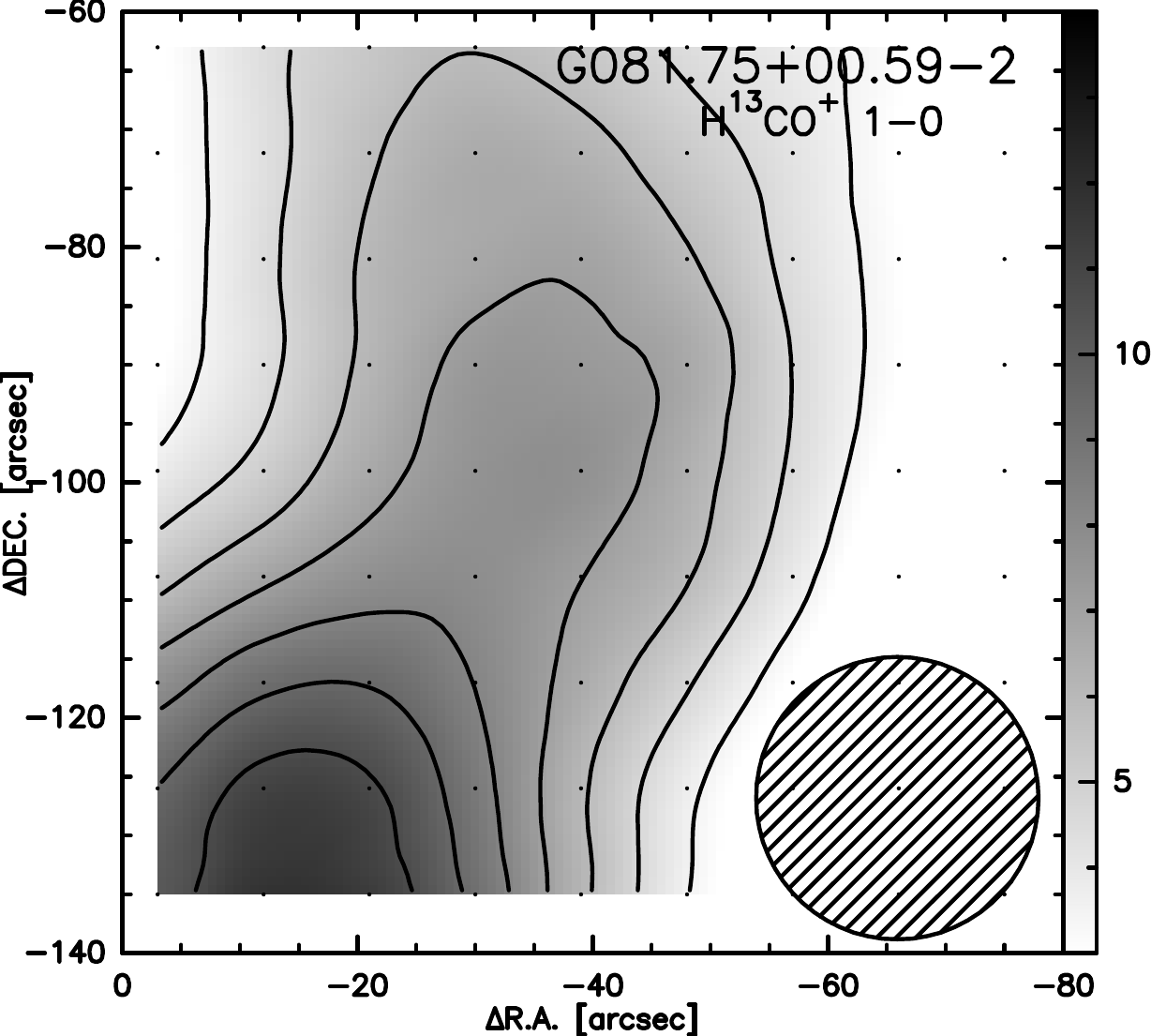} 
  
\includegraphics[width=0.22\columnwidth]{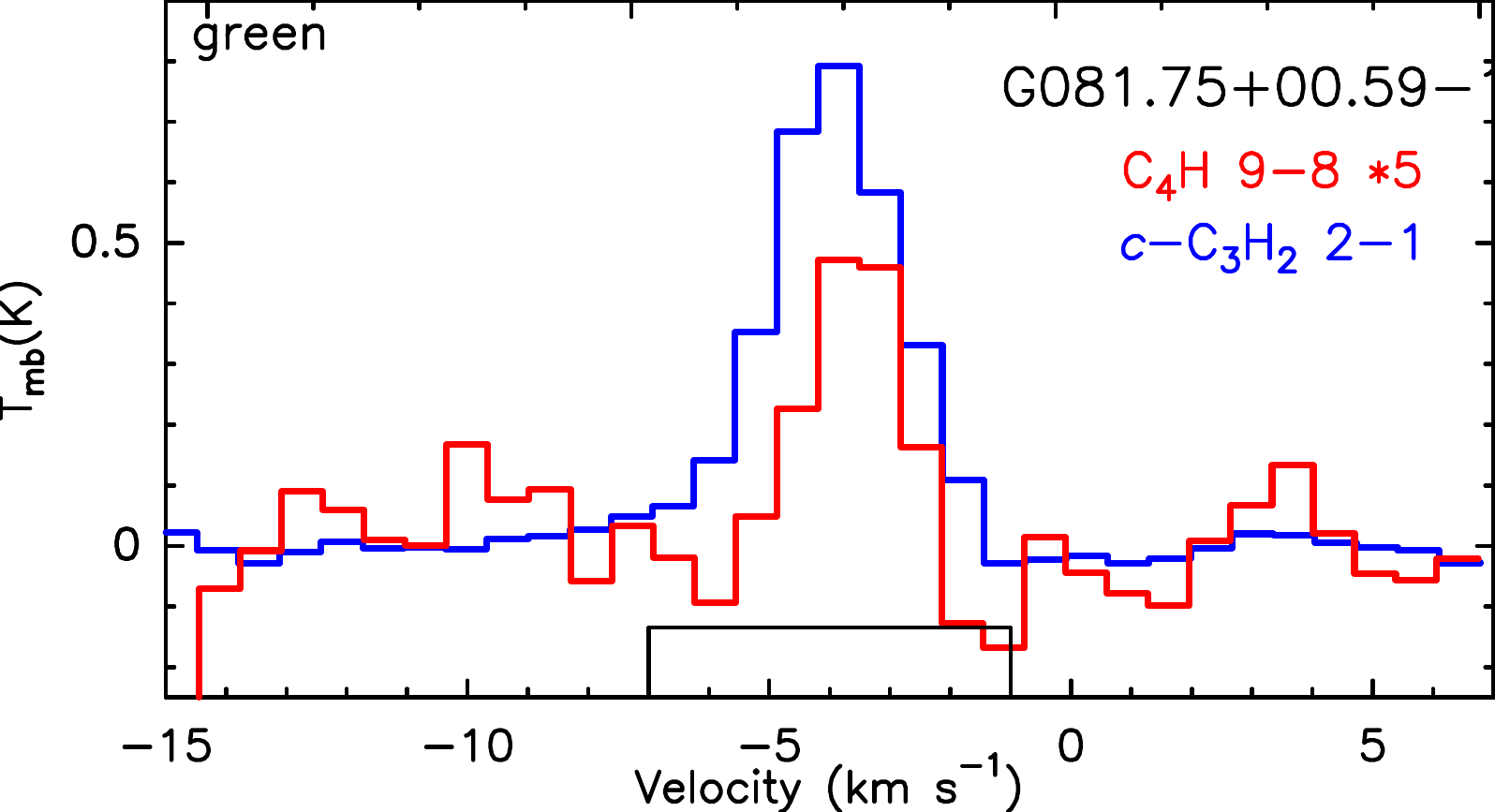}
\includegraphics[width=0.22\columnwidth]{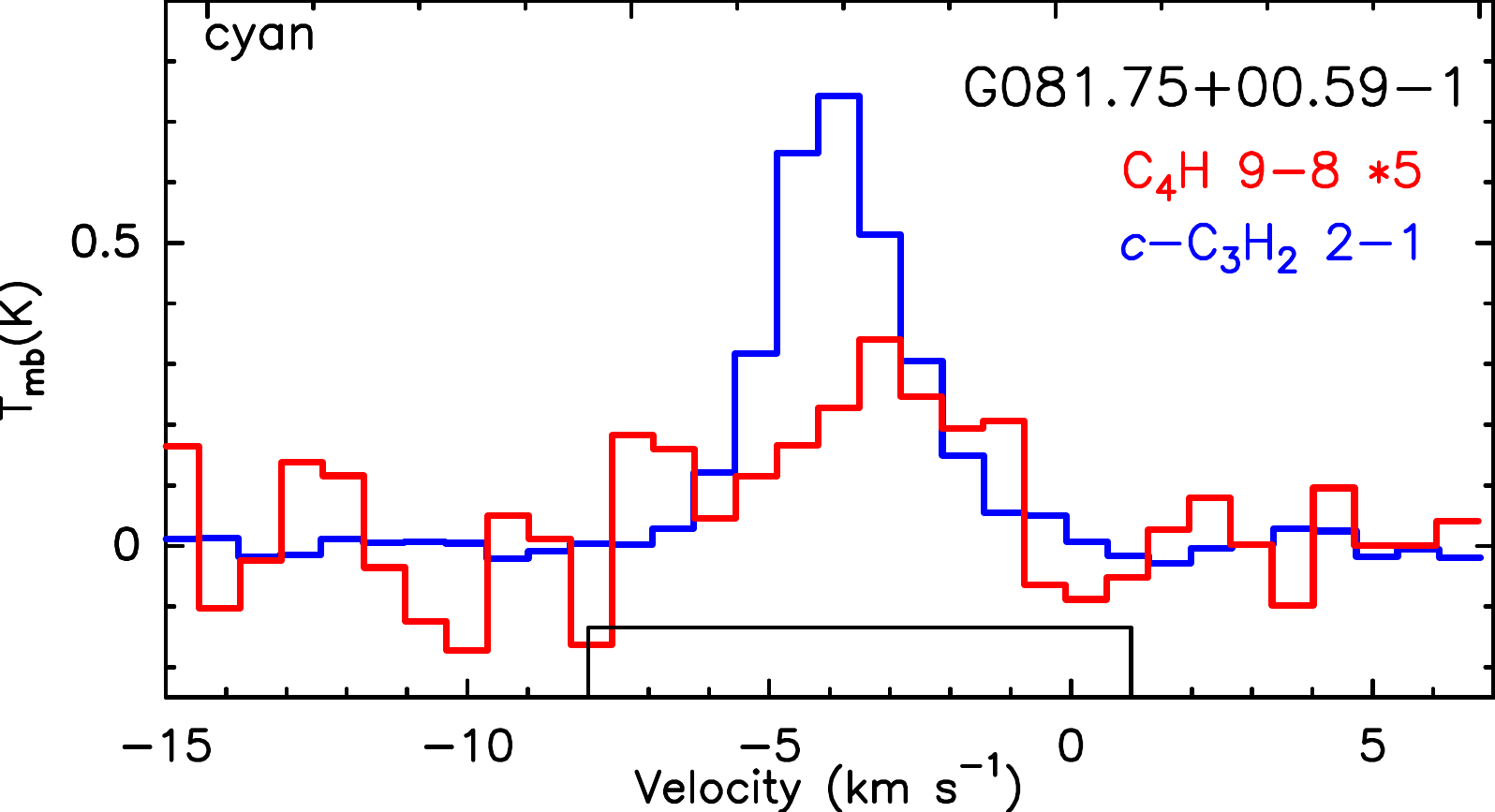}
\includegraphics[width=0.22\columnwidth]{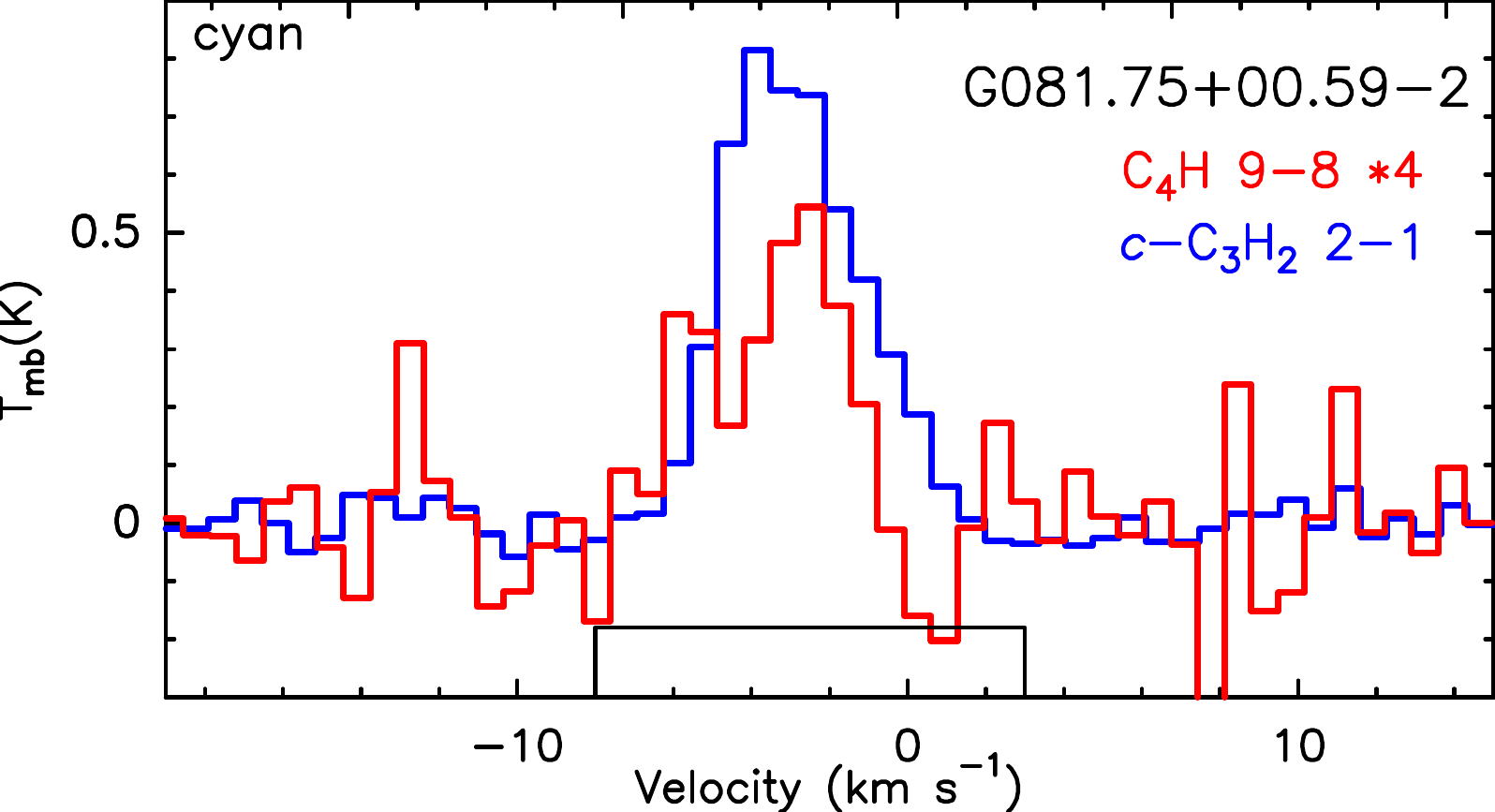}
\includegraphics[width=0.22\columnwidth]{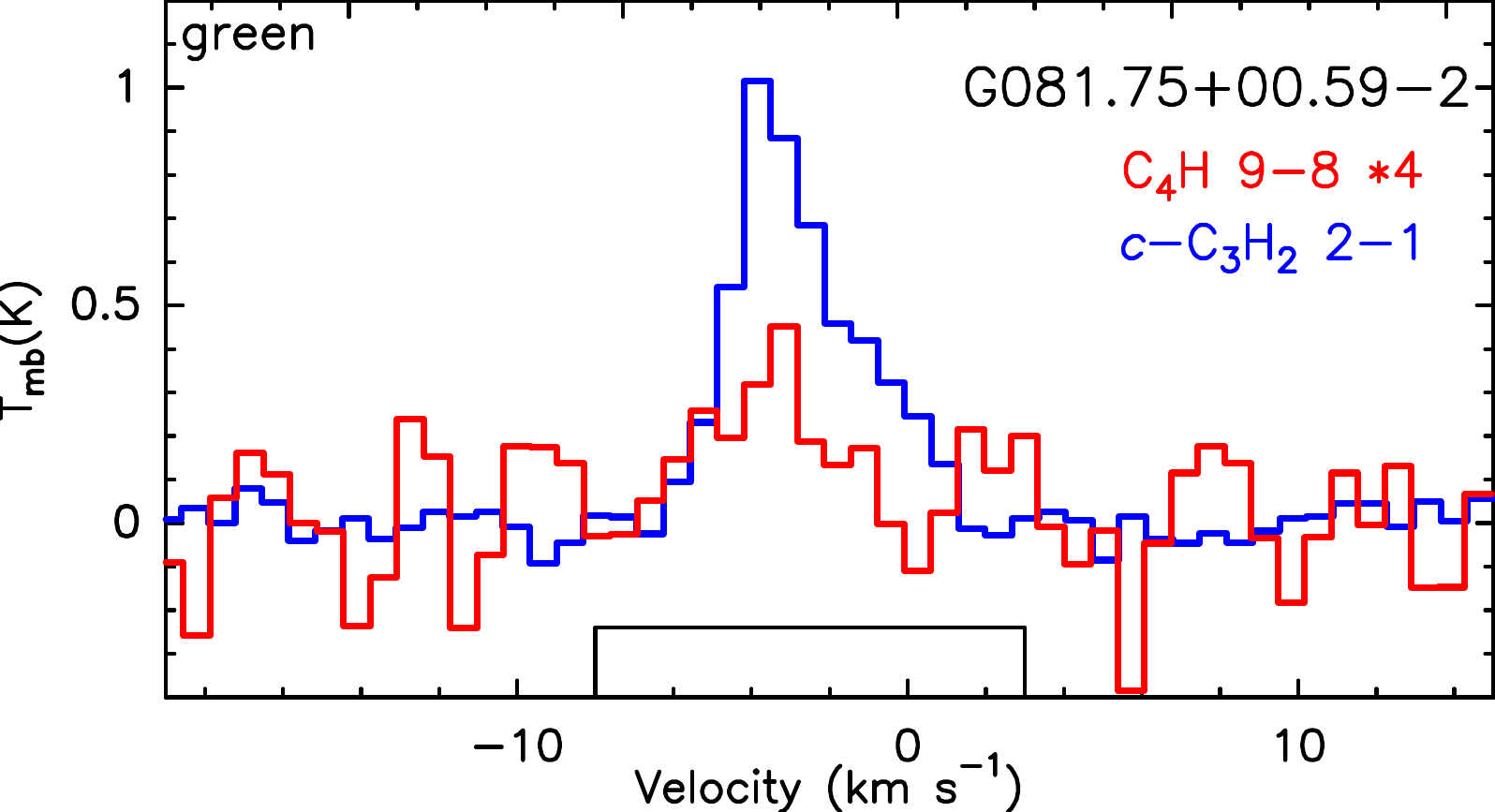}

\includegraphics[width=0.22\columnwidth]{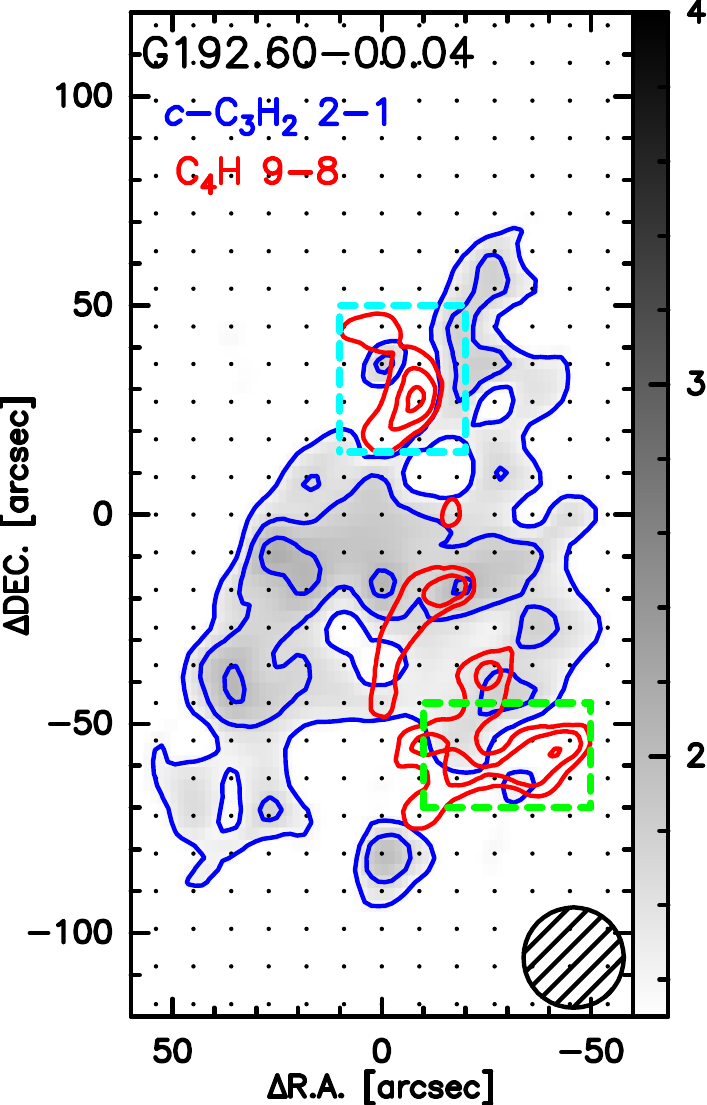}   
\includegraphics[width=0.22\columnwidth]{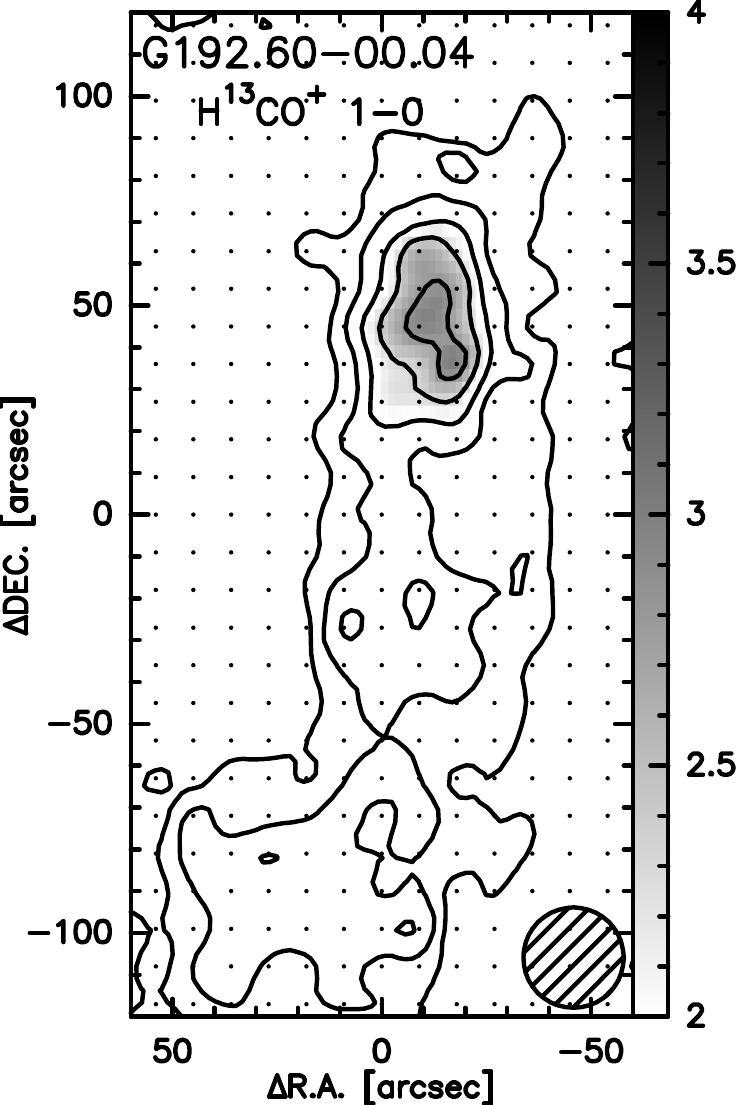} 
\includegraphics[width=0.22\columnwidth]{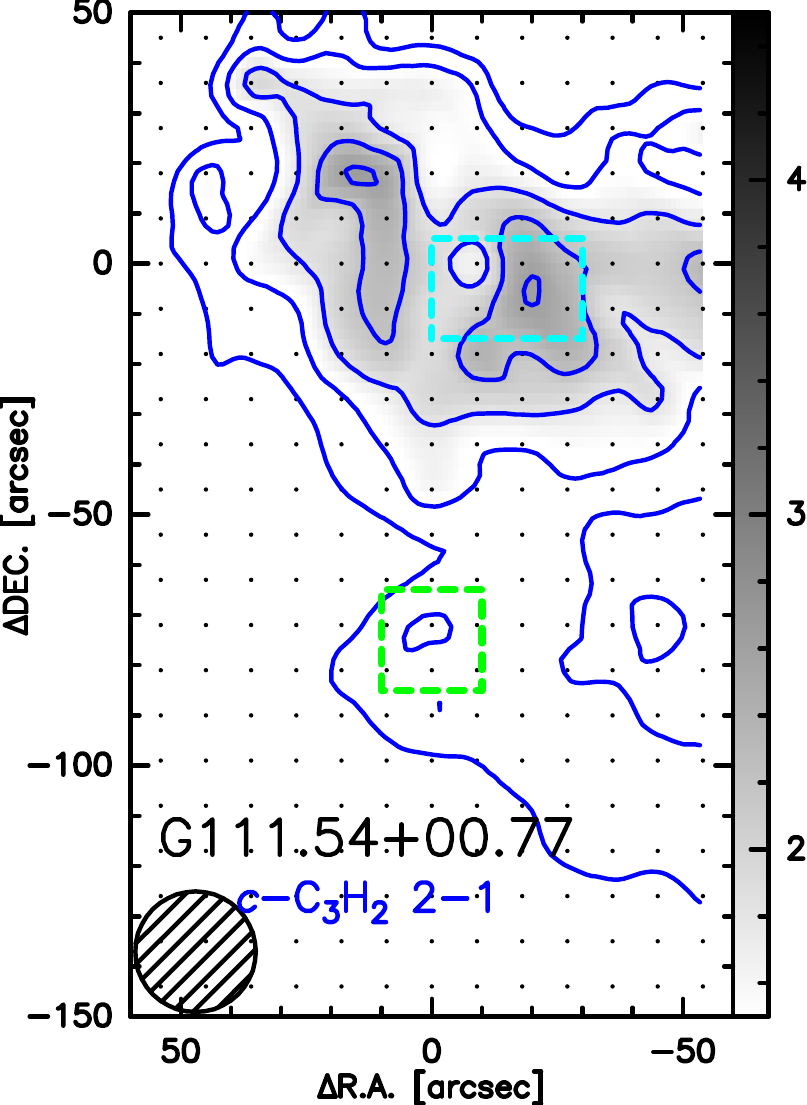}
\includegraphics[width=0.22\columnwidth]{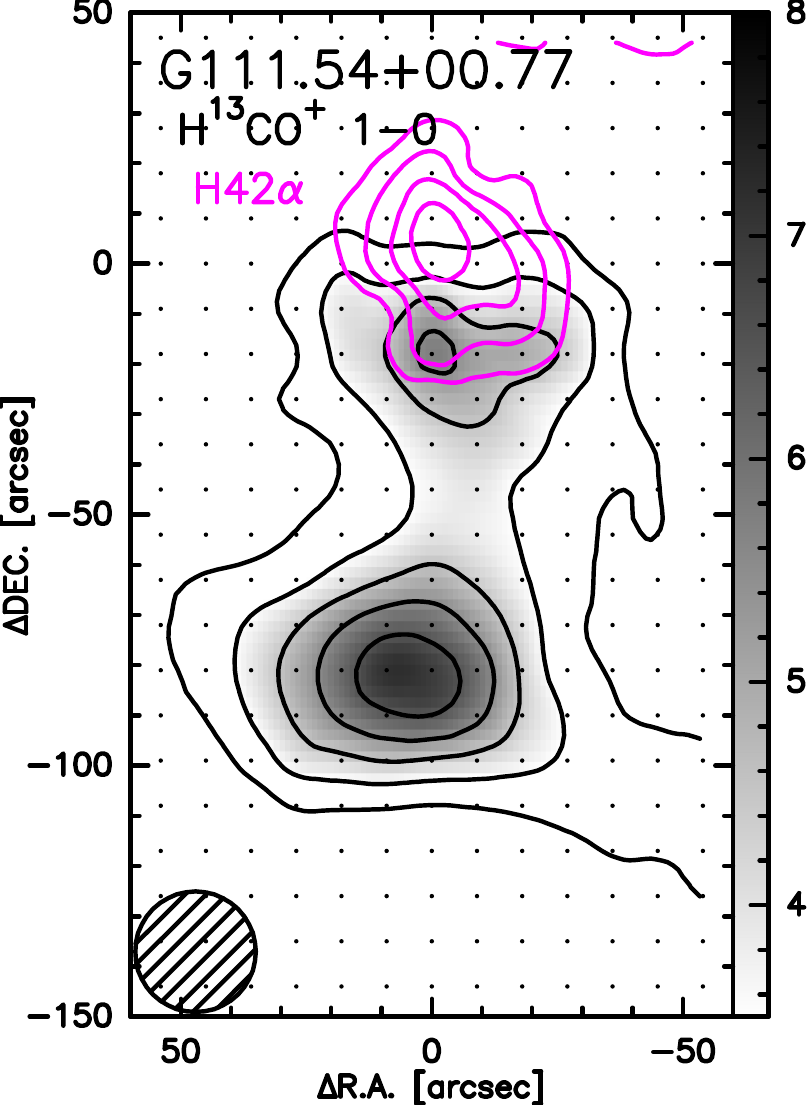}

\includegraphics[width=0.22\columnwidth]{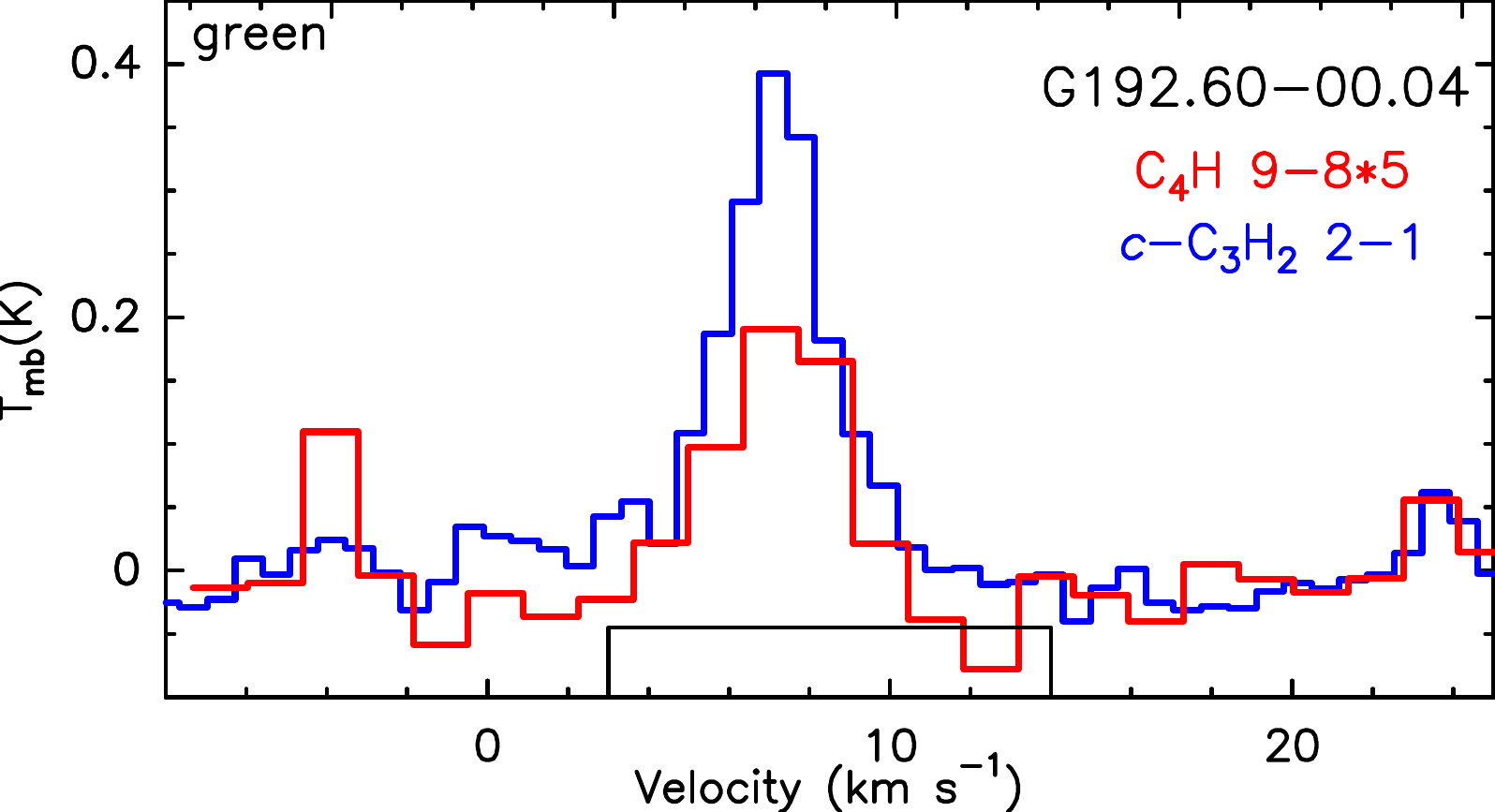}
\includegraphics[width=0.22\columnwidth]{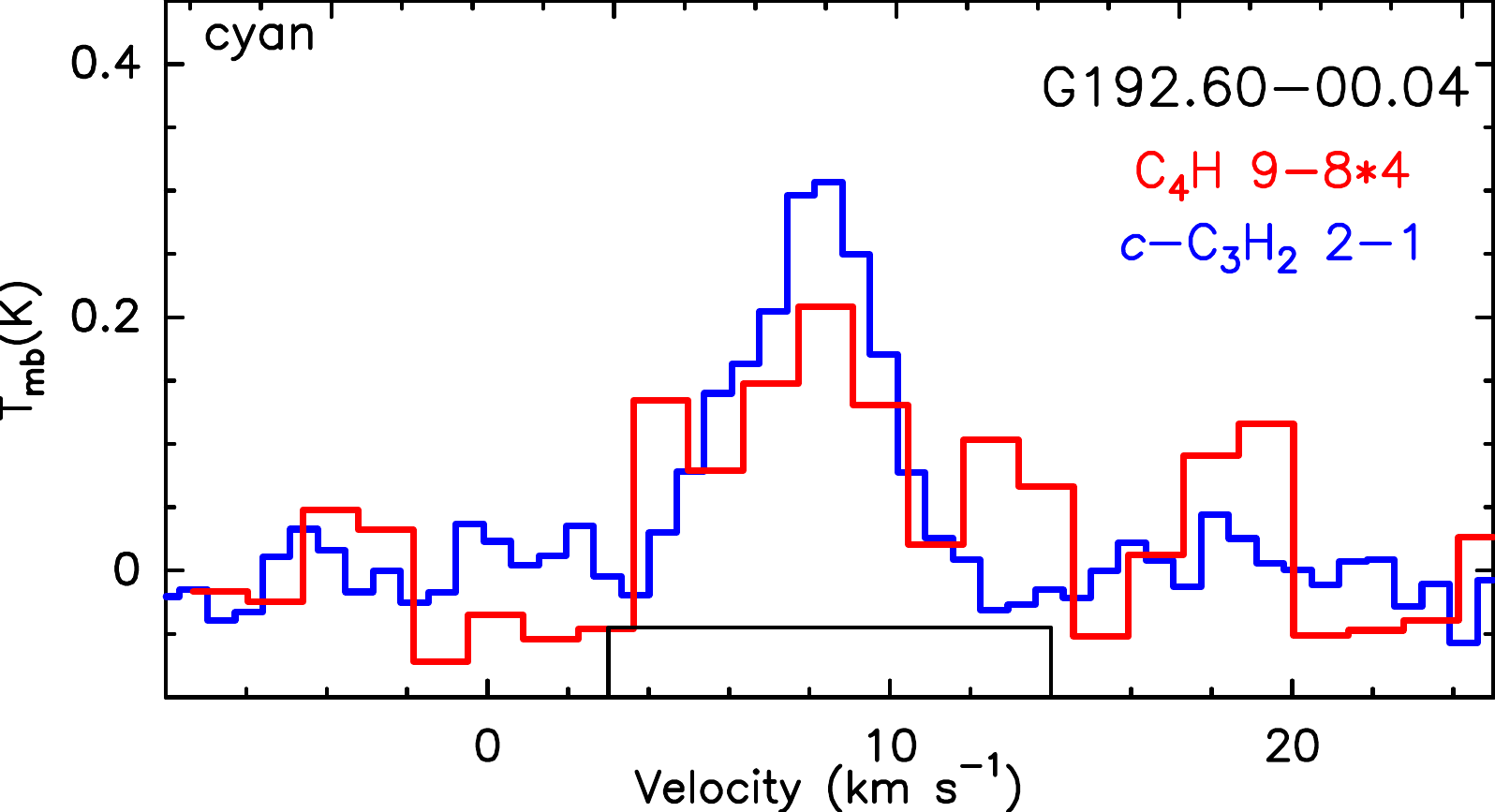}
 \includegraphics[width=0.22\columnwidth]{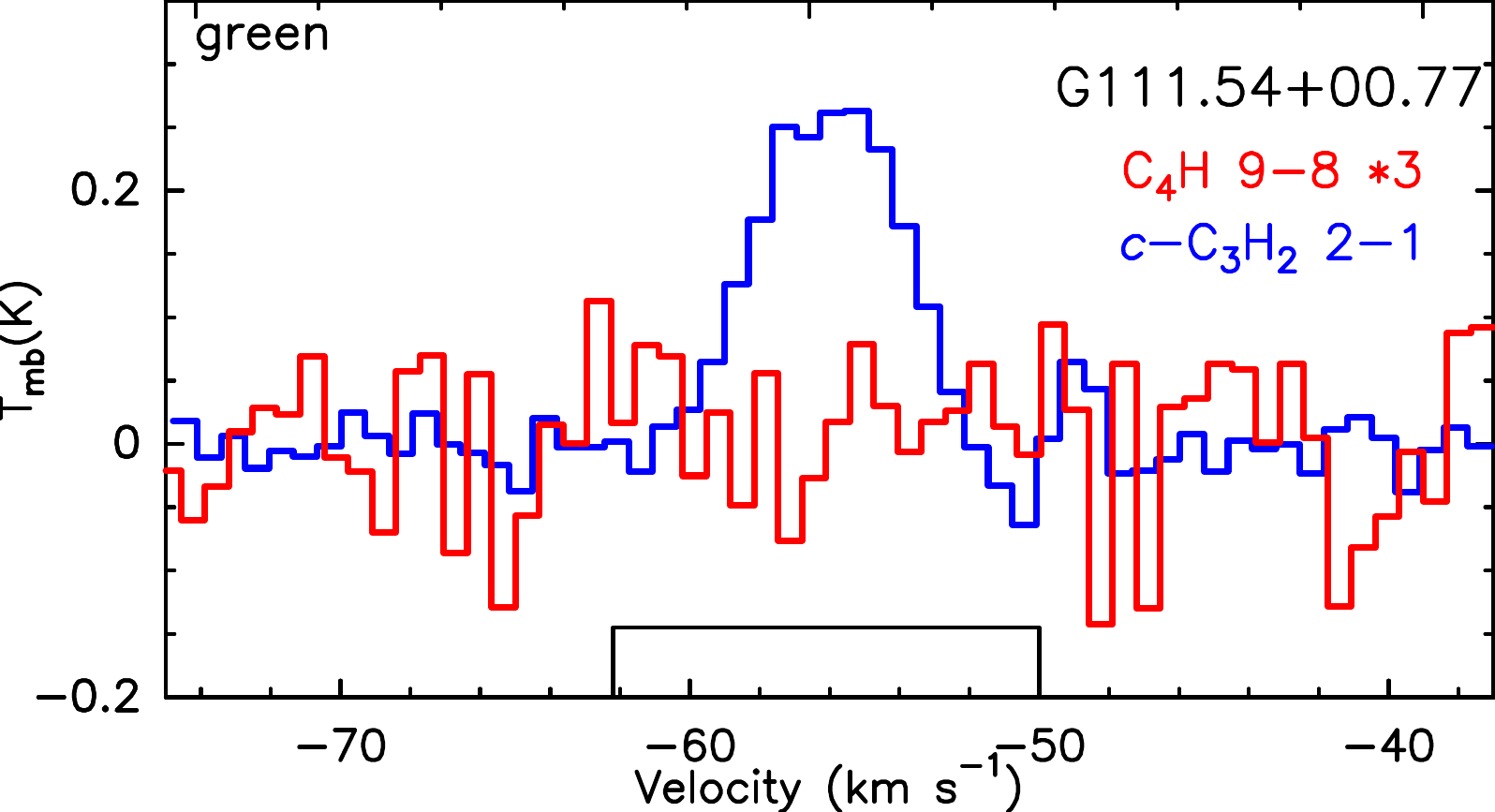}
 \includegraphics[width=0.22\columnwidth]{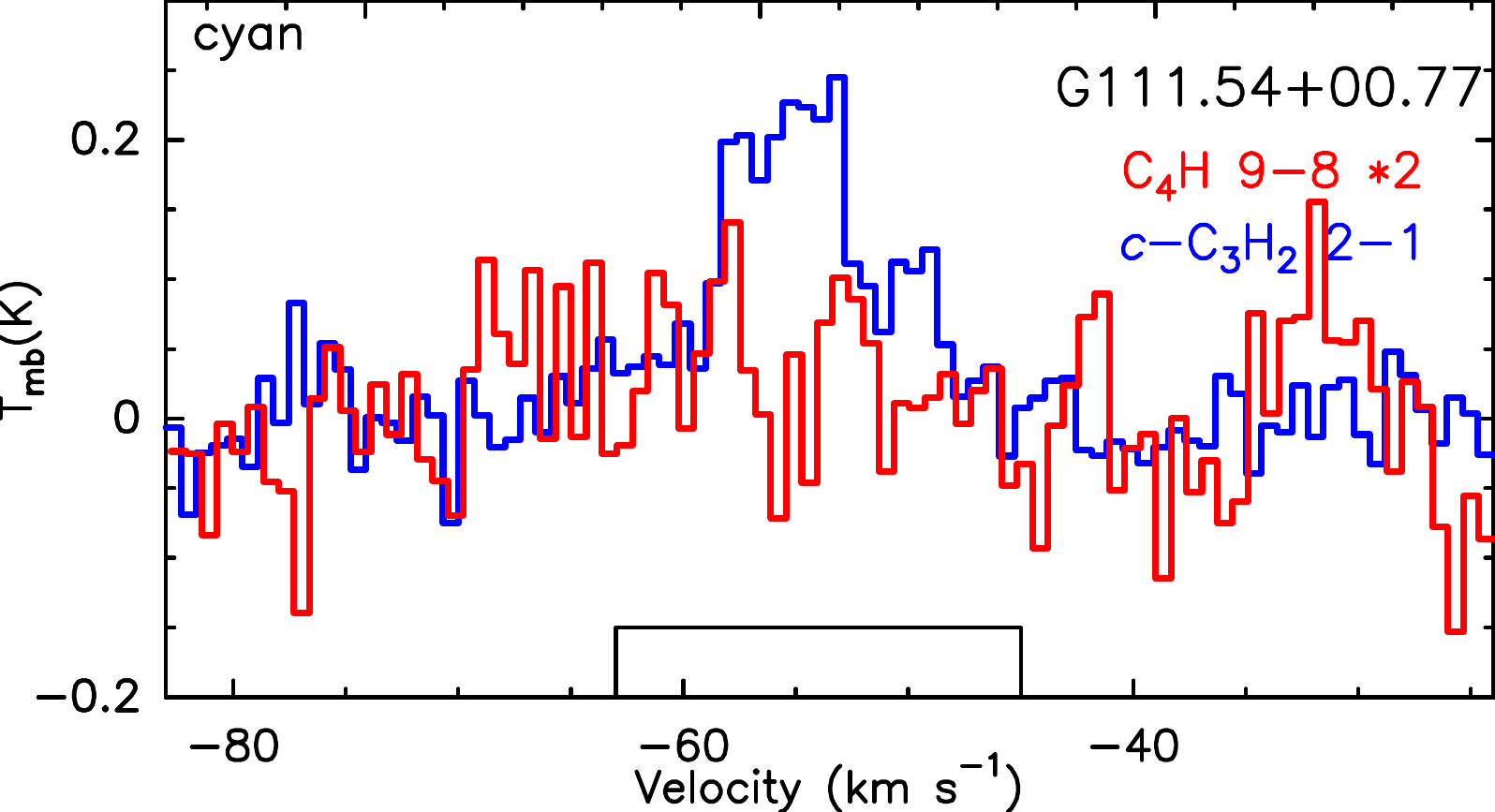}

     \includegraphics[width=0.20\columnwidth]{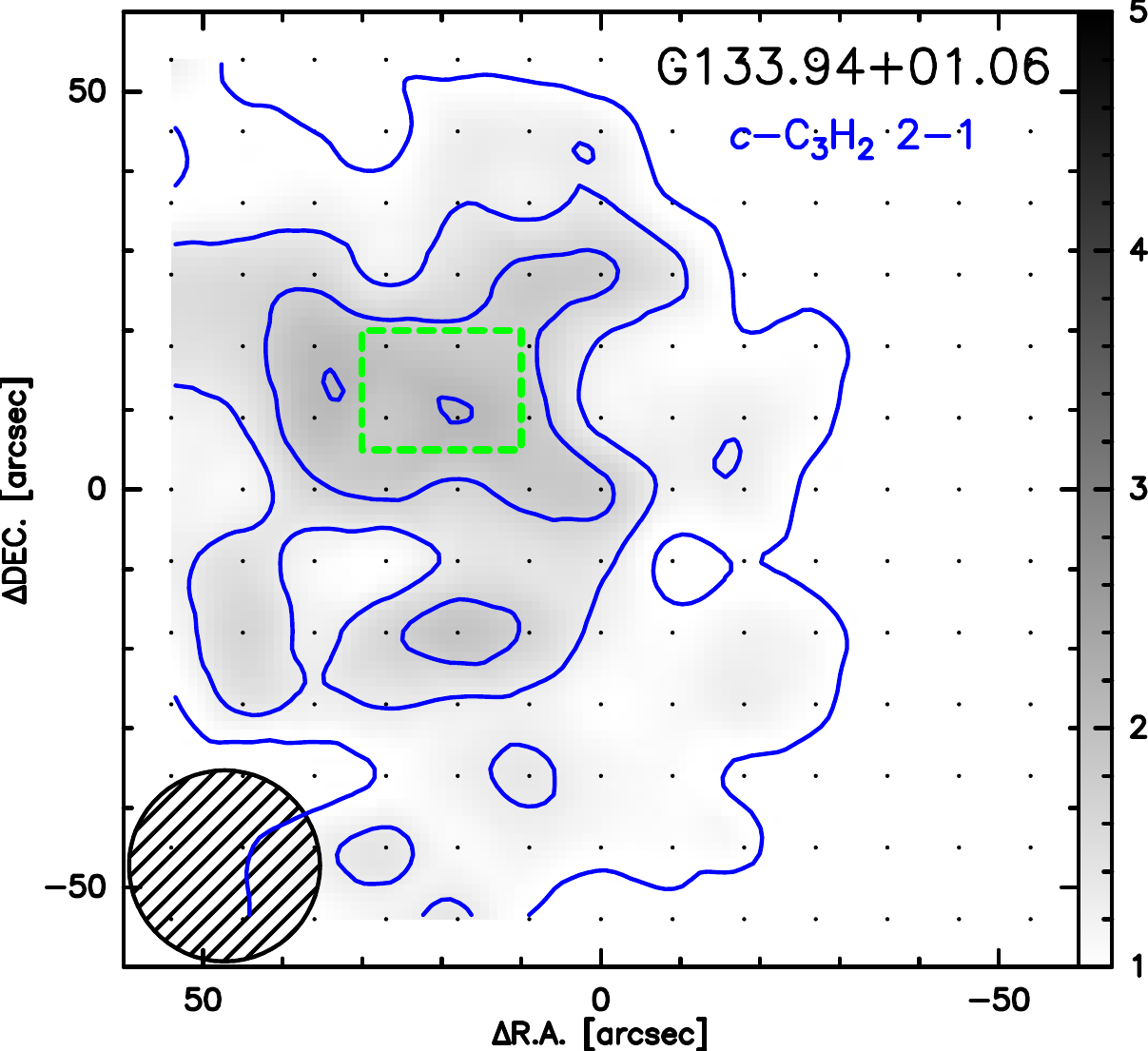}
 \includegraphics[width=0.20\columnwidth]{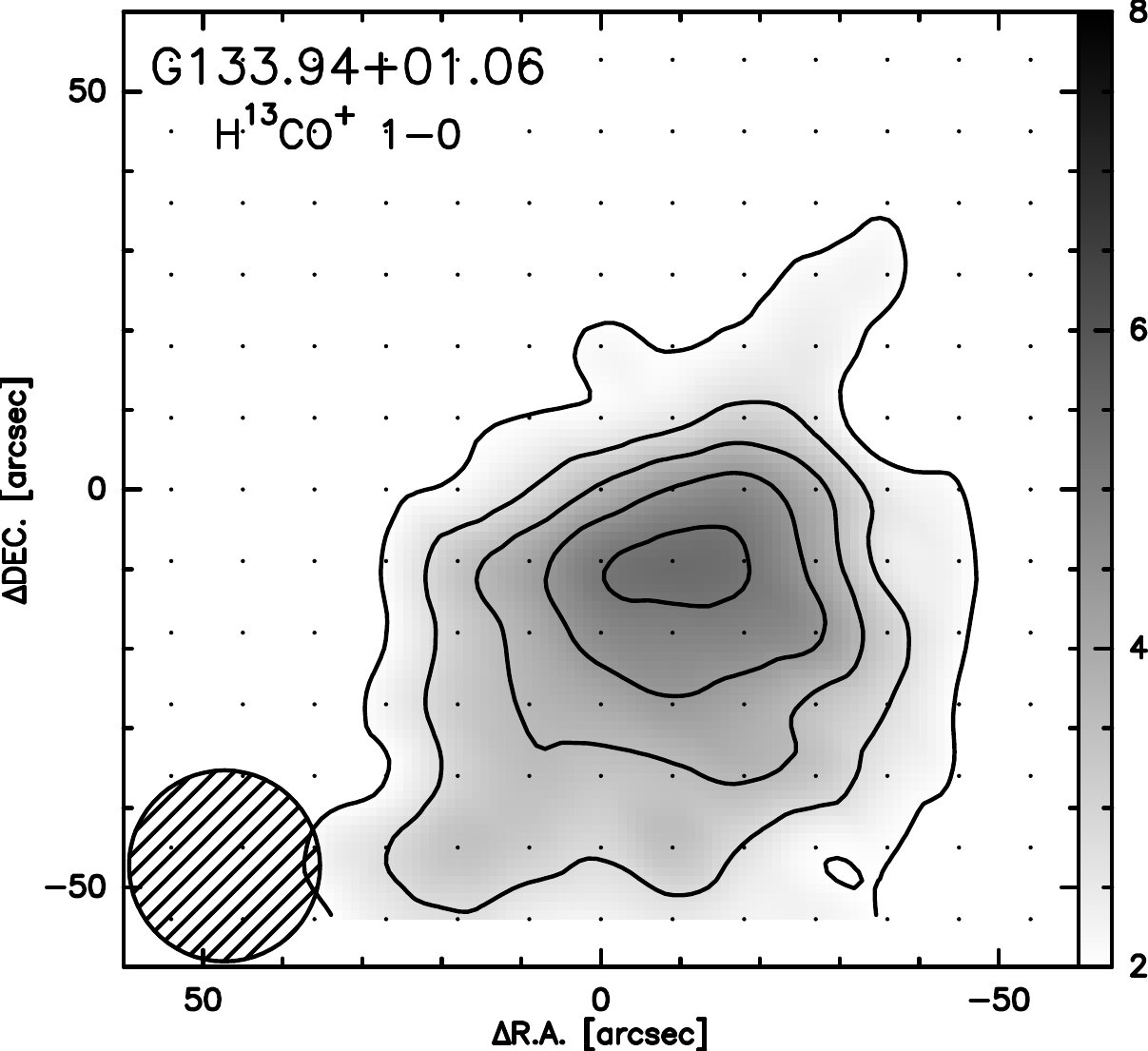}
    \includegraphics[width=0.25\columnwidth]{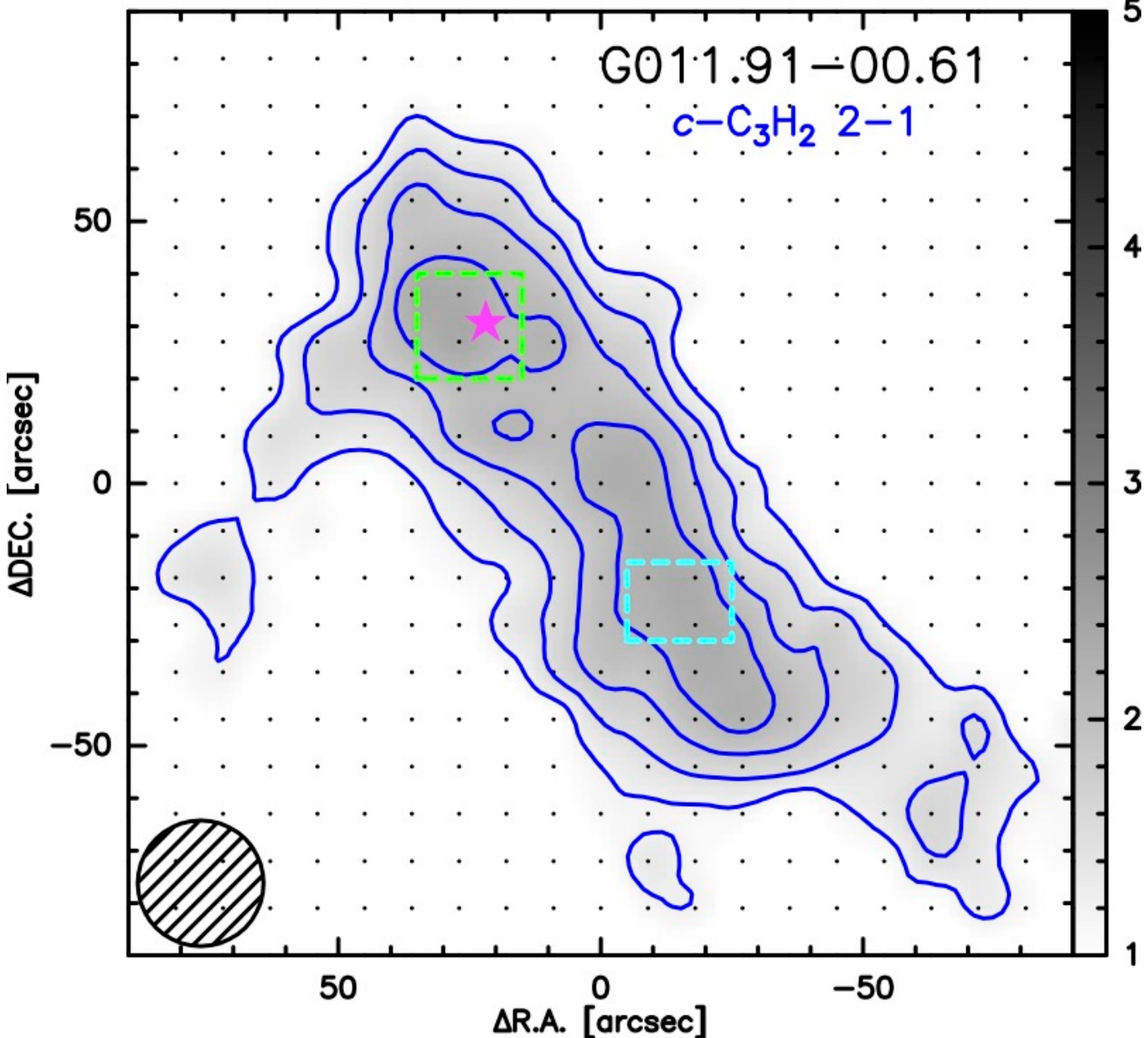}
 \includegraphics[width=0.20\columnwidth]{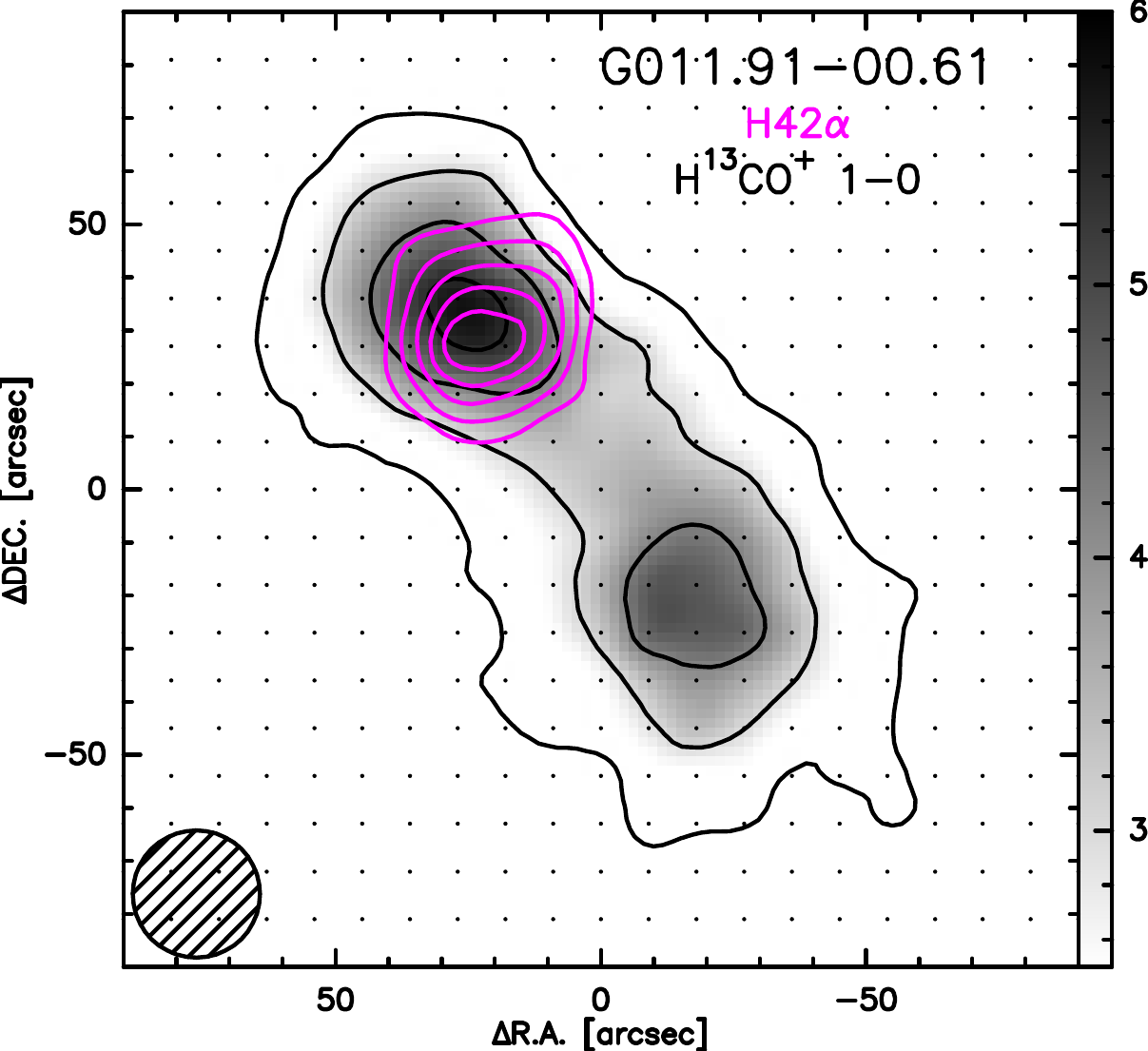}

   \includegraphics[width=0.22\columnwidth]{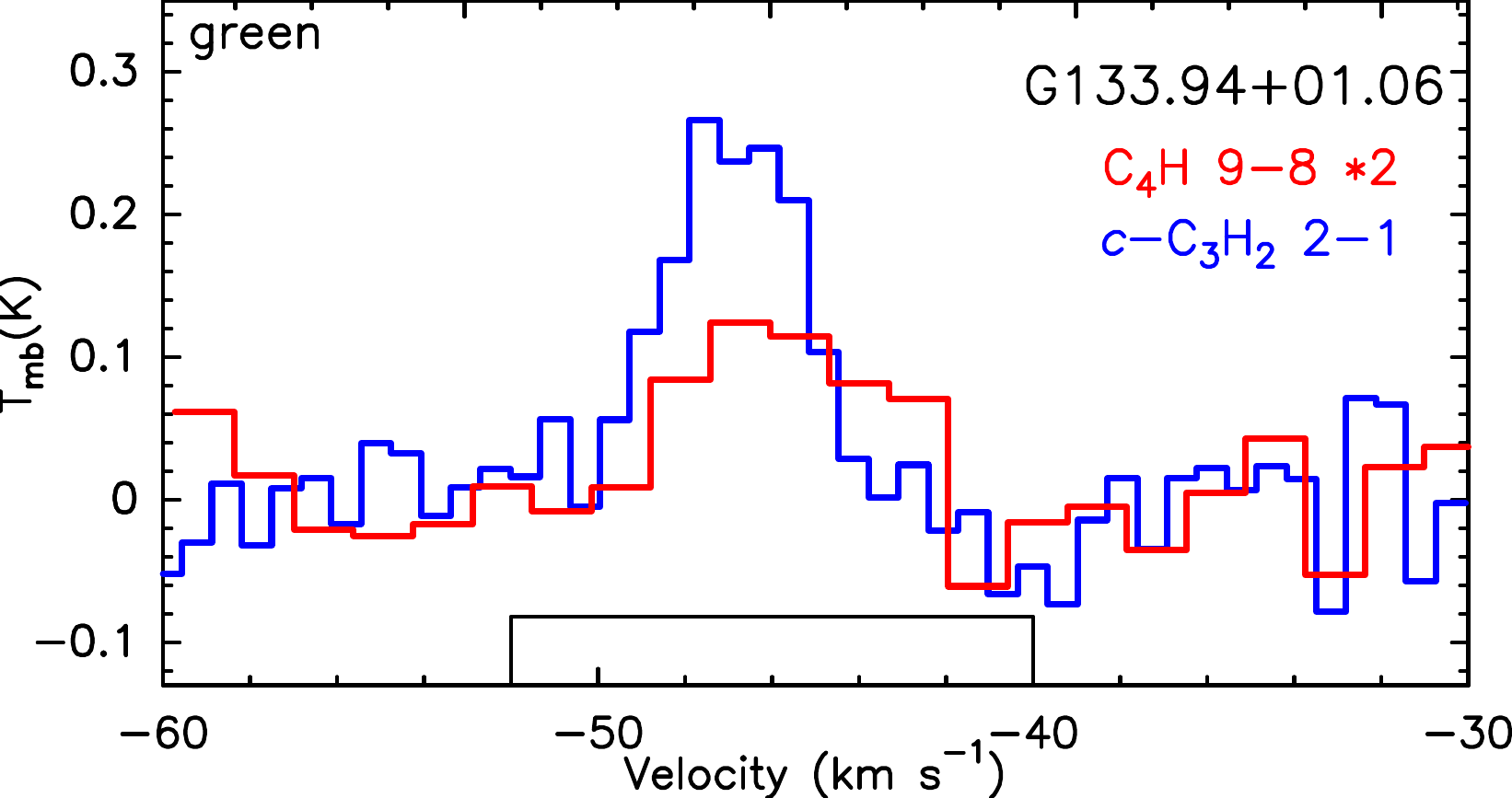}
\includegraphics[width=0.22\columnwidth]{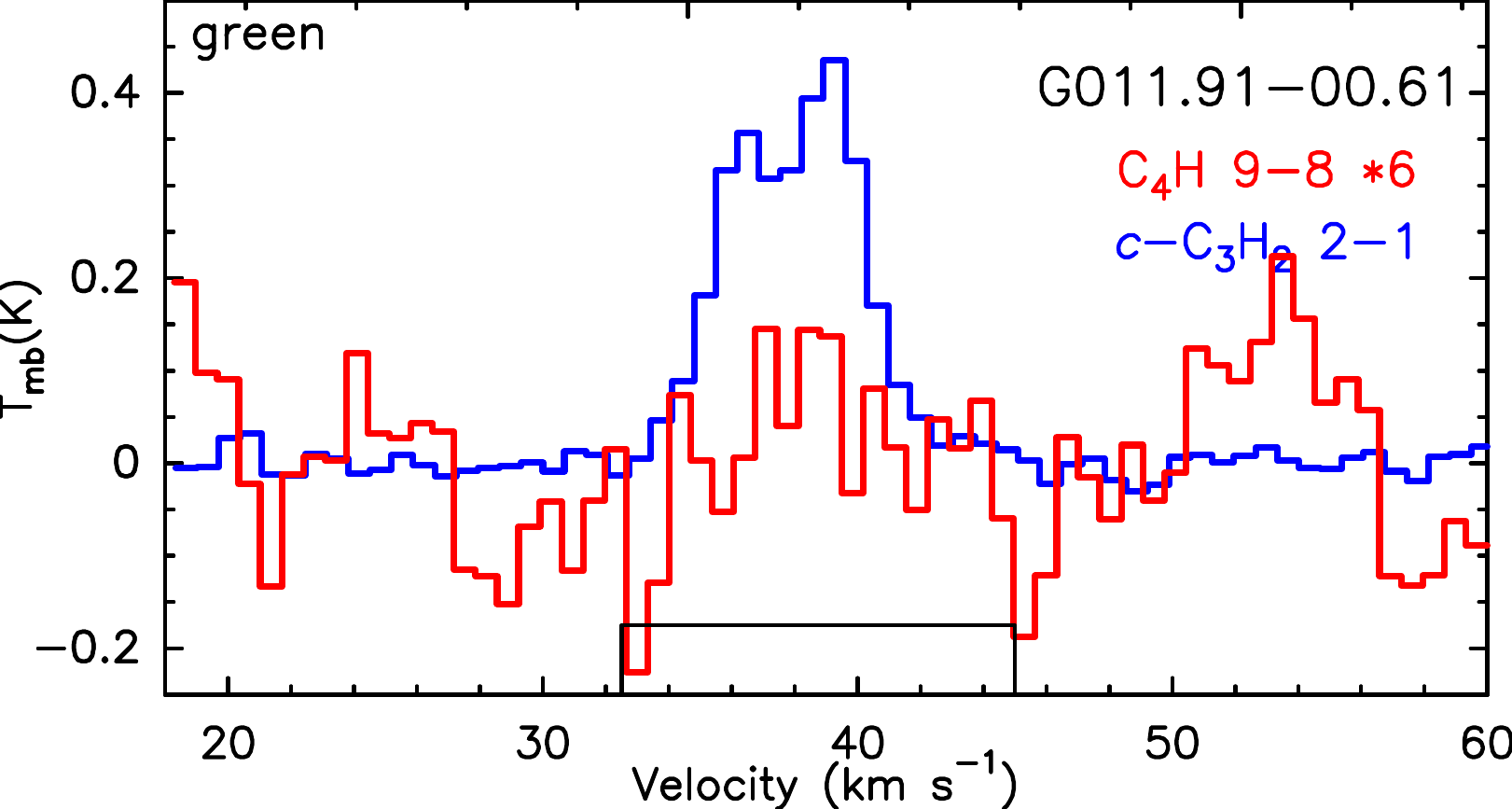}
 \includegraphics[width=0.22\columnwidth]{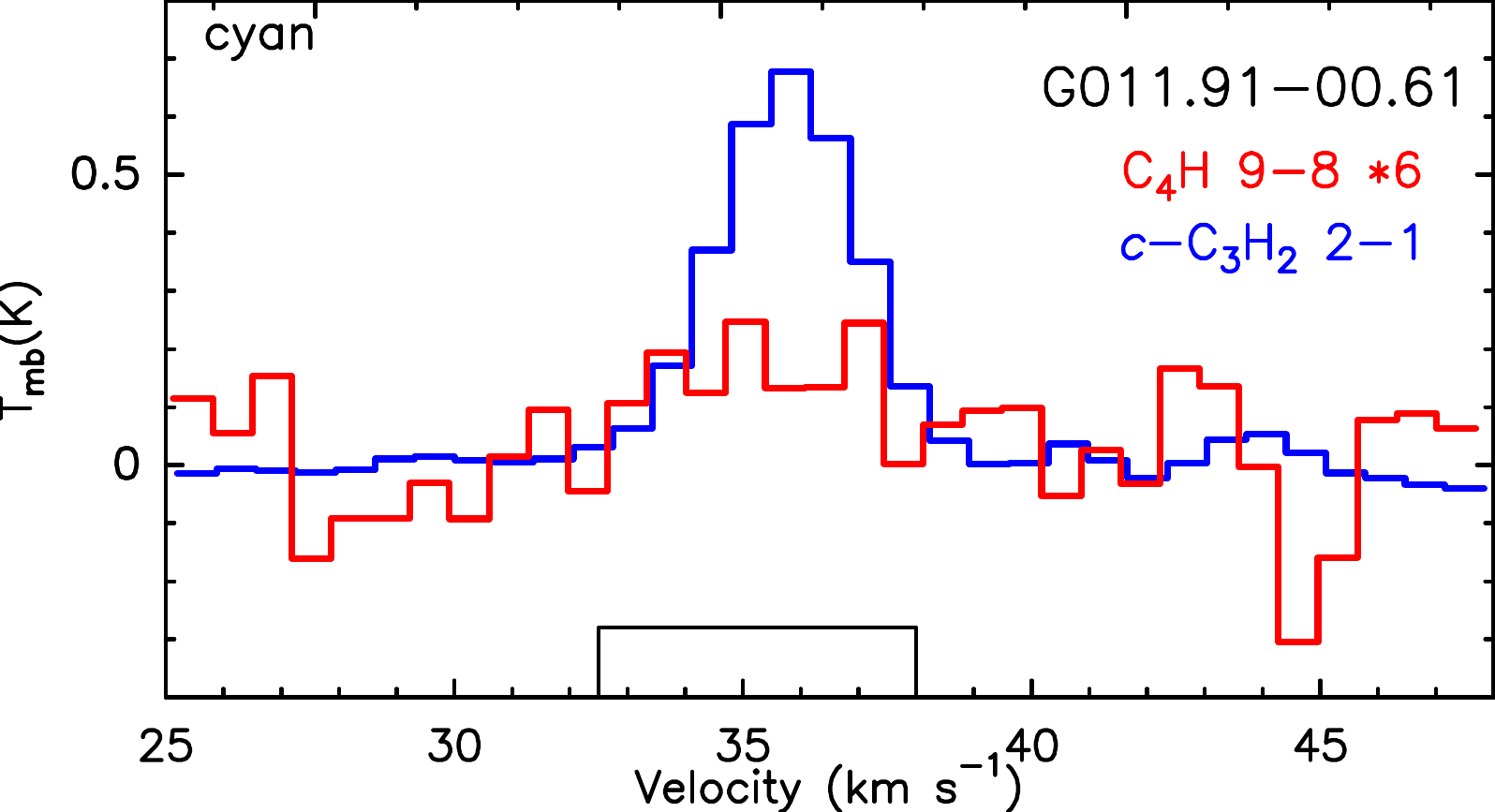}

 \centering 
\caption{Continued.}    
\label{Appendix-1}
\end{center}
\end{figure}

 \begin{figure}[h]
\begin{center}

\centering 
\addtocounter{figure}{-1}

\includegraphics[width=0.22\columnwidth]{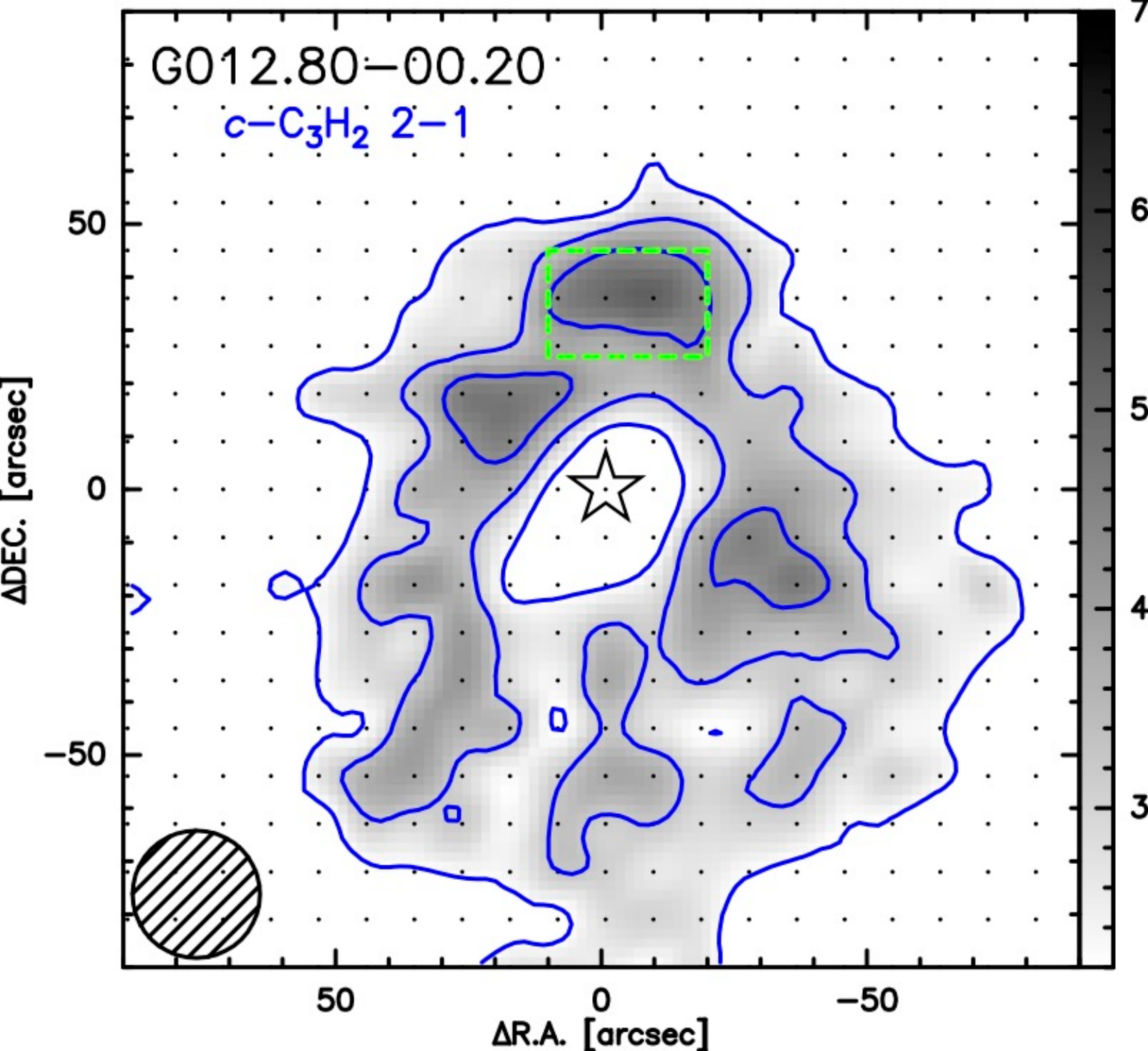}
\includegraphics[width=0.22\columnwidth]{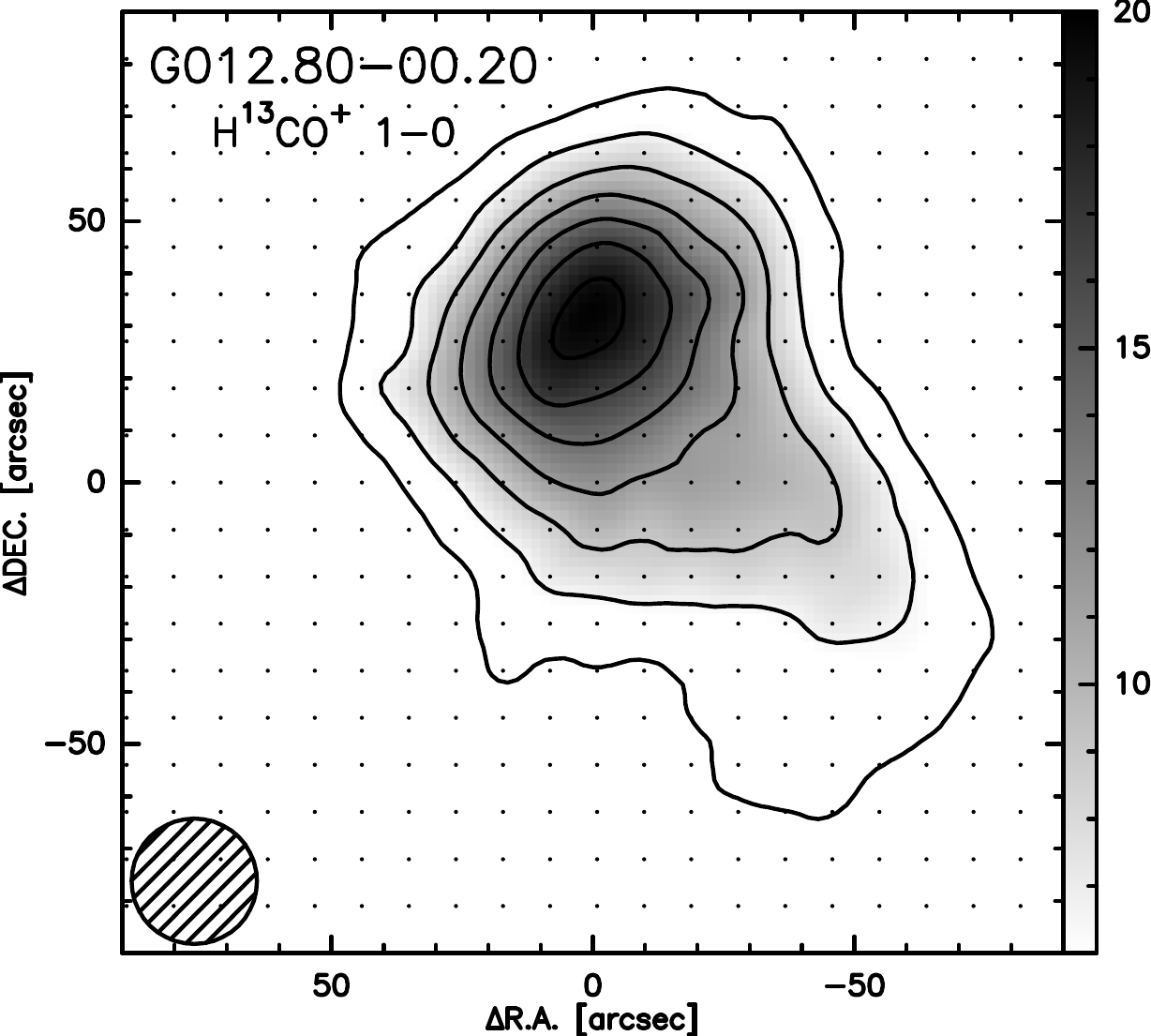}
 \includegraphics[width=0.22\columnwidth]{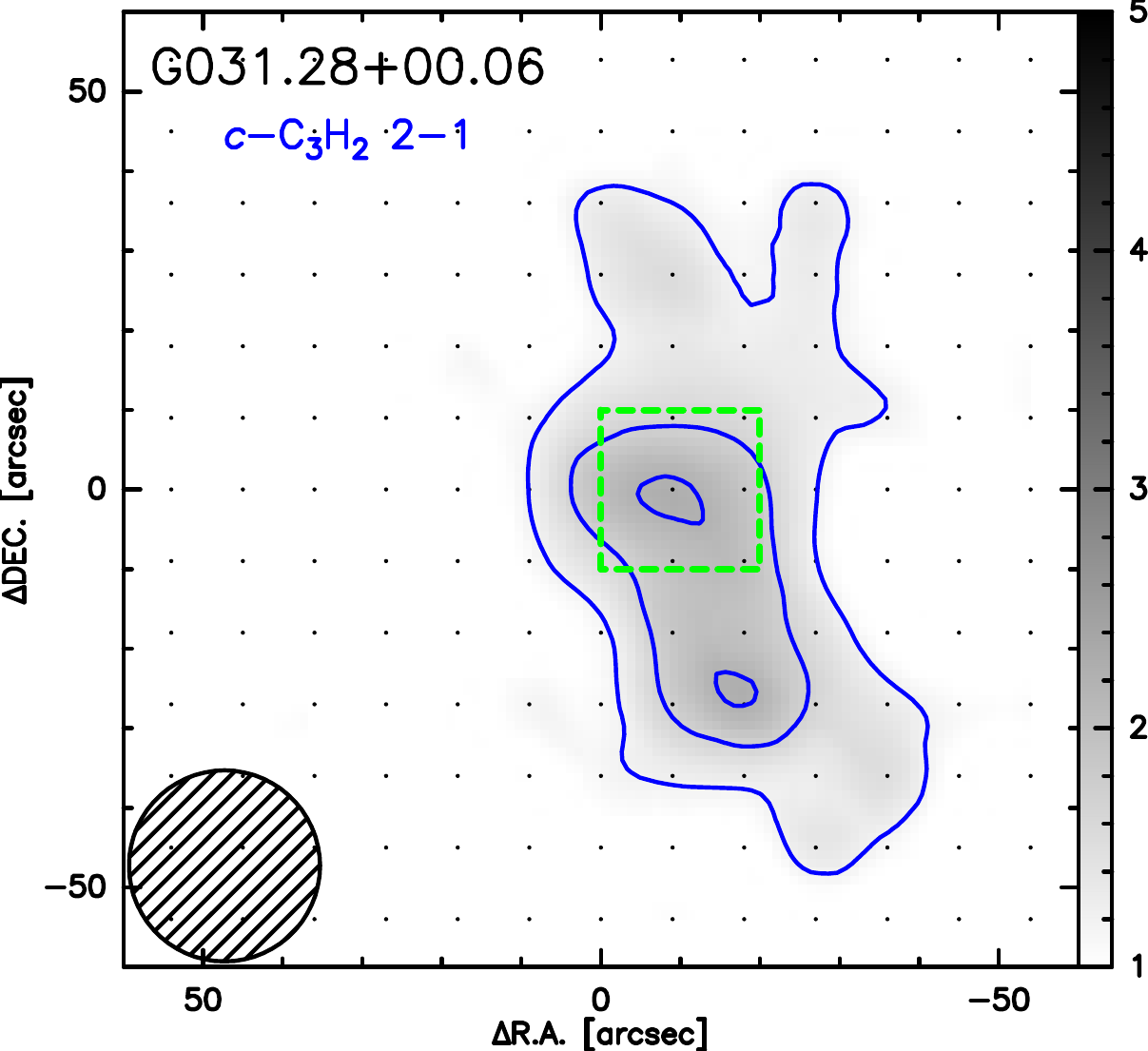}
  \includegraphics[width=0.22\columnwidth]{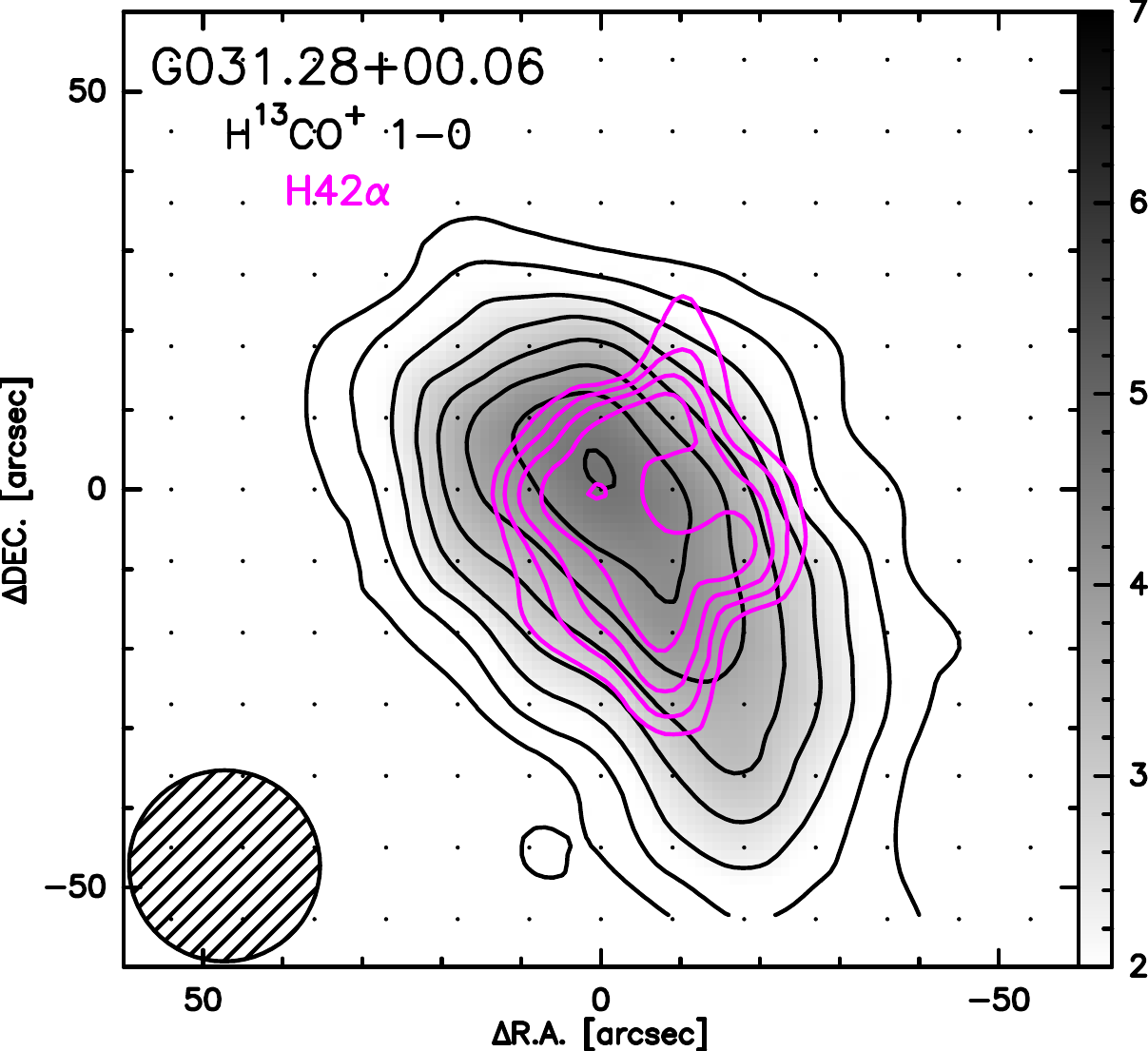}
  
 \includegraphics[width=0.22\columnwidth]{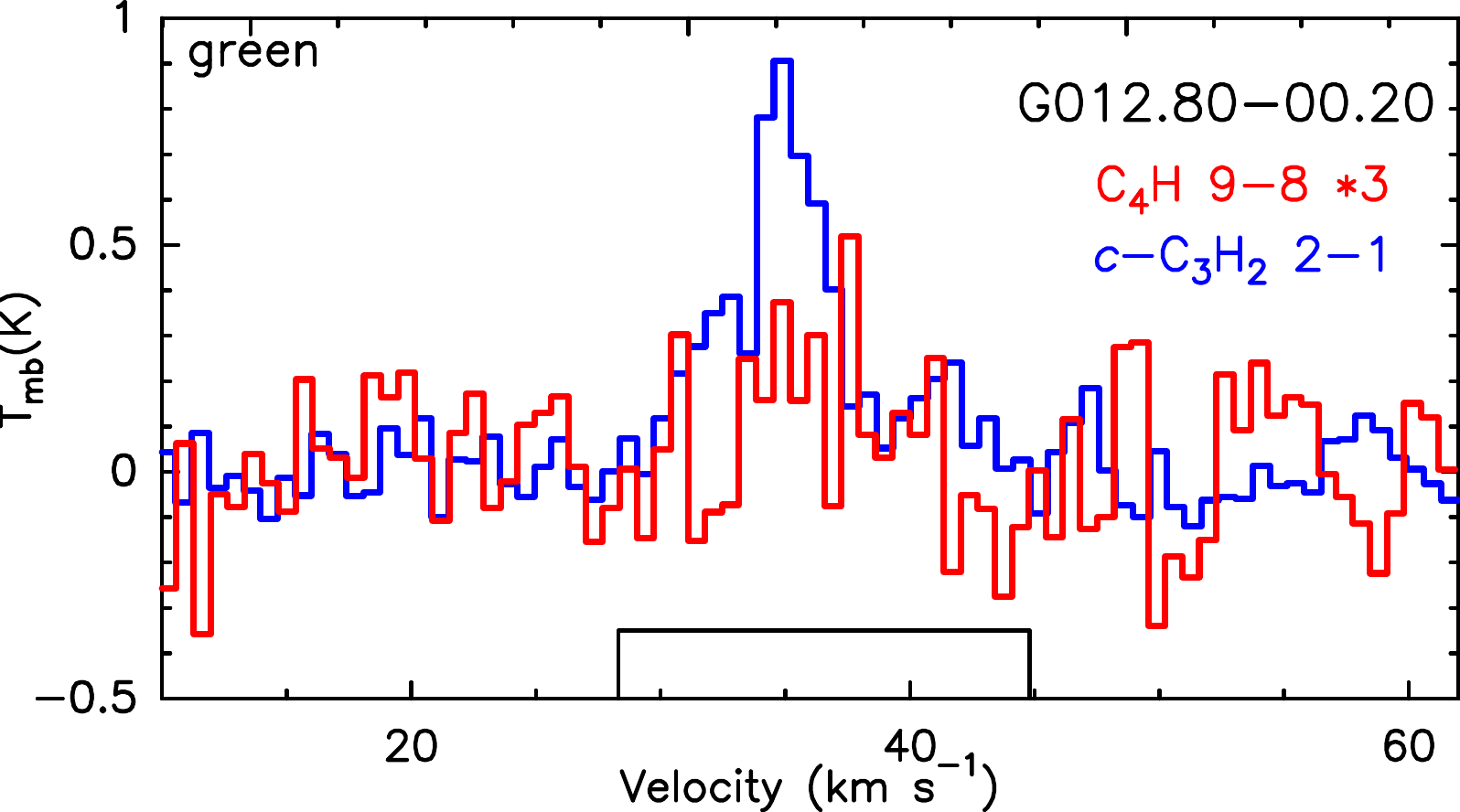}
\includegraphics[width=0.22\columnwidth]{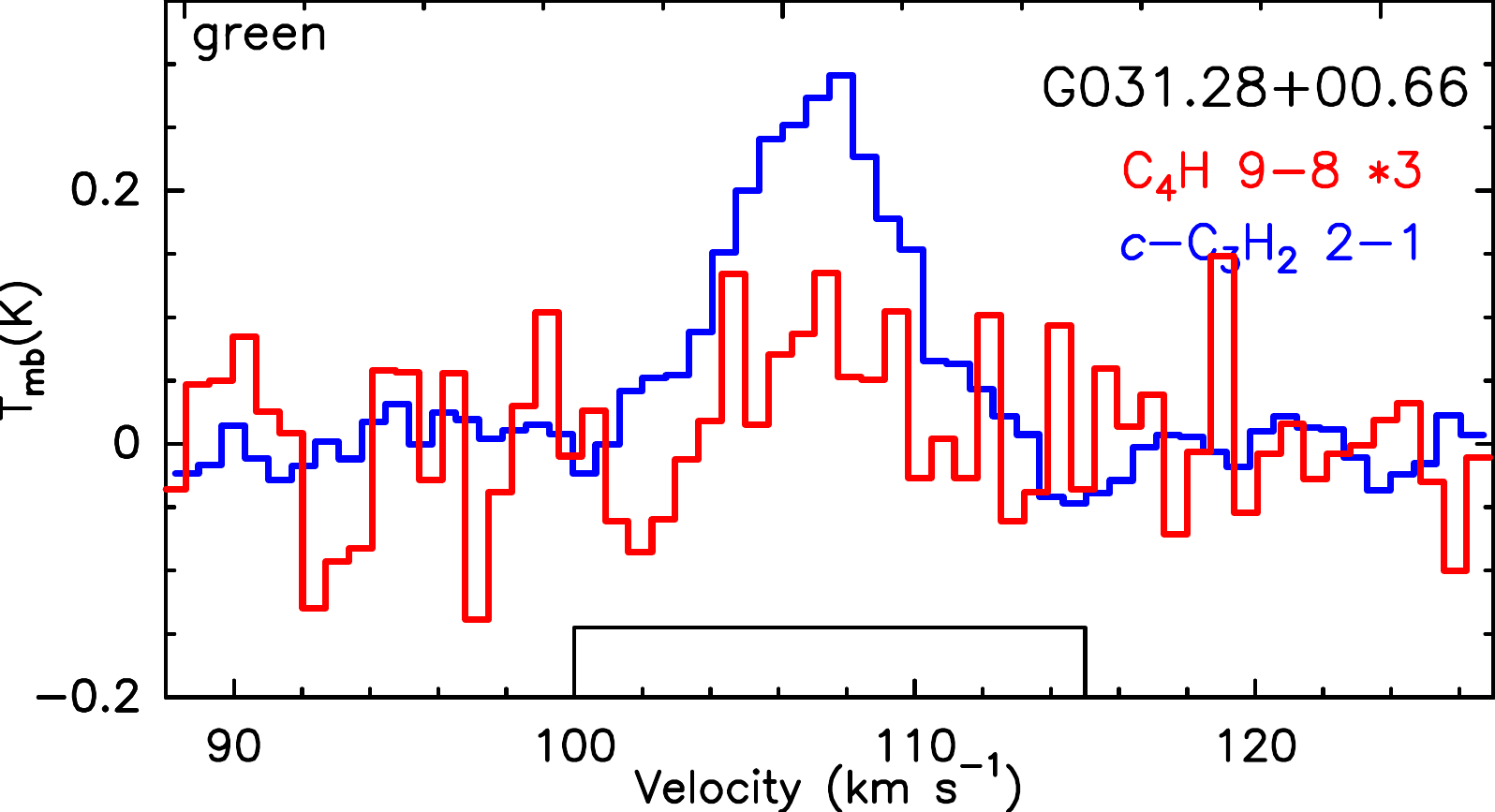}
\includegraphics[width=0.22\columnwidth]{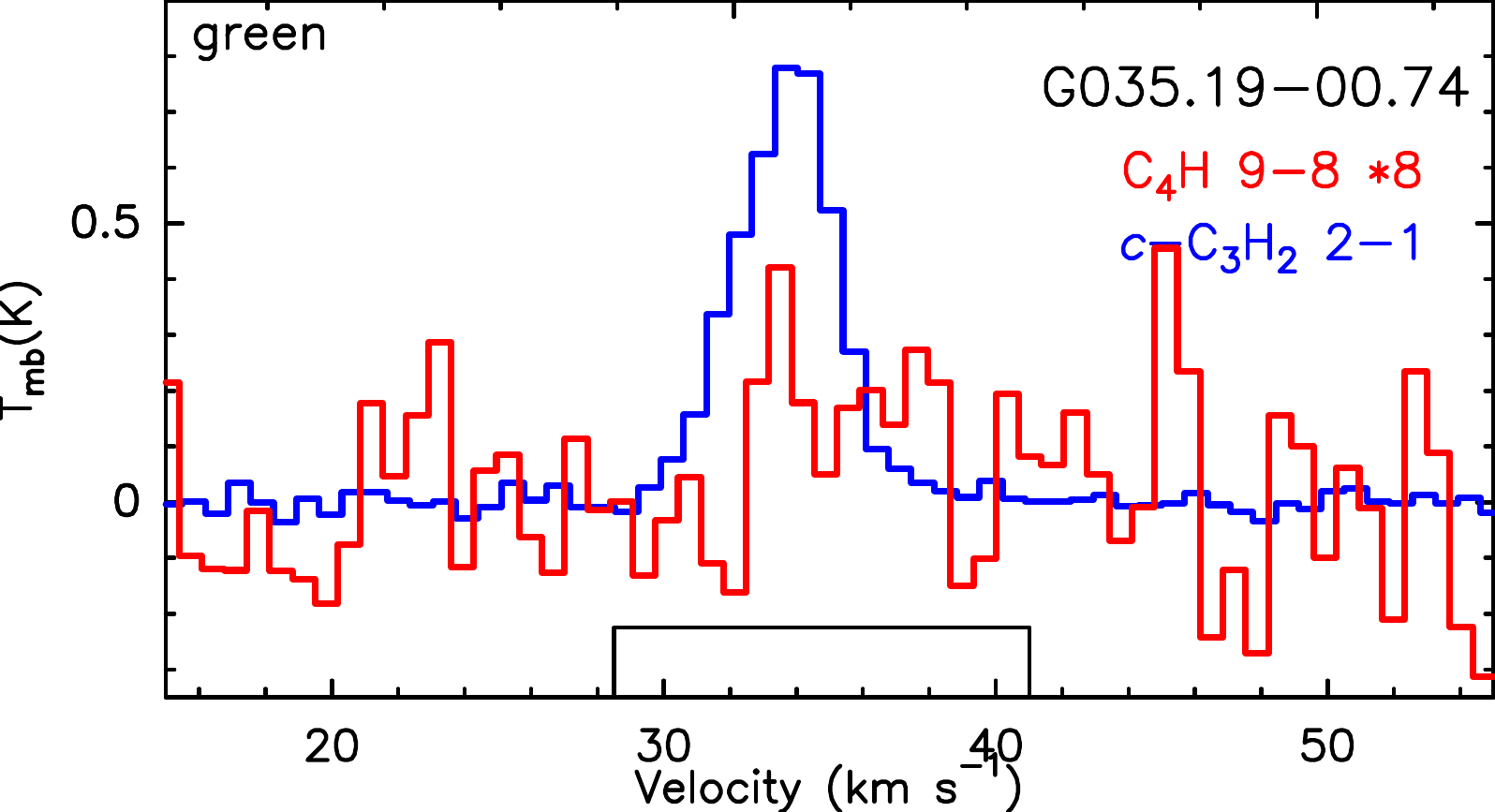}
\includegraphics[width=0.22\columnwidth]{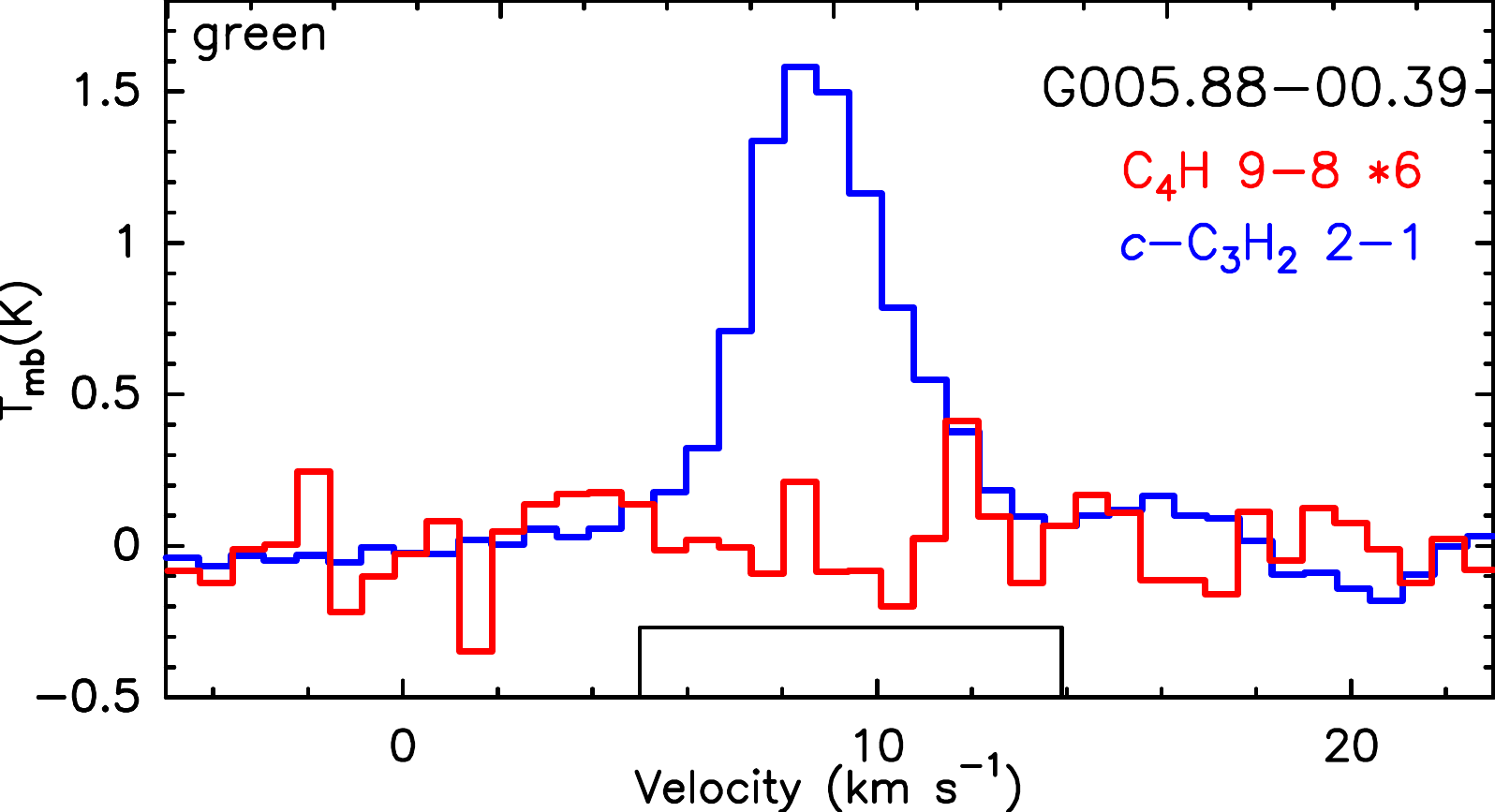}

\includegraphics[width=0.22\columnwidth]{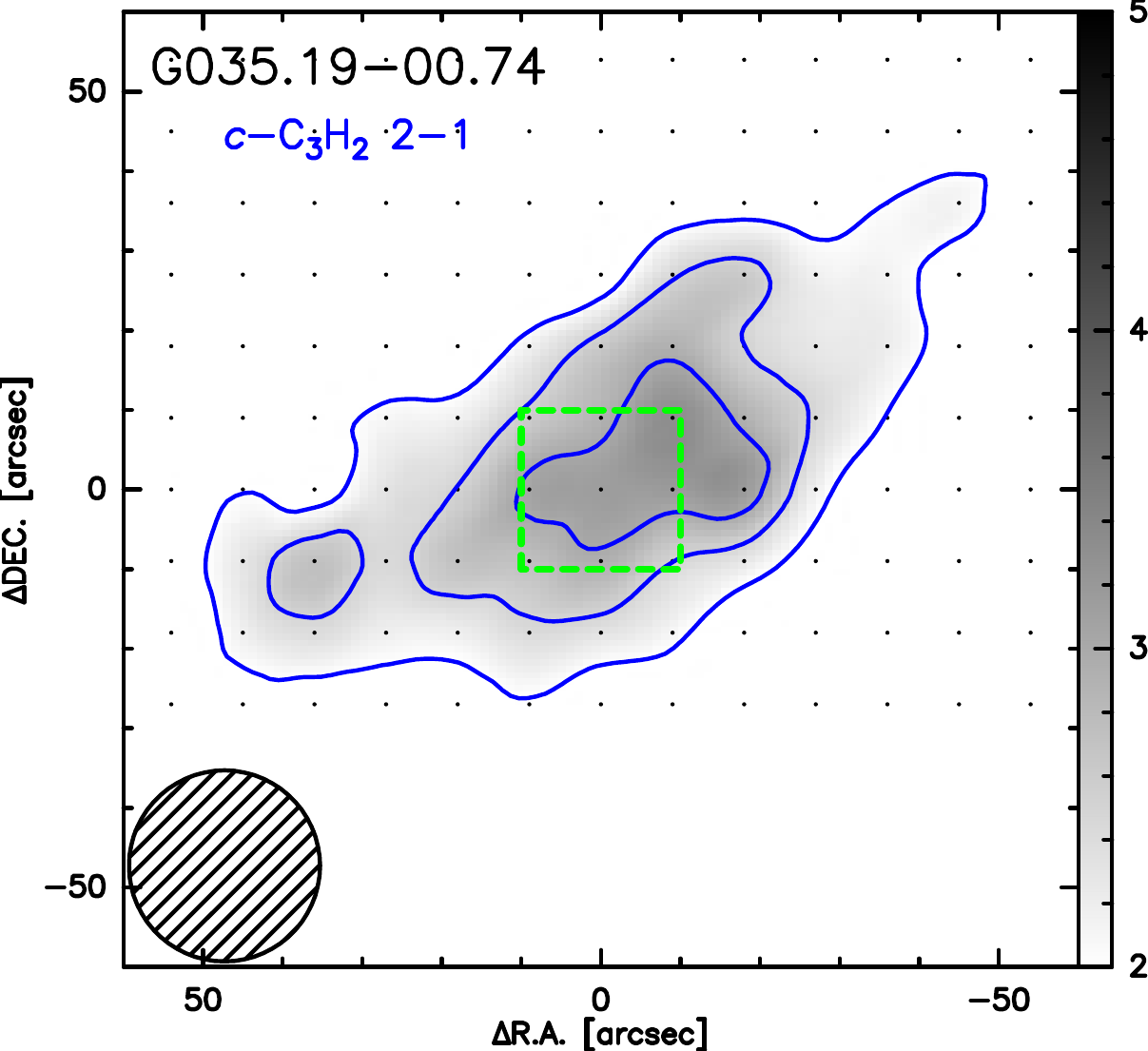}
\includegraphics[width=0.22\columnwidth]{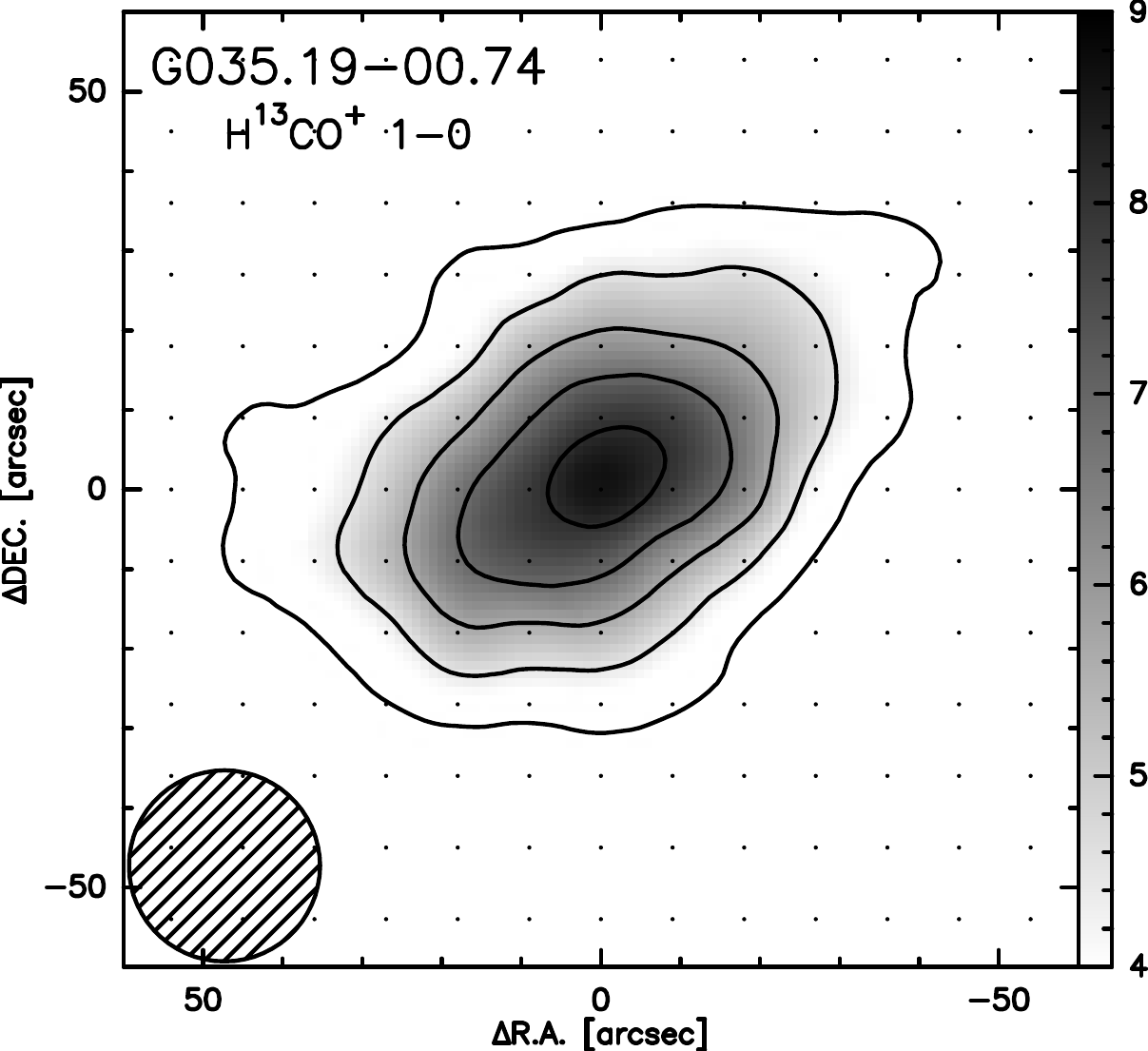}
\includegraphics[width=0.28\columnwidth]{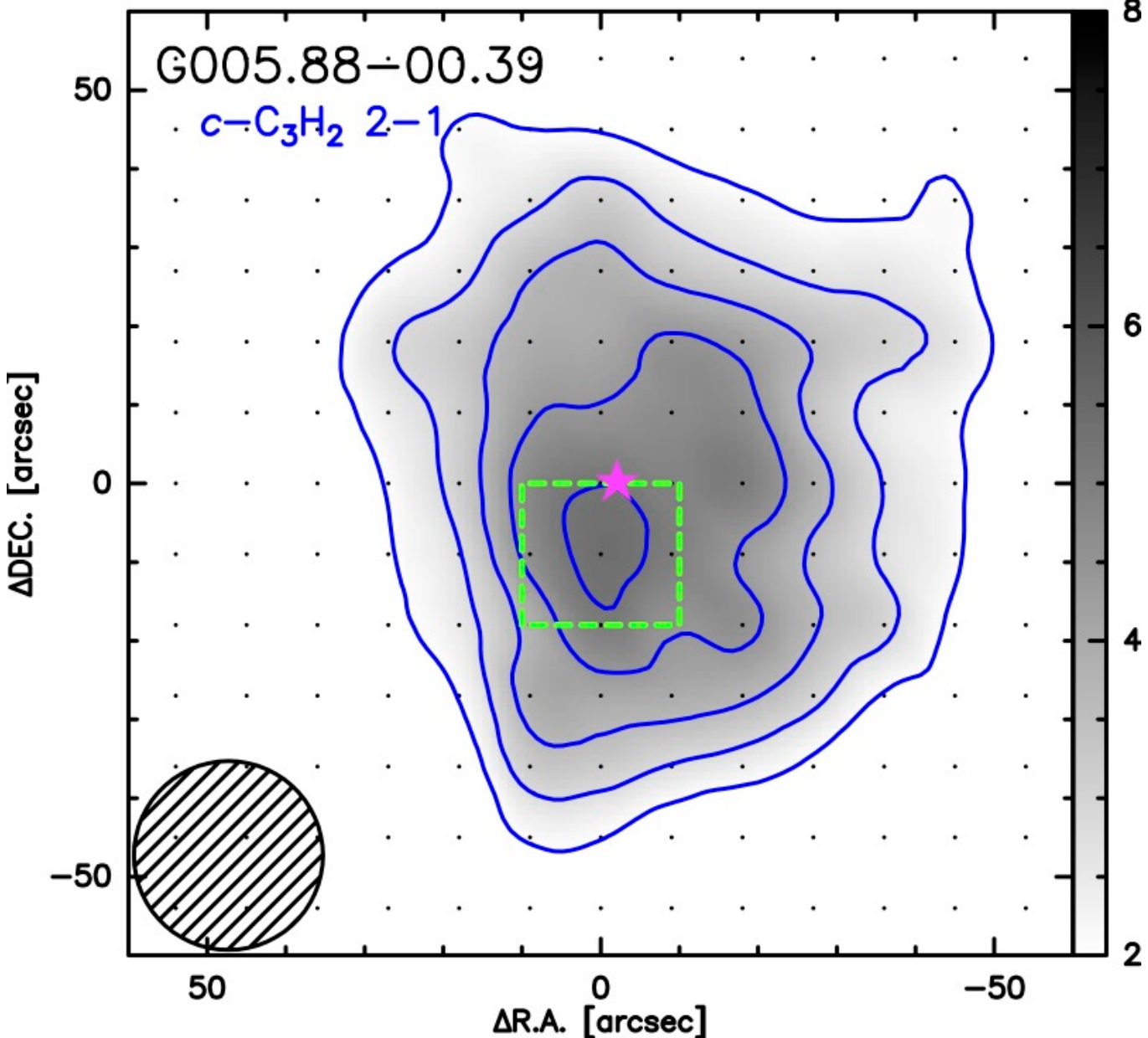}
\includegraphics[width=0.22\columnwidth]{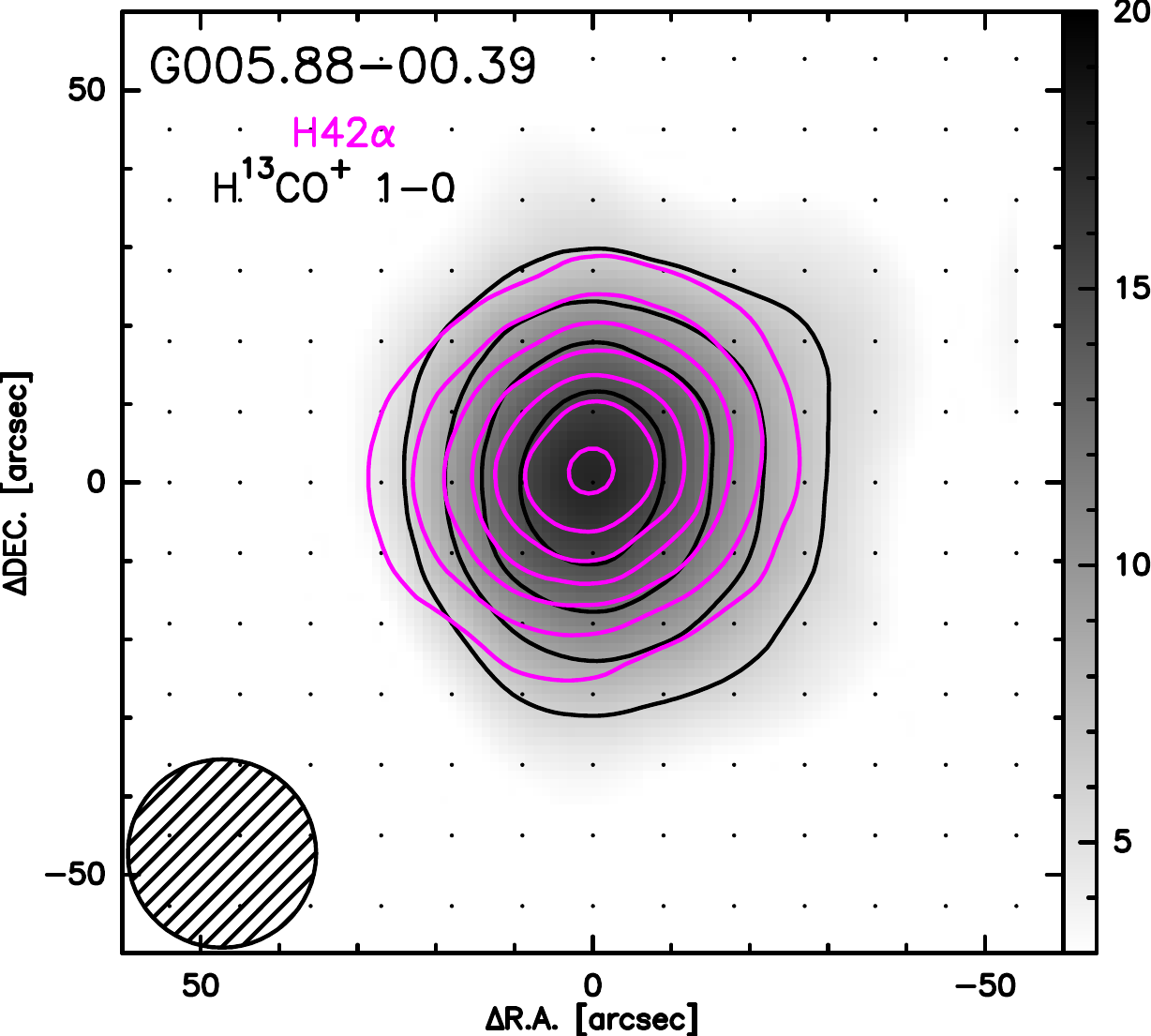}

 \includegraphics[width=0.22\columnwidth]{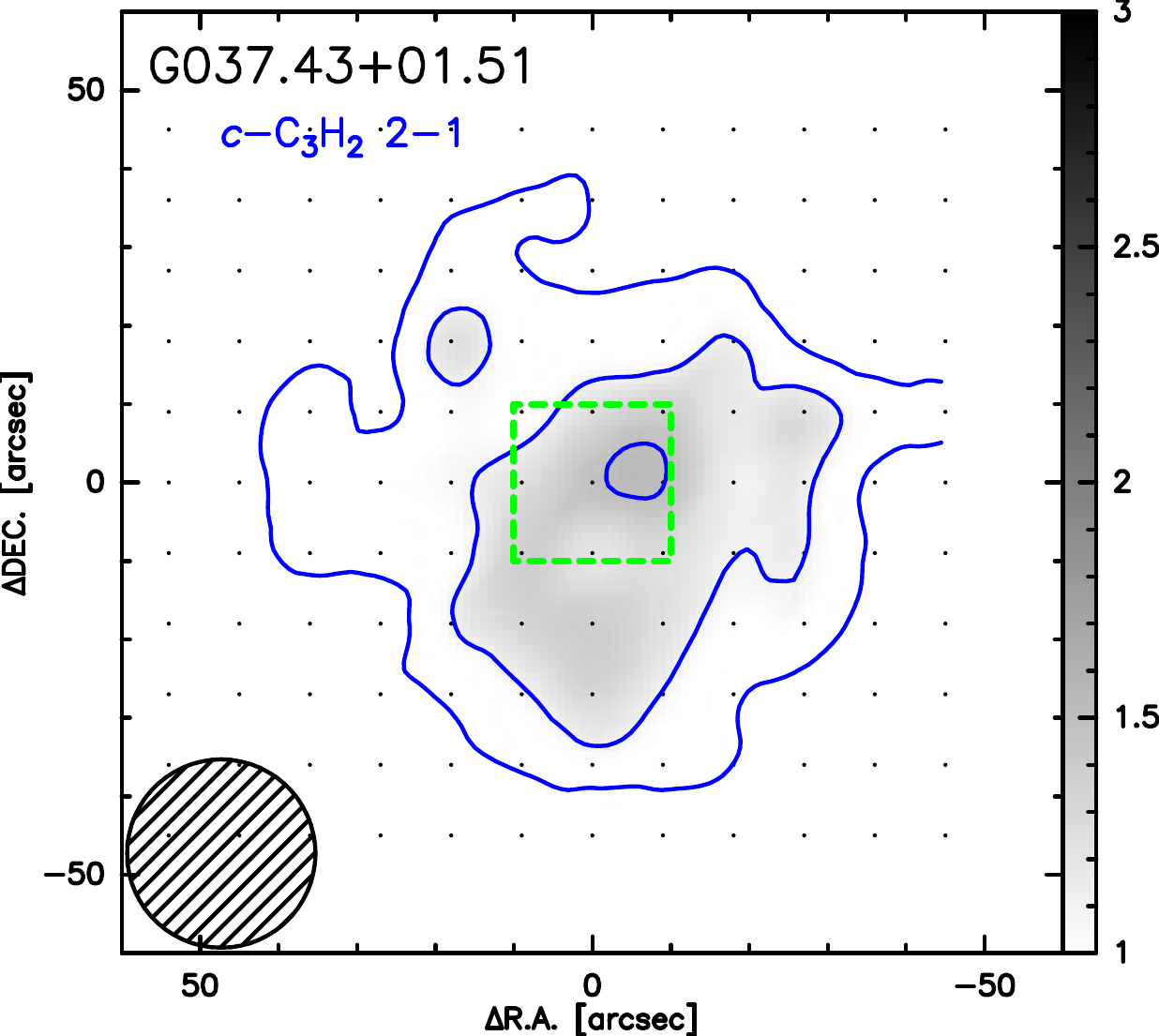}
 \includegraphics[width=0.22\columnwidth]{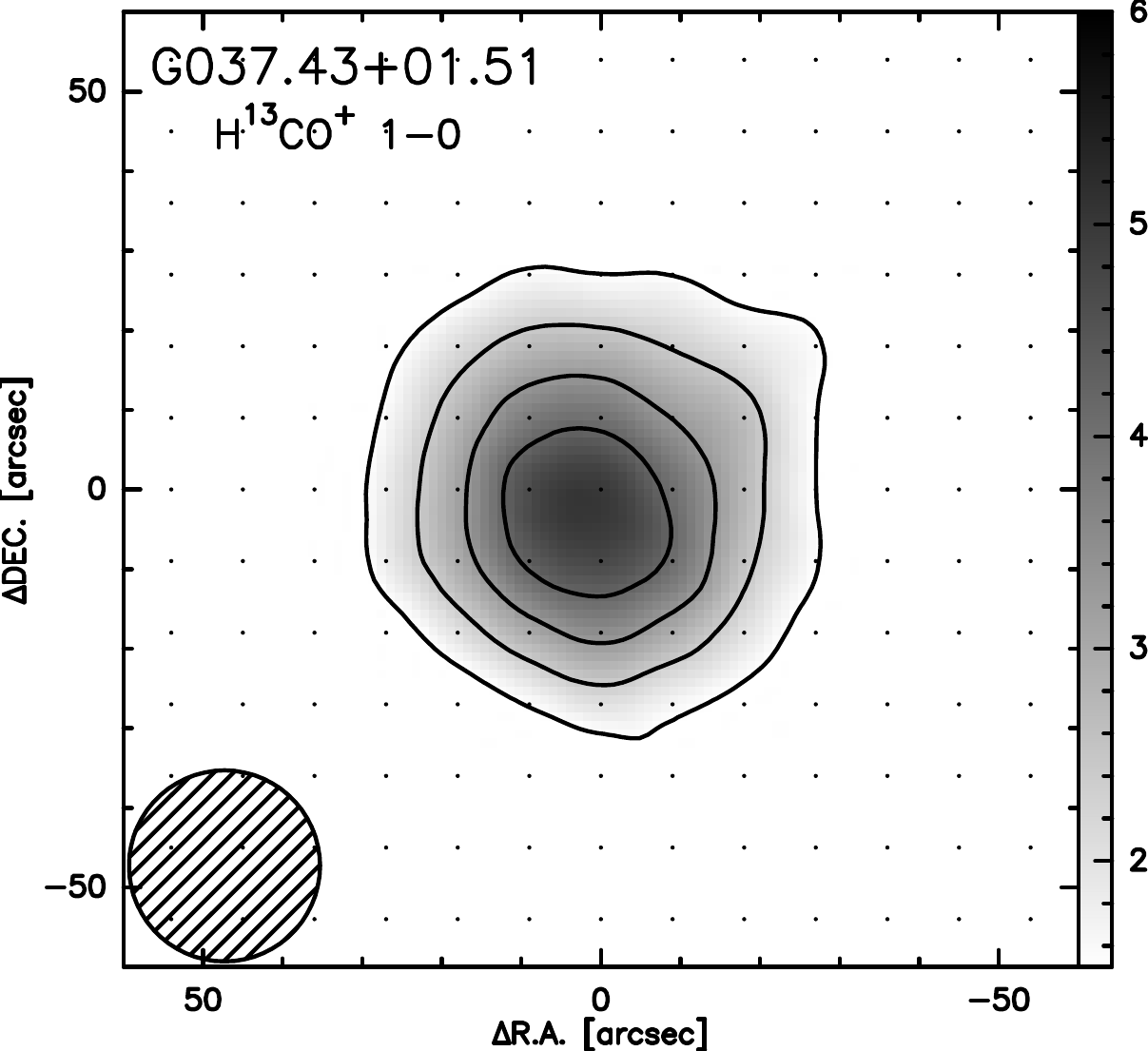}
  \includegraphics[width=0.22\columnwidth]{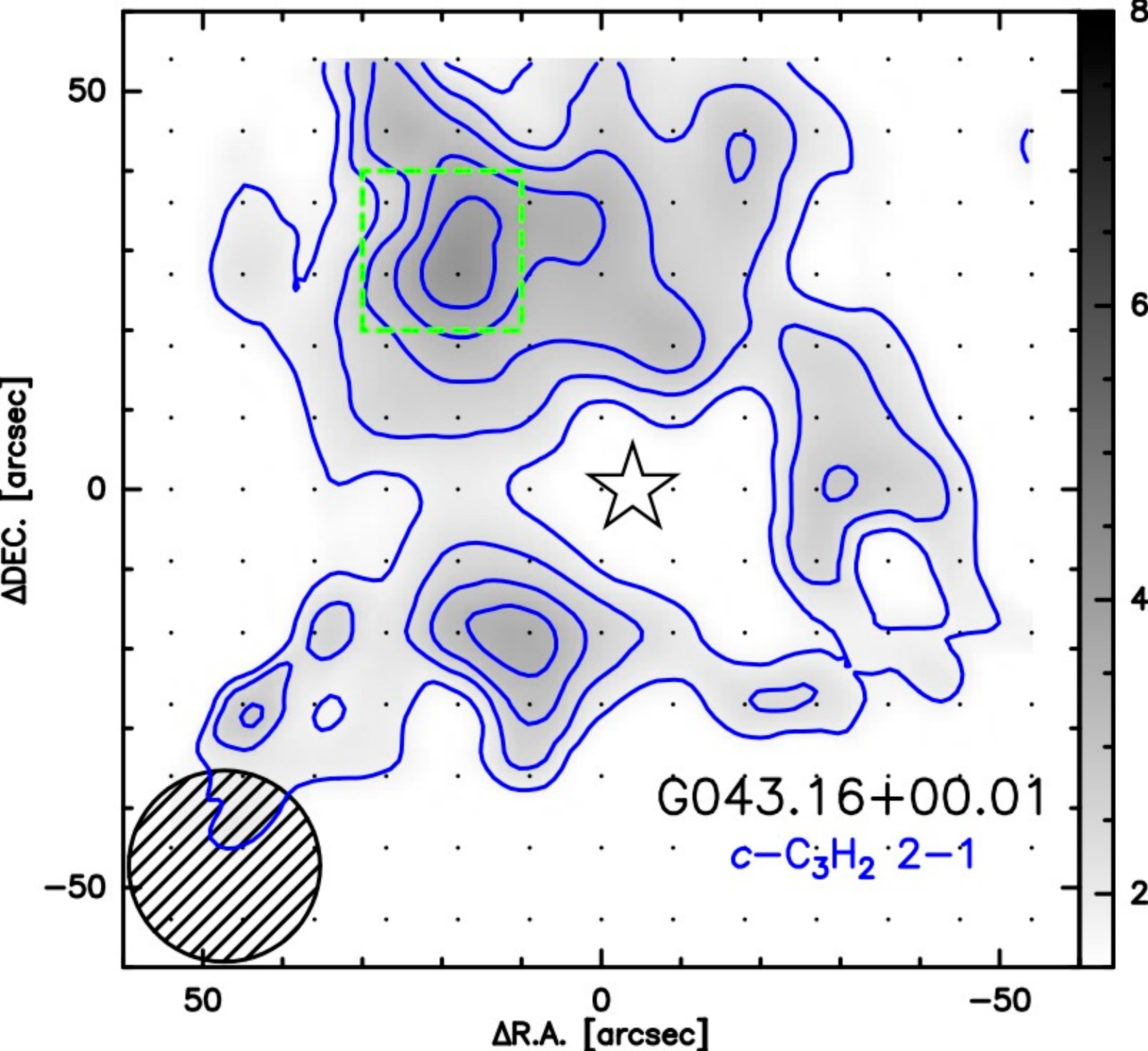}
 \includegraphics[width=0.22\columnwidth]{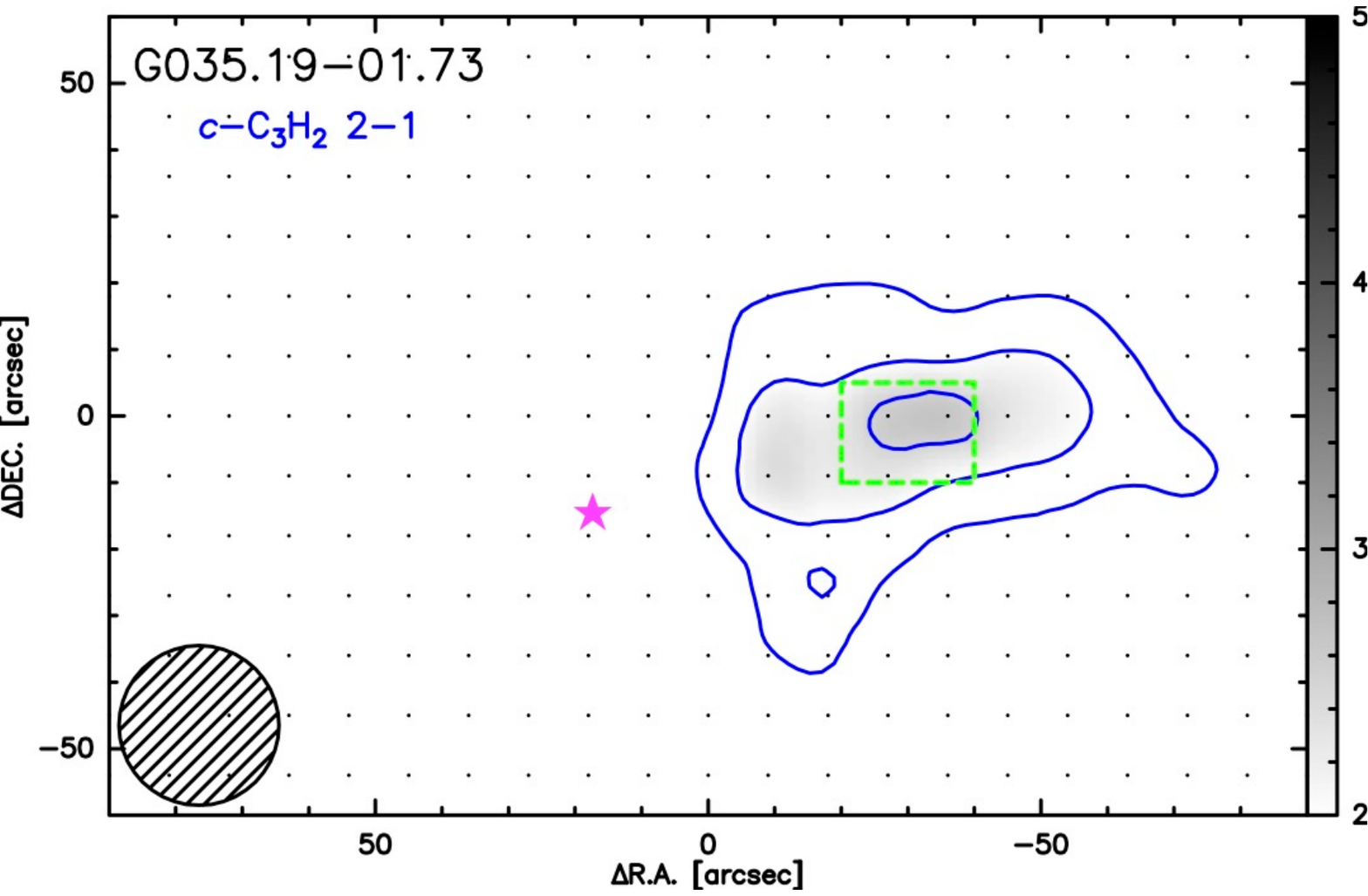}
 
 \includegraphics[width=0.22\columnwidth]{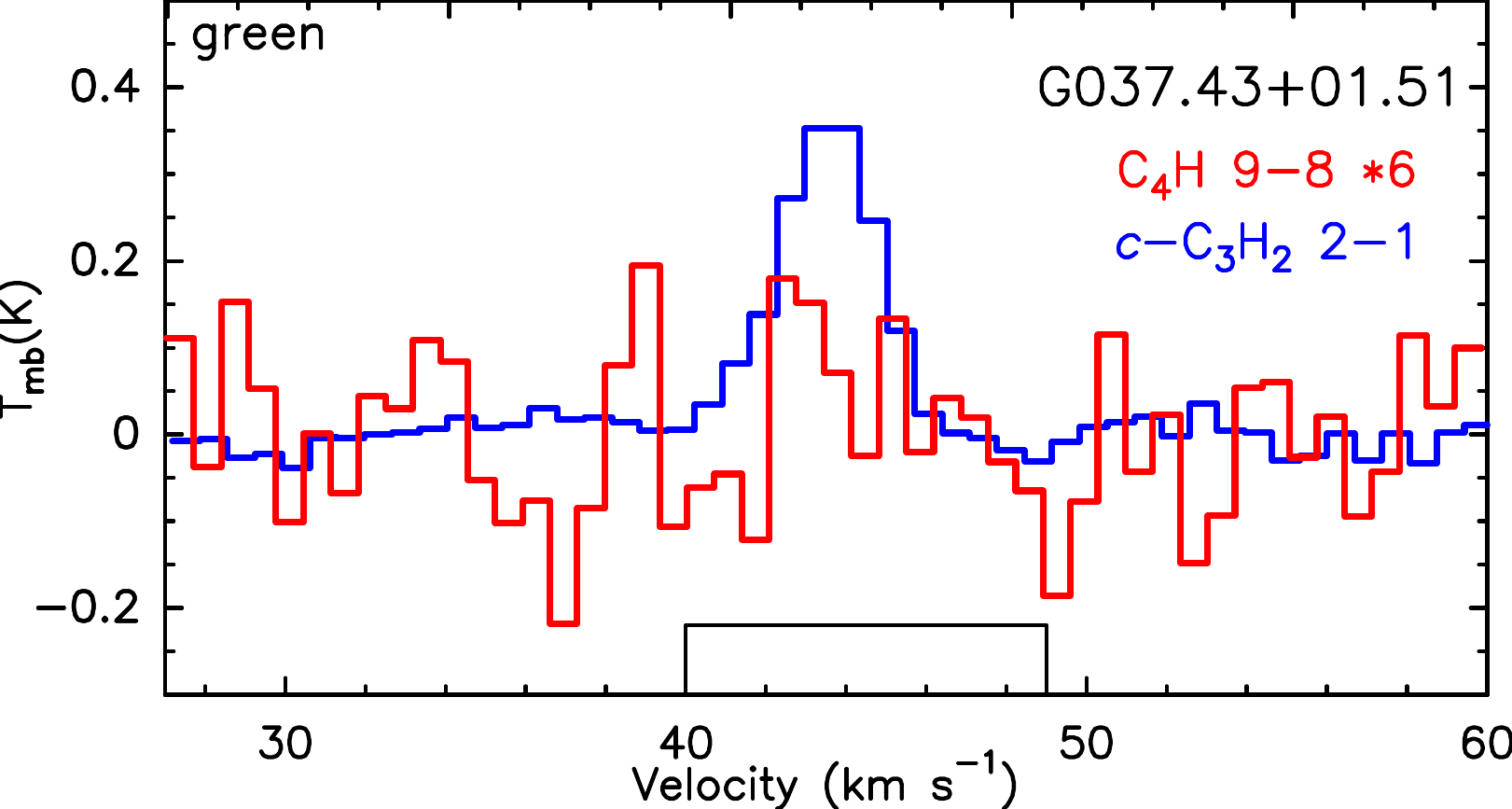}
 \includegraphics[width=0.22\columnwidth]{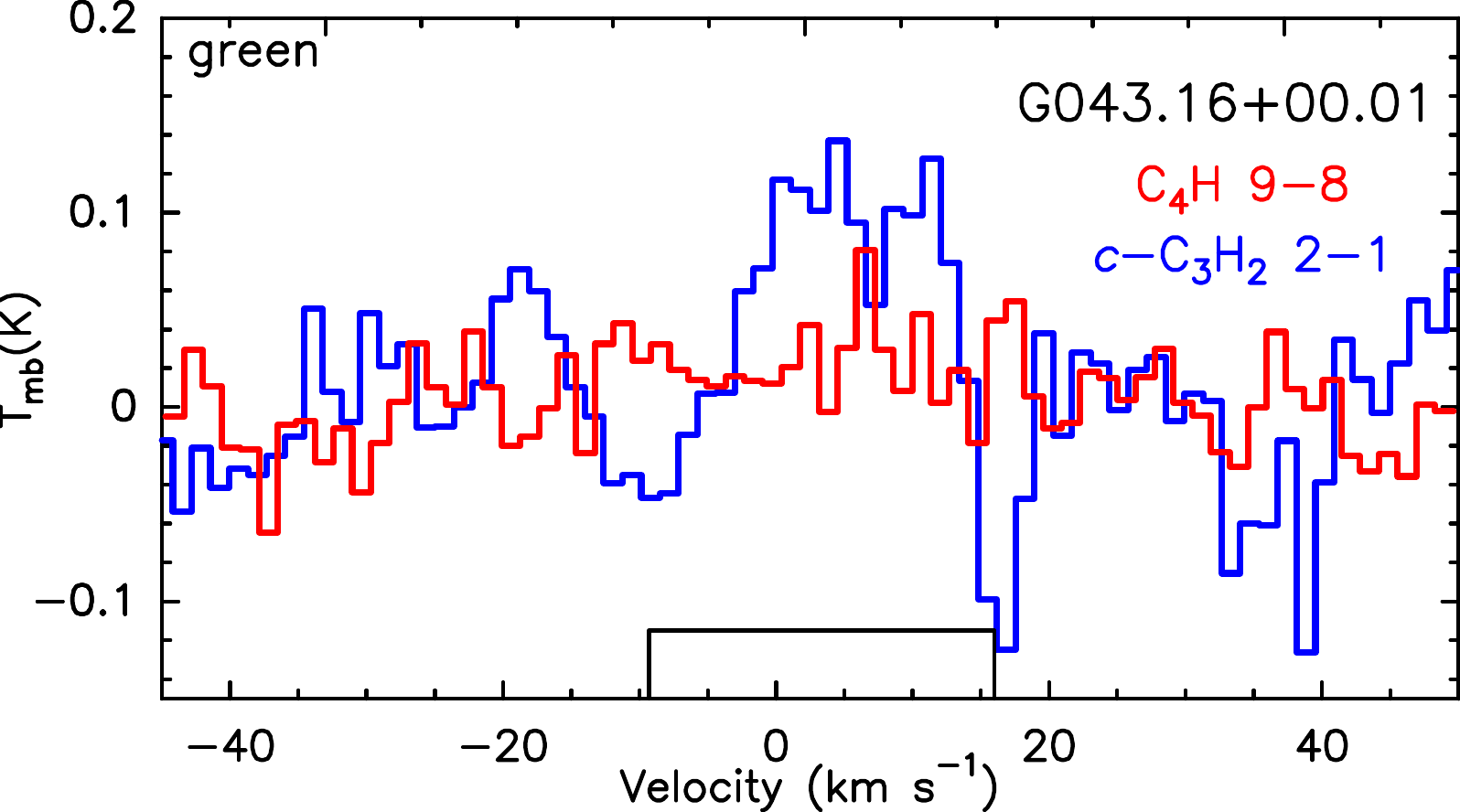}
  \includegraphics[width=0.22\columnwidth]{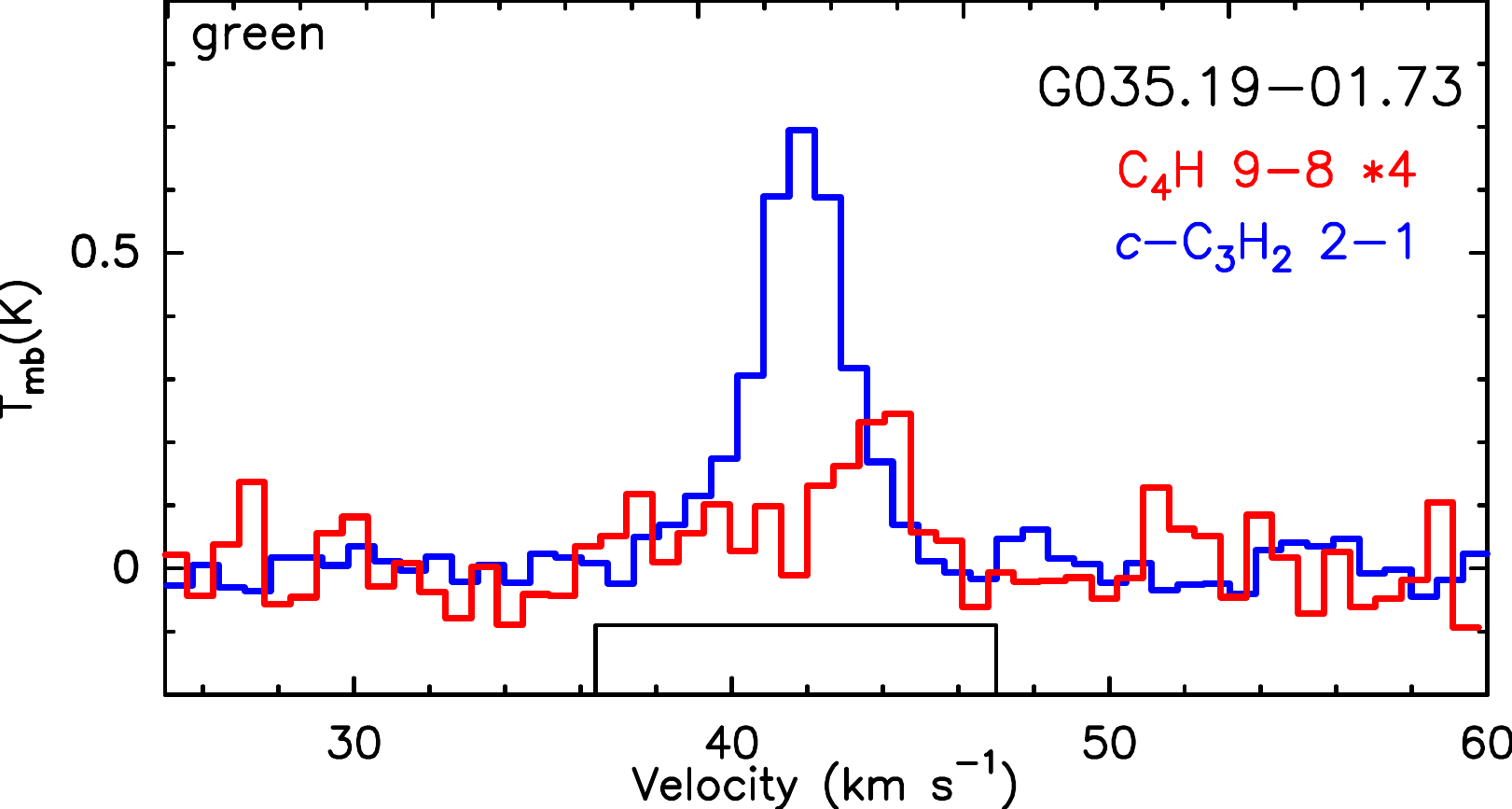}
 \includegraphics[width=0.22\columnwidth]{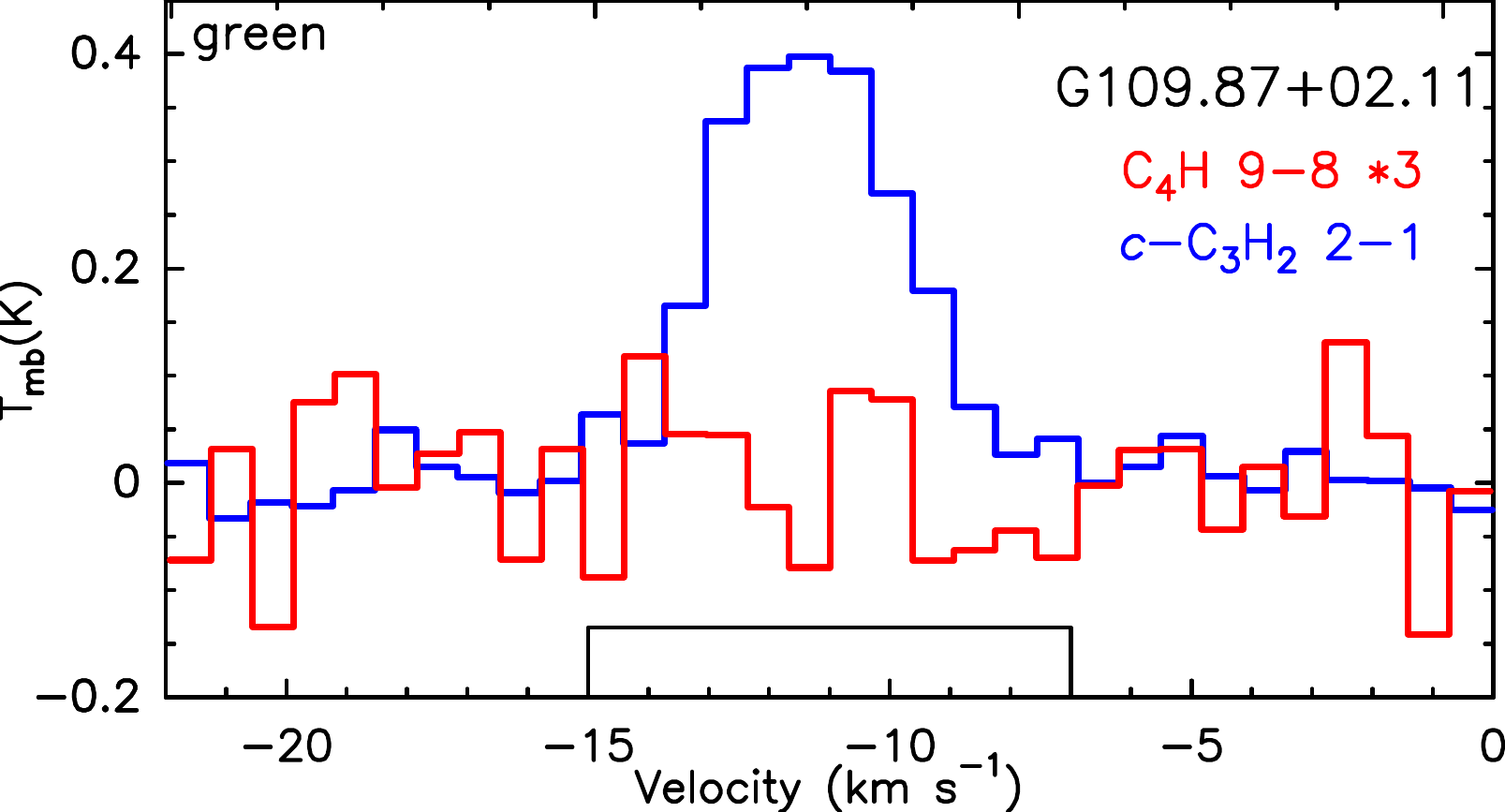}

 \includegraphics[width=0.22\columnwidth]{45}
 \includegraphics[width=0.22\columnwidth]{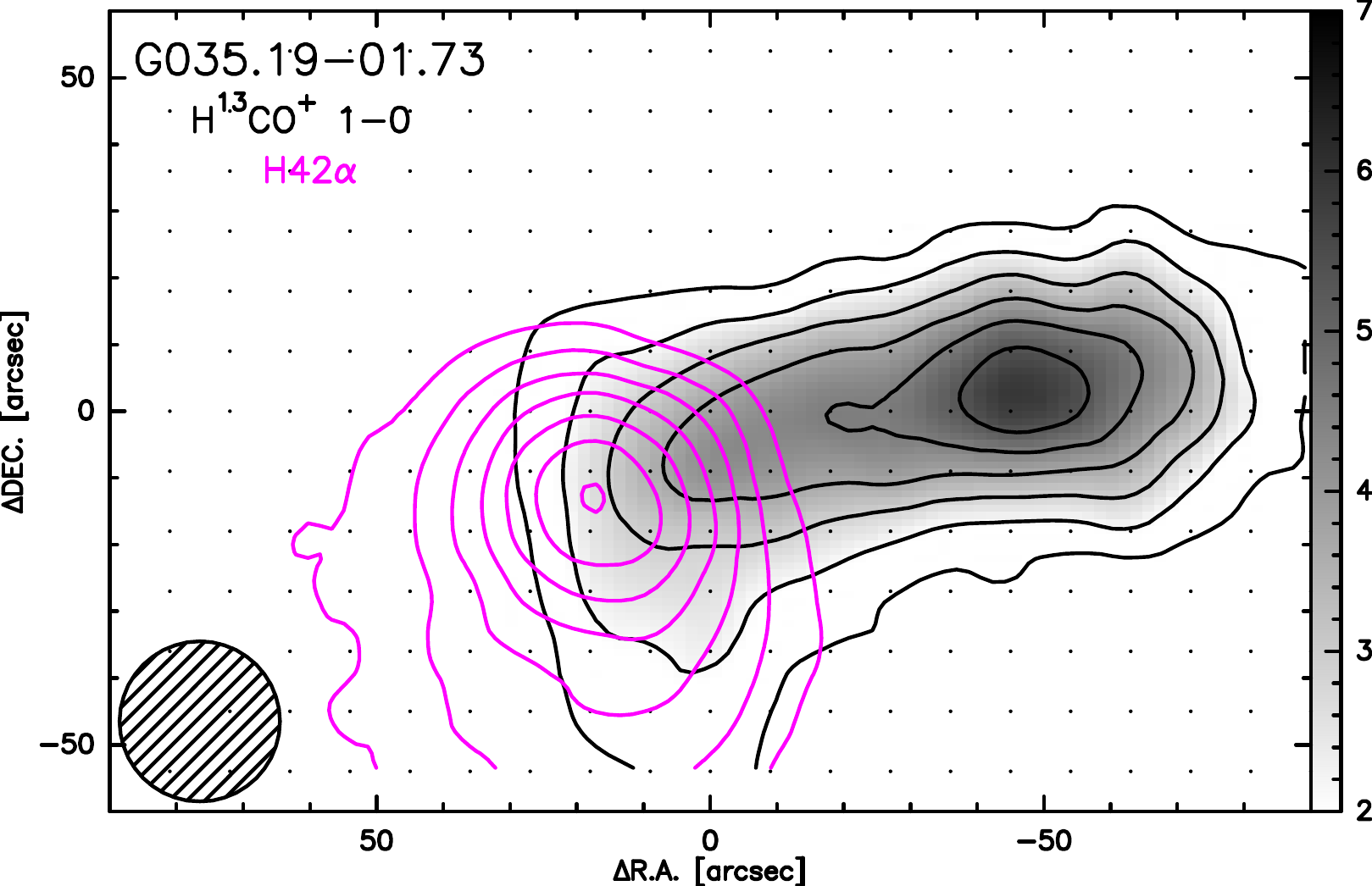}
\includegraphics[width=0.22\columnwidth]{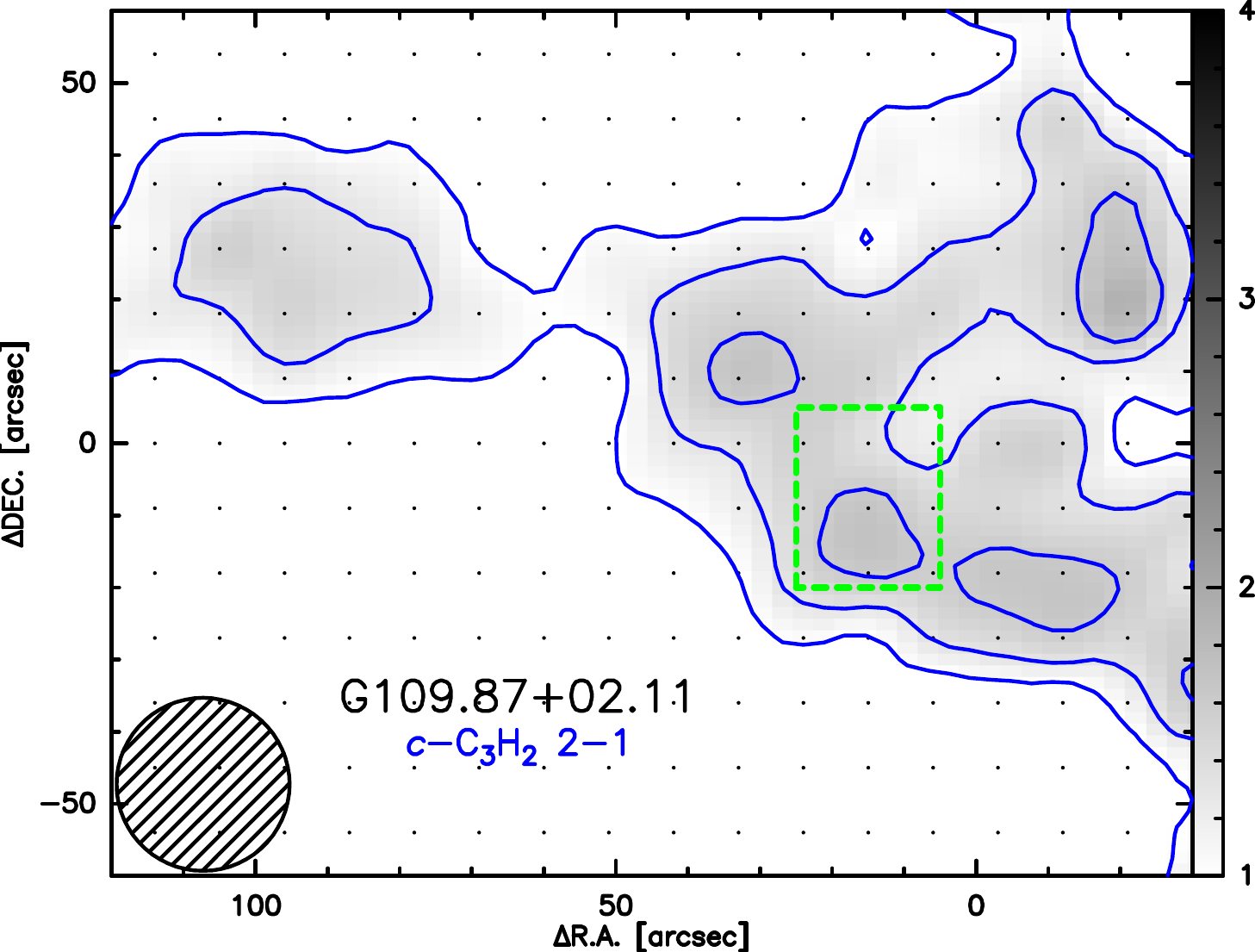}
 \includegraphics[width=0.22\columnwidth]{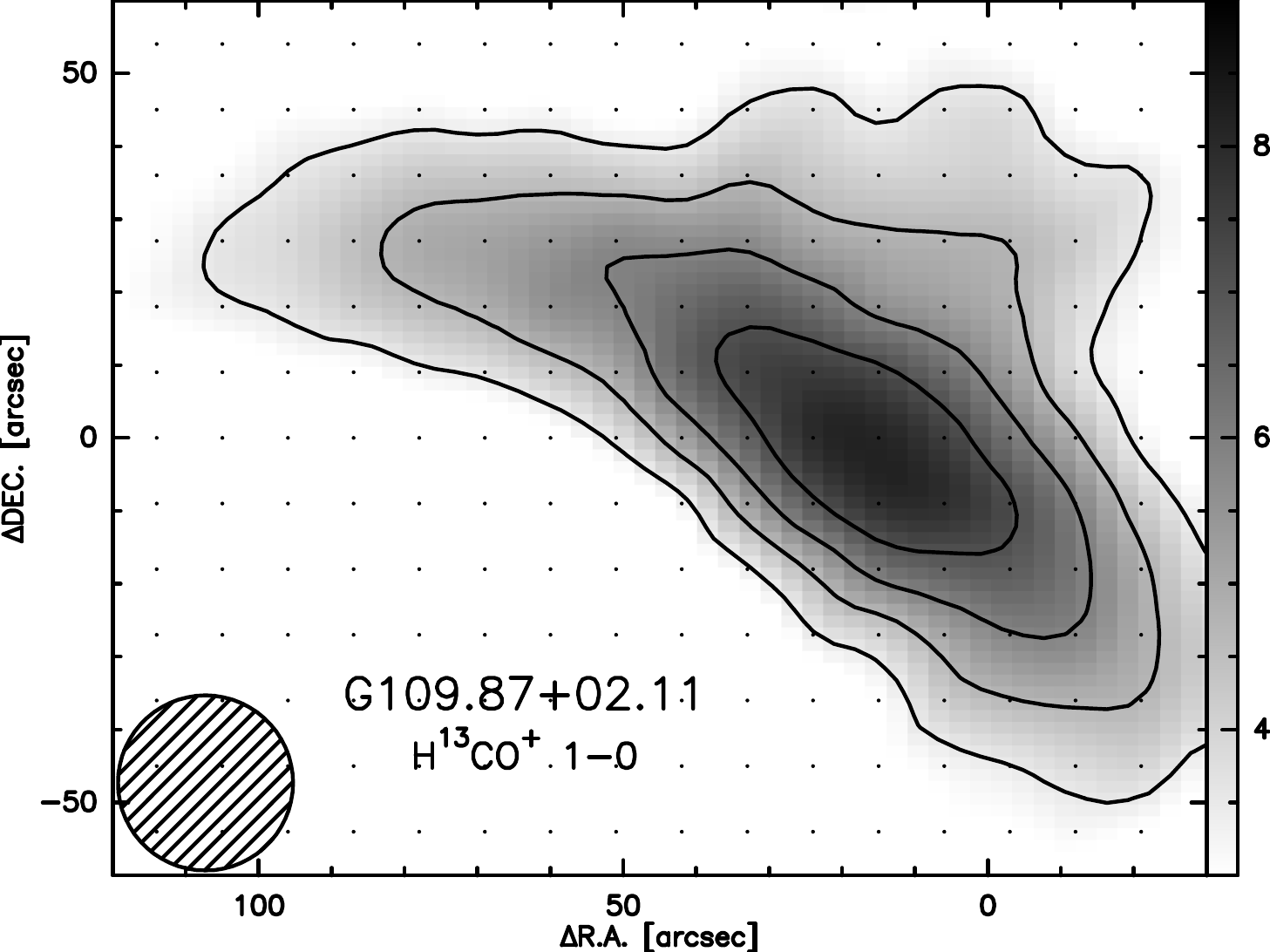}
  
  \includegraphics[width=0.22\columnwidth]{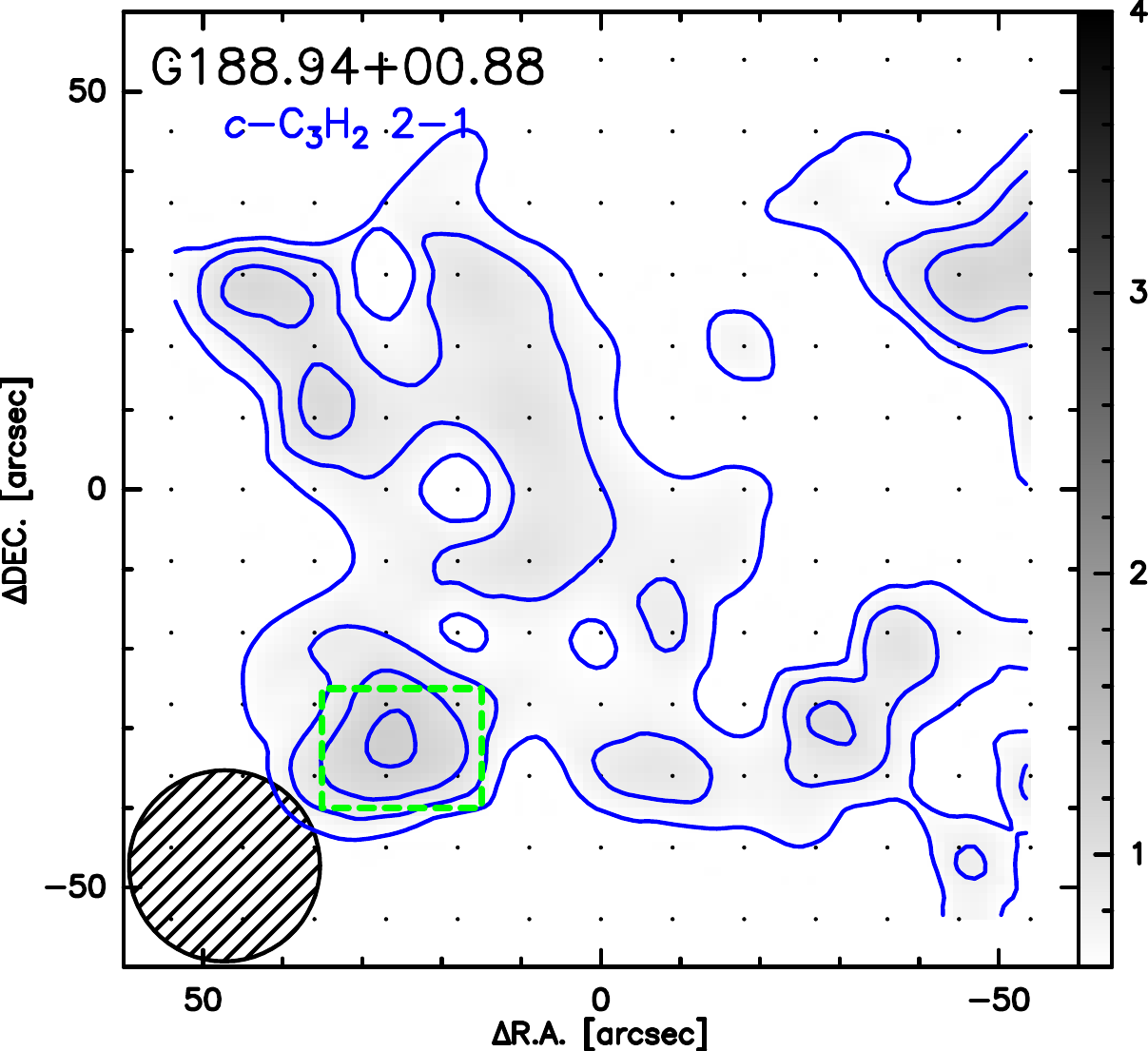}
\includegraphics[width=0.22\columnwidth]{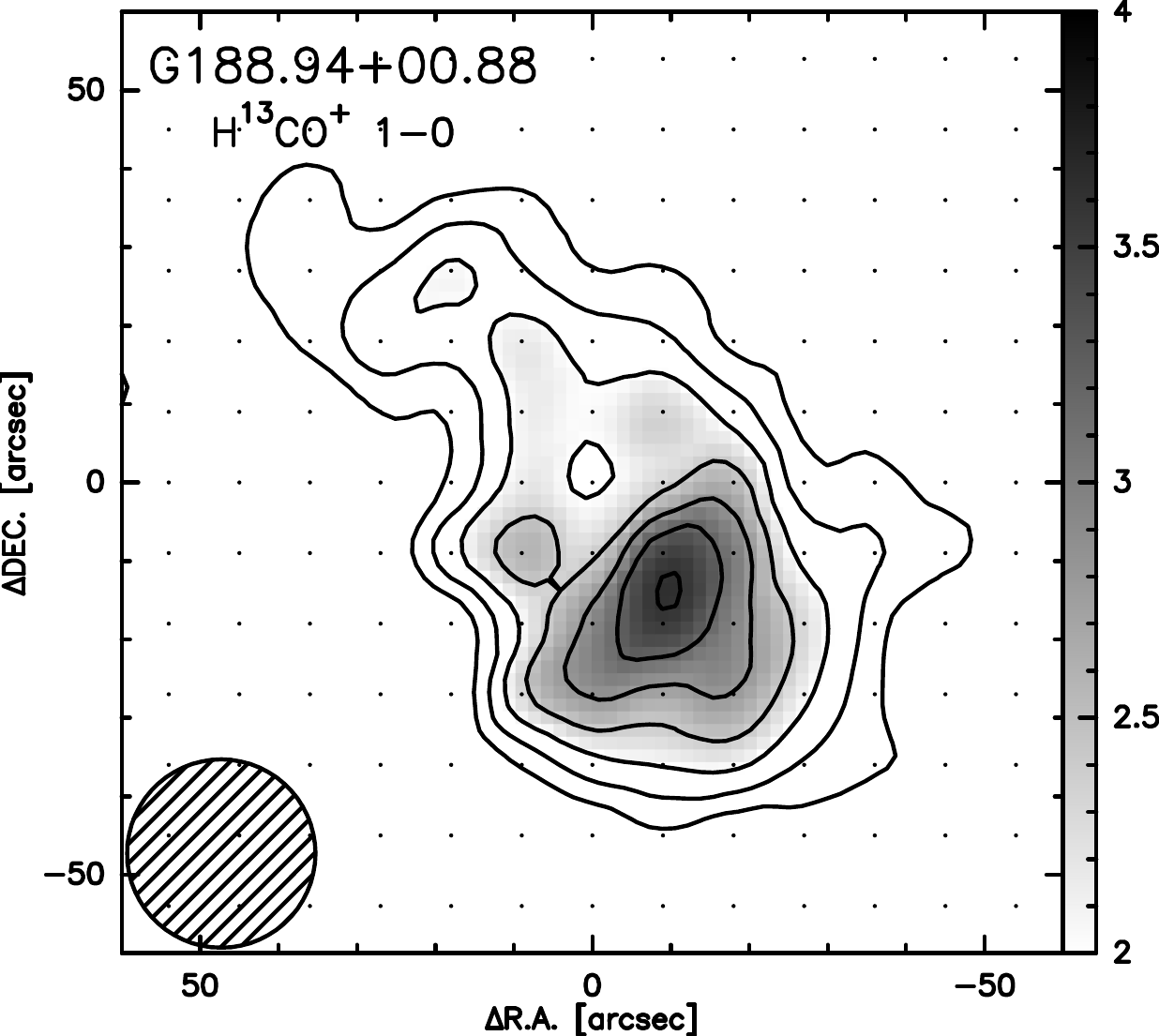}
\includegraphics[width=0.22\columnwidth]{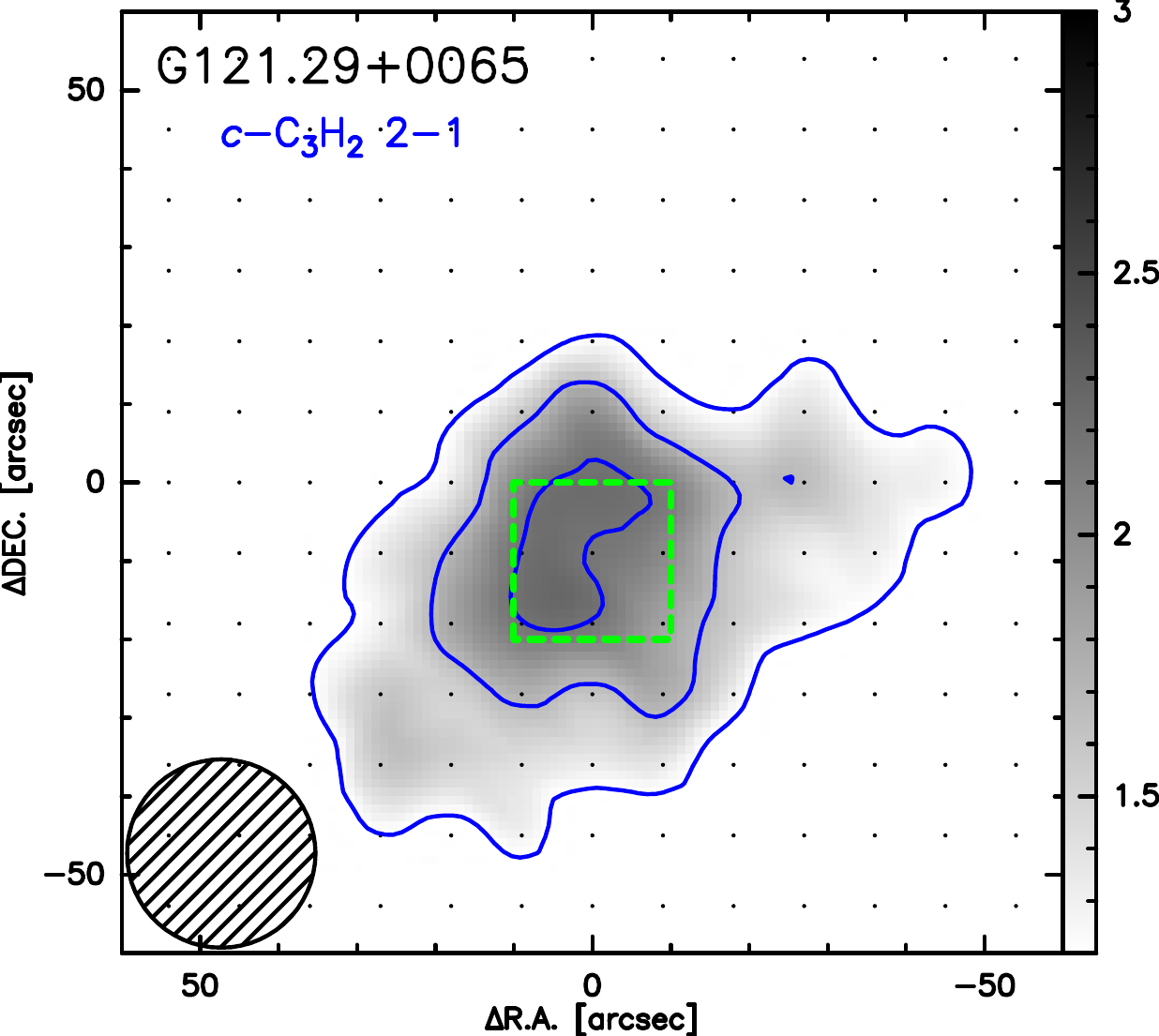}
\includegraphics[width=0.22\columnwidth]{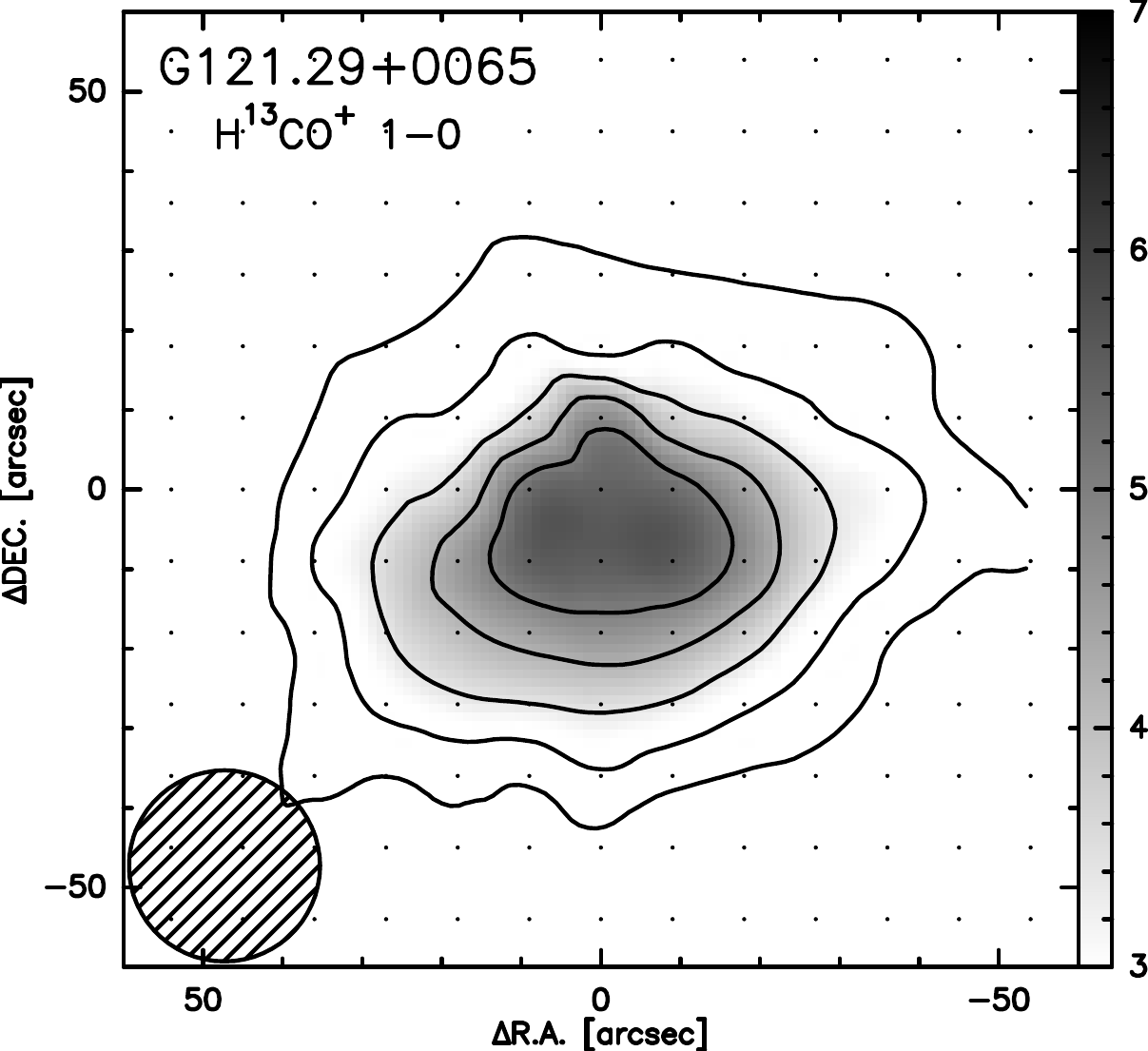}

 \includegraphics[width=0.22\columnwidth]{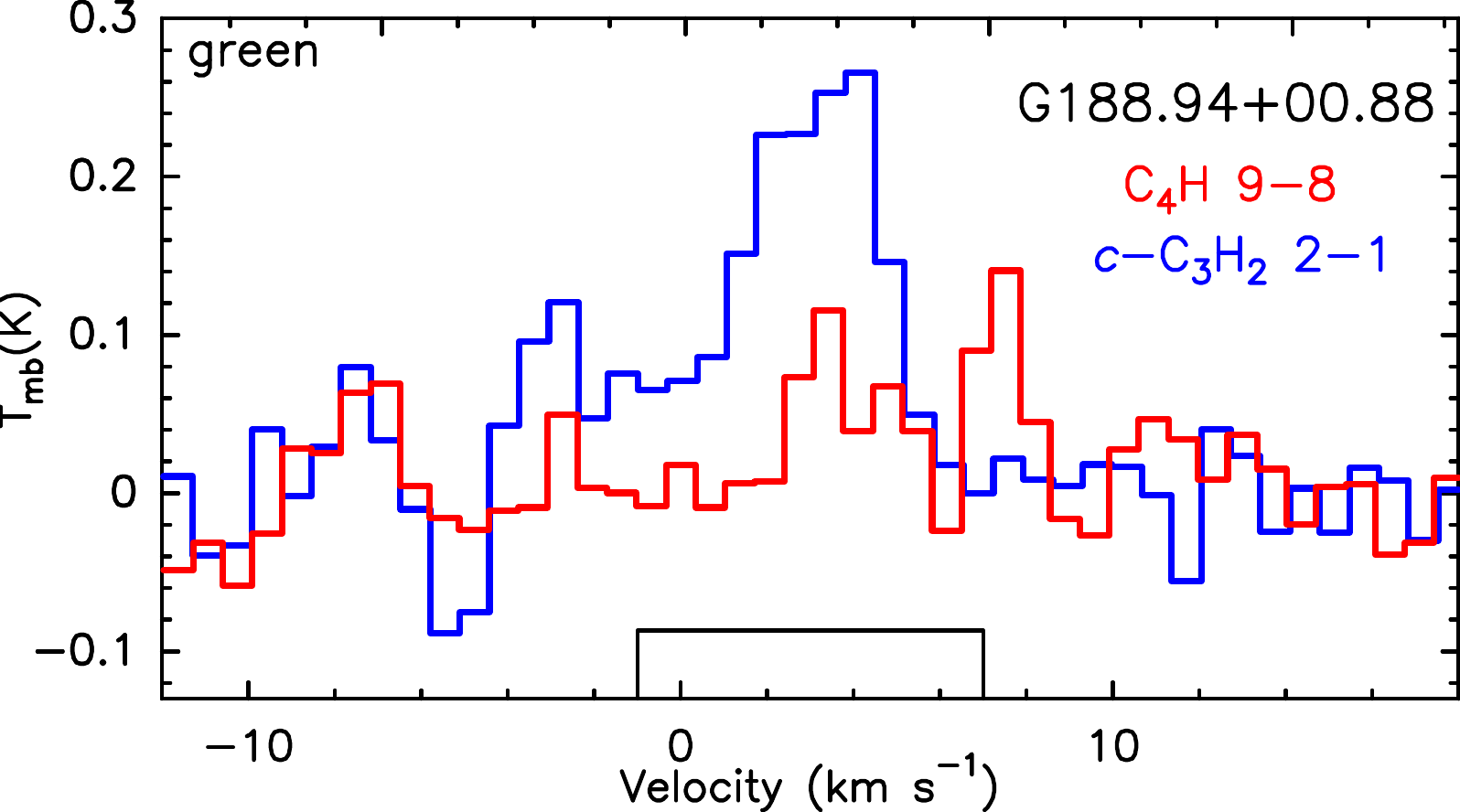}
\includegraphics[width=0.22\columnwidth]{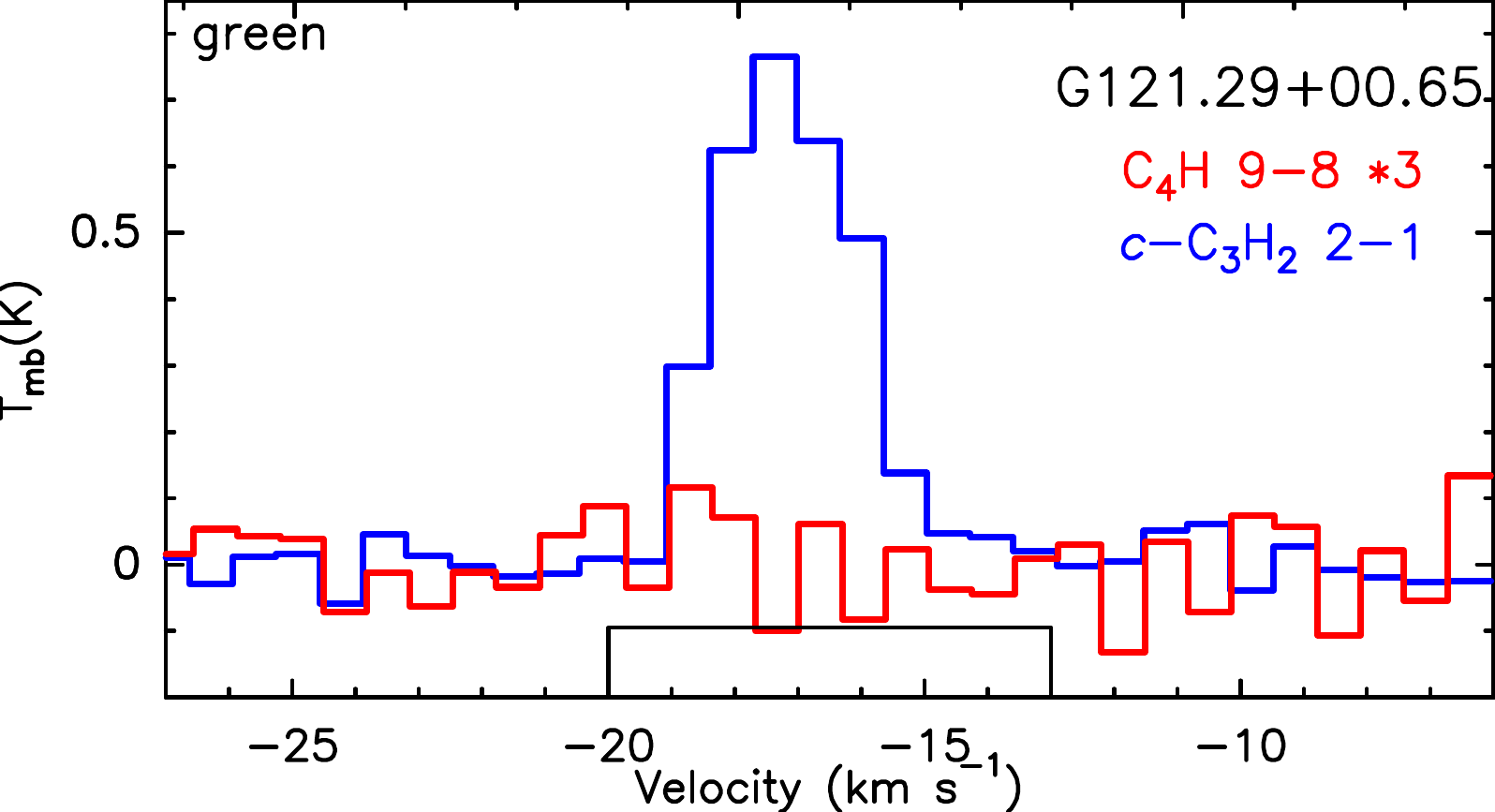}

 \caption{Continued.}

\label{Appendix-1}
\end{center}
\end{figure}

\clearpage
\section{Source information and observing parameters in hot cores} 

\begin{table*}[!htbp]  
\centering
\small  
\setlength{\tabcolsep}{0.1in}  
\vspace{-2mm}

\caption{Source information and observing parameters in hot cores.}
\label{table:source}
\begin{tabular}{cccccccl}
  \hline \hline
Source name  & RA   & Decl.        & rms & Mapping size & $v_{\rm LSR}$  &   \\
Hot cores    & (hh:mm:ss)  & (dd:mm:ss)  & (10$^{-2}$K)  & $('')$   & (km\,s$^{-1}$) &   \\
\hline
G005.88-00.39 & 18:00:30.31 & -24:04:04.50 & 4.5 & 120 $\times$ 120 & 9.0 & \\
G011.91-00.61 & 18:13:59.72 & -18:53:50.30 & 3.2 & 180 $\times$ 180 & 37.0 & \\
G012.80-00.20 & 18:14:14.23 & -17:55:40.50 & 3.8 & 180 $\times$ 180 & 34.0 & \\
G015.03-00.67 & 18:20:22.01 & -16:12:11.30 & 2.5 & 240 $\times$ 240 & 22.0 & \\
G023.43-00.18 & 18:34:39.29 & -08:31:25.40 & 3.1 & 120 $\times$ 120 & 97.0 & \\
G031.28+00.06 & 18:48:12.39 & -01:26:30.70 & 4.4 & 120 $\times$ 120 & 109.0 & \\
G034.39+00.22 & 18:53:19.00 & +01:24:50.80 & 5.3 & 120 $\times$ 120 & 57.0 & \\
G035.19-00.74 & 18:58:13.05 & +01:40:35.70 & 2.4 & 120 $\times$ 120 & 30.0 & \\
G035.19-01.73 & 19:01:45.54 & +01:13:32.50 & 3.5 & 180 $\times$ 120 & 42.0 & \\
G037.43+01.51 & 18:54:14.35 & +04:41:41.70 & 4.7 & 120 $\times$ 120 & 41.0 & \\
G043.16+00.01 & 19:10:13.41 & +09:06:12.80 & 3.4 & 120 $\times$ 120 & 10.0 & \\
G049.48-00.36 & 19:23:39.82 & +14:31:05.00 & 4.1 & 120 $\times$ 120 & 56.0 & \\
G049.48-00.38 & 19:23:43.87 & +14:30:29.50 & 4.3 & 120 $\times$ 120 & 58.0 & \\
G075.76+00.33 & 20:21:41.09 & +37:25:29.30 & 3.4 & 120 $\times$ 120 & -9.0 & \\
G081.75+00.59 & 20:39:01.99 & +42:24:59.30 & 2.7 & 120 $\times$ 180 & -3.0 & \\
G081.87+00.78 & 20:38:36.43 & +42:37:34.80 & 5.5 & 120 $\times$ 120 & 7.0 & \\
G109.87+02.11 & 22:56:18.10 & +62:01:49.50 & 5.2 & 150 $\times$ 120 & -7.0 & \\
G111.54+00.77 & 23:13:45.36 & +61:28:10.60 & 3.7 & 120 $\times$ 200 & -57.0 & \\
G121.29+00.65 & 00:36:47.35 & +63:29:02.20 & 4.8 & 120 $\times$ 120 & -23.0 & \\
G133.94+01.06 & 02:27:03.82 & +61:52:25.20 & 4.6 & 120 $\times$ 120 & -47.0 & \\
G188.94+00.88 & 06:08:53.35 & +21:38:28.70 & 6.3 & 120 $\times$ 120 & 8.0 & \\
G192.60-00.04 & 06:12:54.02 & +17:59:47.30 & 4.2 & 120 $\times$ 240 & 6.0 & \\
\hline
\end{tabular}

\vspace{4mm}  

\caption{Spectral detection information of C$_4$H 9$-8$, $c$-C$_3$H$_2$ 2$-1$, and H$^{13}$CO$^+$ 1$-0$.}
\label{table:spectral detection}
\setlength{\tabcolsep}{0.14in}
\begin{tabular}{cccccccccl}

  \hline \hline
Source name          & \multicolumn{1}{c}{C$_4$H(9-8)} & \multicolumn{1}{c}{$c$-C$_3$H$_2$(2-1)} & \multicolumn{1}{c}{H$^{13}$CO$^+$(1-0)} & Colour box & \\
\hline
G005.88-00.39 & ... & $\surd$ & $\surd$ & green   \\
G011.91-00.61 & ... & $\surd$ & $\surd$ & green   \\
G011.91-00.61 & ... & $\surd$ & $\surd$ & cyan    \\
G012.80-00.20 & ... & $\surd$ & $\surd$ & green   \\
G015.03-00.67 & $\surd$ & $\surd$ & $\surd$ & green   \\
G015.03-00.67 & $\surd$ & $\surd$ & $\surd$ & cyan    \\
G023.43-00.18 & $\surd$ & $\surd$ & $\surd$ & green   \\
G031.28+00.66 & ... & $\surd$ & $\surd$ & green   \\
G034.39+00.22 & $\surd$ & $\surd$ & $\surd$ & green   \\
G035.19-00.74 & ... & $\surd$ & $\surd$ & green   \\
G035.19-01.73 & ... & $\surd$ & $\surd$ & green   \\
G037.43+01.51 & ... & $\surd$ & $\surd$ & green   \\
G043.16+00.01 & ... & $\surd$ & $\surd$ & green   \\
G049.48-00.36 & $\surd$ & $\surd$ & $\surd$ & green   \\
G049.48-00.36 & $\surd$ & $\surd$ & $\surd$ & cyan    \\
G049.48-00.38 & $\surd$ & $\surd$ & $\surd$ & green   \\
G049.48-00.38 & $\surd$ & $\surd$ & $\surd$ & cyan    \\
G075.76+00.33 & $\surd$ & $\surd$ & $\surd$ & green   \\
G081.75+00.59-1 & $\surd$ & $\surd$ & $\surd$ & green   \\
G081.75+00.59-1 & $\surd$ & $\surd$ & $\surd$ & cyan    \\
G081.75+00.59-2 & $\surd$ & $\surd$ & $\surd$ & green   \\
G081.75+00.59-2 & $\surd$ & $\surd$ & $\surd$ & cyan    \\
G081.87+00.78 & $\surd$ & $\surd$ & $\surd$ & green   \\
G109.87+02.11 & ... & $\surd$ & $\surd$ & green   \\
G111.54+00.77 & ... & $\surd$ & $\surd$ & green   \\
G111.54+00.77 & ... & $\surd$ & $\surd$ & cyan    \\
G121.29+00.65 & ... & $\surd$ & $\surd$ & green   \\
G133.94+01.06 & $\surd$ & $\surd$ & $\surd$ & green   \\
G188.94+00.88 & ... & $\surd$ & $\surd$ & green   \\
G192.60-00.04 & $\surd$ & $\surd$ & $\surd$ & green   \\
G192.60-00.04 & $\surd$ & $\surd$ & $\surd$ & cyan    \\
\hline
\end{tabular}
\end{table*}

\begin{table*}
\centering
\small
\setlength{\tabcolsep}{0.0000002in}
\small
\centering
\caption{\label{table:step}Information parameters for the map setting of  C$_4$H 9$-8$, $c$-C$_3$H$_2$ 2$-1$, and H$^{13}$CO$^+$ 1$-0$ in hot cores.}

\vspace{-0.5mm}
\begin{tabular}{cccccccccccccl}
  \hline
    \hline
             
  \multirow{3}{*}{Source name}          &   \multicolumn{3}{c}{\cfh(9-8)}   &   \multicolumn{3}{c}{\cctht(2-1) }      &   \multicolumn{3}{c}{H$^{13}$CO$^+$ (1$-0$) }  &   \multicolumn{3}{c}{ H42$\alpha$  }     \\
        &  1 $\sigma$    & starting  & step &  1 $\sigma$      &  starting    & step  &  1 $\sigma$      &  starting    & step &  1 $\sigma$      &  starting    & step   \\
        &  K km\,s$^{-1}$ &  K km\,s$^{-1}$ &  K km\,s$^{-1}$  &  K km\,s$^{-1}$ &  K km\,s$^{-1}$   &  K km\,s$^{-1}$  &  K km\,s$^{-1}$ &  K km\,s$^{-1}$   &  K km\,s$^{-1}$  &  K km\,s$^{-1}$ &  K km\,s$^{-1}$   &  K km\,s$^{-1}$    \\
                 
\hline

G005.88-00.39           &       0.06    &       ...     &       ...     &       0.15         &       3.8     &       2.0     & 0.2   &       6.0     &       3.0     &0.25         &       4 .0    &       5.0     &       \\
G011.91-00.61           &       0.06    &       ...     &       ...     &       0.10         &       2.5     &       0.55    & 0.17  &       1.8     &       1.2         &0.2    &       1.8     &       0.8     &       \\
G012.80-00.20           &       0.07    &       ...     &       ...     &       0.17         &       4.5     &       2.0     &0.15   &       4.0     &       2.5         &...    &       ...     &       ...     &       \\
G015.03-00.67           &       0.26    &       1.4     &       0.3     &       0.20         &       4.0     &       2.0     &0.22   &       4.0     &       3.0     &0.25   &       5.0         &       3.0     &       \\
G023.43-00.18   &       0.22    &       0.5     &       0.2     &       0.18         &       1.5     &       0.8     &       0.15    &       2.0     &       1.2         &...    &       ...     &       ...     &\\
G031.28+00.06           &       0.045   &       ...     &       ...     &       0.16         &       1.8     &       0.8     &0.13   &       1.2     &       0.5     &       0.12    &       0.5         &       0.2     &\\
G034.39+00.22           &       0.21    &       0.7     &       0.2     &       0.32         &       3.2     &       0.8     &       0.21    &       3.0&    1.0         &...    &       ...     &       ...     &\\
G035.19-00.74           &       0.06    &       ...     &       ...     &       0.16         &       3.3     &       1.2     &       0.24    &       3.3     &       1.2         &...    &       ...     &       ...     &\\
G035.19-01.73   &       0.05    &       ...     &       ...     &       0.13         &       1.3     &       0.8     &       0.15    &       1.3     &       0.8     &0.25         &       3.5     &       2.2     &\\
G037.43+01.51           &       0.053   &       ...     &       ...     &       0.13         &       1.5     &       0.9     &0.13   &       1.5     &       0.9     &...    &       ...     &       ...     &       \\
G043.16+00.01           &       0.05    &       ...     &       ...     &       0.48         &       3.0     &       1.5     &0.22   &       3.2     &       1.2         &       ...     &       ...     &       ...     &\\
G049.48-00.36           &       0.23    &       0.8     &       0.3     &       0.24         &       4.0     &       2.3     &0.24   &       4.0     &       2.3     &0.21   &       4.0&    2.0         &       \\
G049.48-00.38           &       0.25    &       0.8     &       0.3     &       0.46         &       4.0     &       2.3     &0.24   &       4.0     &       2.3&0.21         &       4.0     &       2.0     &       \\
G075.76+00.33           &       0.09    &       0.3     &       0.1     &       0.25         &       1.9     &       0.8     &0.15   &       1.2     &       1.0     &0.17   &       2.0         &       1.0     &       \\
G081.75+00.59-1         &       0.06    &       ...     &       ...     &       0.22         &       2.4     &       1.2     &0.23   &       2.5&    1.2     &...    &       ...     &       ...         &       \\
G081.75+00.59-2         &       0.16    &       0.8     &       0.2     &       0.18         &       3.2     &       0.8     &0.19   &       3.2     &       1.2         &...    &       ...     &       ...     &       \\
G081.87+00.78           &       0.16    &       2       &       0.3     &       0.18         &       2.7     &       1.0     &0.15   &       2.2     &       1.3         &...    &       ...     &       ...     &       \\
G109.87+02.11           &       0.06    &       ...     &       ...     &       0.22         &       3.2     &       1.2     &0.28   &       3.5     &       1.2&...         &       ...     &       ...     &       \\
G111.54+00.77           &       0.06    &       ...     &       ...     &       0.17         &       2.5     &       1.0     &       0.16    &       2.5     &       1.0     &0.12         &       1.2     &       0.8     &\\
G121.29+00.65           &       0.04    &       ...     &       ...     &       0.25         &       2.5     &       0.8     &       0.15    &       1.8     &       0.8         &...    &       ...     &       ... &\\
G133.94+01.06           &       0.24    &       0.8     &       0.2     &       0.13         &       2.0     &       0.7     &0.21   &       2.0     &       0.8         &...    &       ...     &       ...     &       \\
G188.94+00.88           &       0.05    &       ...     &       ...     &       0.25         &       1.8     &       0.5     &0.15   &       1.2     &       0.4         &...    &       ...     &       ...     &       \\
G192.60-00.04           &       0.20    &       0.65    &       0.15    &       0.17         &       1.0     &       0.6     &0.12   &       0.8     &       0.5         &...&   ...     &       ...&    \\
\hline
\end{tabular}
\\

\end{table*}

\begin{table*}
\centering
\setlength{\tabcolsep}{0.1in}
\centering
\caption{Distribution information of C$_4$H 9$-8$, $c$-C$_3$H$_2$ 2$-1$, and H$^{13}$CO$^+$ 1$-0$.} \label{table:distribution}

\vspace{-0.5mm}
\begin{tabular}{cccccccccl}
  \hline
    \hline
             
  \multirow{2}{*}{Source Name}          &   \multicolumn{2}{c}{\cfh(9-8)}   &   \multicolumn{2}{c}{\cctht(2-1) }         & \multicolumn{1}{c}{Difference between}   \\
        & Detection   &Clear feature  &  Detection  & Clear  feature      &  \cfh (9-8) and  \cctht (2-1)    \\
                         
\hline
 
 G005.88-00.39  &       ...     &       ...     &       $\surd$ &       $\surd$ &       ...     &       \\
G011.91-00.61   &       ...     &       ...     &       $\surd$ &       $\surd$ &       ...     &       \\
G012.80-00.20   &       ...     &       ...     &       $\surd$ &       $\surd$ &       ...     &       \\
G015.03-00.67   &       $\surd$ &       $\surd$ &       $\surd$ &       $\surd$ &       $\surd$ &       \\
G023.43-00.18   &       $\surd$ &       $\surd$ &       $\surd$ &       $\surd$ &       $\surd$ &       \\
G031.28+00.66   &       ...     &       ...     &       $\surd$ &       $\surd$ &       ...     &       \\
G034.39+00.22   &       $\surd$ &       ...     &       $\surd$ &       $\surd$ &       ...     &       \\
G035.19-00.74   &       ...     &       ...     &       $\surd$ &       $\surd$ &       ...     &       \\
G035.19-01.73   &       ...     &       ...     &       $\surd$ &       $\surd$ &       ...     &       \\
G037.43+01.51   &       ...     &       ...     &       $\surd$ &       $\surd$ &       ...     &       \\
G043.16+00.01   &       ...     &       ...     &       $\surd$ &       $\surd$ &       ...     &       \\
G049.48-00.36   &       $\surd$ &       $\surd$ &       $\surd$ &       $\surd$ &       $\surd$ &       \\
G049.48-00.38   &       $\surd$ &       $\surd$ &       $\surd$ &       $\surd$ &       $\surd$ &       \\
G075.76+00.33   &       $\surd$ &       ...     &       $\surd$ &       $\surd$ &       ...     &       \\
G081.75+00.59   &       $\surd$ &       $\surd$ &       $\surd$ &       $\surd$ &       $\surd$ &       \\
G081.87+00.78   &       ...     &       ...     &       $\surd$ &       $\surd$ &       ...     &       \\
G109.87+02.11   &       ...     &       ...     &       $\surd$ &       $\surd$ &       ...     &       \\
G111.54+00.77   &       ...     &       ...     &       $\surd$ &       $\surd$ &       ...     &       \\
G121.29+00.65   &       ...     &       ...     &       $\surd$ &       $\surd$ &       ...     &       \\
G133.94+01.06   &       ...     &       ...     &       $\surd$ &       $\surd$ &       ...     &       \\
G188.94+00.88   &       ...     &       ...     &       $\surd$ &       $\surd$ &       ...     &       \\
G192.60-00.04   &       $\surd$ &       $\surd$ &       $\surd$ &       $\surd$ &       $\surd$ &       \\
 \hline
\end{tabular}
\\

\end{table*}

\clearpage

\small
\setlength{\tabcolsep}{0.14in}
\begin{longtable}{ccccccl}

\caption{\label{table:Observed date} Observed data for the C$_4$H, $c$-C$_3$H$_2$, and H$^{13}$CO$^+$ lines.}\\
\hline
 \hline
\multicolumn{1}{l}{Source name} & \multicolumn{2}{c}{Molecular line} & \multicolumn{1}{c}{$\int T_{\rm mb}\rm  d\rm{v}$} & \multicolumn{1}{c}{FWHM} & $T_{\rm peak}$  & Box colour \\

\multicolumn{1}{l}{  } & \multicolumn{2}{c}{ } & \multicolumn{1}{c}{(K·km\,s$^{-1}$)} & \multicolumn{1}{c}{ (km\,s$^{-1}$) } &  (K)   &  \\

\hline
\endfirsthead
\caption{Continued.}\\
\hline
\hline
\multicolumn{1}{l}{Source name} & \multicolumn{2}{c}{Molecular Line} & \multicolumn{1}{c}{$\int T_{\rm mb}\rm  d\rm{v}$} & \multicolumn{1}{c}{FWHM} & $T_{\rm peak}$  & Box colour \\
\multicolumn{1}{l}{  } & \multicolumn{2}{c}{ } & \multicolumn{1}{c}{(K·km\,s$^{-1}$)} & \multicolumn{1}{c}{ (km\,s$^{-1}$) } &  (K)   &  \\
\hline
\endhead
\hline
\endfoot
 G005.88-00.39  & C$_4$H  & N=9-8   J=19/2-17/2 &        $\le$  0.18                    &       ...                     &       ...     &       green   \\
        & C$_4$H  & N=9-8   J=17/2-15/2 &        $\le$  0.22                    &       ...                     &       ...     &               \\
        &   $c$-C$_3$H$_2$ & J=2(1,2)-1(0,1) &          5.65    $\pm$   0.09         &       3.49    $\pm$   0.06    &       1.52    &               \\
        & H$^{13}$CO$^+$   & J=1-0 &            13.60   $\pm$   0.19    &       3.92         $\pm$   0.07    &       3.26    &               \\
G011.91-00.61   & C$_4$H  & N=9-8   J=19/2-17/2 &        $\le$  0.15                    &       ...                     &       ...     &       green   \\
        & C$_4$H  & N=9-8   J=17/2-15/2 &        $\le$  0.18                    &       ...                     &       ...     &               \\
        &   $c$-C$_3$H$_2$ & J=2(1,2)-1(0,1) &          2.14    $\pm$   0.04         &       5.26    $\pm$   0.10    &       0.38    &               \\
        & H$^{13}$CO$^+$   & J=1-0 &            5.00    $\pm$   0.04    &       4.59         $\pm$   0.04    &       1.02    &               \\
G011.91-00.61   & C$_4$H  & N=9-8   J=19/2-17/2 &        $\le$  0.16                    &       ...                     &       ...     &       cyan    \\
        & C$_4$H  & N=9-8   J=17/2-15/2 &        $\le$  0.16                    &       ...                     &       ...     &               \\
        &   $c$-C$_3$H$_2$ & J=2(1,2)-1(0,1) &          1.96    $\pm$   0.05         &       2.79    $\pm$   0.08    &       0.66    &               \\
        & H$^{13}$CO$^+$   & J=1-0 &            4.05    $\pm$   0.06    &       3.84         $\pm$   0.07    &       0.99    &               \\
G012.80-00.20   & C$_4$H  & N=9-8   J=19/2-17/2 &        $\le$  0.23                    &       ...                     &       ...     &       green   \\
        & C$_4$H  & N=9-8   J=17/2-15/2 &        $\le$  0.21                    &       ...                     &       ...     &               \\
        &   $c$-C$_3$H$_2$ & J=2(1,2)-1(0,1) &          3.87    $\pm$   0.37         &       5.01    $\pm$   0.67    &       0.73    &               \\
        & H$^{13}$CO$^+$   & J=1-0 &            6.59    $\pm$   0.12    &       4.81         $\pm$   0.10    &       1.29    &               \\
G015.03-00.67   & C$_4$H  & N=9-8   J=19/2-17/2 &               0.83    $\pm$   0.08         &       5.38    $\pm$   0.60    &       0.15    &       green   \\
        & C$_4$H  & N=9-8   J=17/2-15/2 &               0.74    $\pm$   0.06         &       4.54    $\pm$   0.42    &       0.15    &               \\
        &   $c$-C$_3$H$_2$ & J=2(1,2)-1(0,1) &          6.42    $\pm$   0.06         &       5.10    $\pm$   0.05    &       1.18    &               \\
        & H$^{13}$CO$^+$   & J=1-0 &            14.37   $\pm$   0.11    &       3.35         $\pm$   0.03    &       4.03    &               \\
G015.03-0067    & C$_4$H  & N=9-8   J=19/2-17/2 &               0.74    $\pm$   0.05         &       4.86    $\pm$   0.42    &       0.14    &       cyan    \\
        & C$_4$H  & N=9-8   J=17/2-15/2 &               0.43    $\pm$   0.05         &       3.46    $\pm$   0.47    &       0.12    &               \\
        &   $c$-C$_3$H$_2$ & J=2(1,2)-1(0,1) &          5.89    $\pm$   0.05         &       4.34    $\pm$   0.05    &       1.28    &               \\
        & H$^{13}$CO$^+$   & J=1-0 &            5.14    $\pm$   0.04    &       3.14         $\pm$   0.03    &       1.54    &               \\
G023.43-00.18   & C$_4$H  & N=9-8   J=19/2-17/2 &               0.29    $\pm$   0.07         &       9.00    $\pm$   1.31    &       0.05    &       green   \\
        & C$_4$H  & N=9-8   J=17/2-15/2 &               0.25    $\pm$   0.06         &       8.89    $\pm$   1.12    &       0.05    &               \\
        &   $c$-C$_3$H$_2$ & J=2(1,2)-1(0,1) &          2.08    $\pm$   0.11         &       6.85    $\pm$   0.45    &       0.29    &               \\
        & H$^{13}$CO$^+$   & J=1-0 &            6.07    $\pm$   0.06    &       4.91         $\pm$   0.06    &       1.16    &               \\
G031.28+00.66   & C$_4$H  & N=9-8   J=19/2-17/2 &        $\le$  0.13                    &       ...                     &       ...     &       green   \\
        & C$_4$H  & N=9-8   J=17/2-15/2 &        $\le$  0.16                    &       ...                     &       ...     &               \\
        &   $c$-C$_3$H$_2$ & J=2(1,2)-1(0,1) &          1.63    $\pm$   0.06         &       5.48    $\pm$   0.24    &       0.28    &               \\
        & H$^{13}$CO$^+$   & J=1-0 &            4.33    $\pm$   0.04    &       4.30         $\pm$   0.04    &       0.95    &               \\
G034.39+00.22   & C$_4$H  & N=9-8   J=19/2-17/2 &               0.24    $\pm$   0.08         &       10.20   $\pm$   1.42    &       0.05    &       green   \\
        & C$_4$H  & N=9-8   J=17/2-15/2 &               0.22    $\pm$   0.11         &       9.35    $\pm$   2.02    &       0.04    &               \\
        &   $c$-C$_3$H$_2$ & J=2(1,2)-1(0,1) &          3.17    $\pm$   0.10         &       5.89    $\pm$   0.23    &       0.51    &               \\
        & H$^{13}$CO$^+$   & J=1-0 &            7.27    $\pm$   0.11    &       4.61         $\pm$   0.08    &       1.48    &               \\
G035.19-00.74   & C$_4$H  & N=9-8   J=19/2-17/2 &        $\le$  0.18                    &       ...                     &       ...     &       green   \\
        & C$_4$H  & N=9-8   J=17/2-15/2 &        $\le$  0.17                    &       ...                     &       ...     &               \\
        &   $c$-C$_3$H$_2$ & J=2(1,2)-1(0,1) &          2.88    $\pm$   0.04         &       3.49    $\pm$   0.06    &       0.77    &               \\
        & H$^{13}$CO$^+$   & J=1-0 &            7.23    $\pm$   0.06    &       4.33         $\pm$   0.04    &       1.57    &               \\
G035.19-01.73   & C$_4$H  & N=9-8   J=19/2-17/2 &        $\le$  0.13                    &       ...                     &       ...     &       green   \\
        & C$_4$H  & N=9-8   J=17/2-15/2 &        $\le$  0.15                    &       ...                     &       ...     &               \\
        &   $c$-C$_3$H$_2$ & J=2(1,2)-1(0,1) &          2.06    $\pm$   0.06         &       2.85    $\pm$   0.11    &       0.68    &               \\
        & H$^{13}$CO$^+$   & J=1-0 &            5.05    $\pm$   0.05    &       2.46         $\pm$   0.03    &       1.93    &               \\
G037.43+01.51   & C$_4$H  & N=9-8   J=19/2-17/2 &        $\le$  0.16                    &       ...                     &       ...     &       green   \\
        & C$_4$H  & N=9-8   J=17/2-15/2 &        $\le$  0.14                    &       ...                     &       ...     &               \\
        &   $c$-C$_3$H$_2$ & J=2(1,2)-1(0,1) &          1.09    $\pm$   0.04         &       2.80    $\pm$   0.11    &       0.37    &               \\
        & H$^{13}$CO$^+$   & J=1-0 &            4.10    $\pm$   0.05    &       2.56         $\pm$   0.04    &       1.51    &               \\
G043.16+00.01   & C$_4$H  & N=9-8   J=19/2-17/2 &        $\le$  0.14                    &       ...                     &       ...     &       green   \\
        & C$_4$H  & N=9-8   J=17/2-15/2 &        $\le$  0.13                    &       ...                     &       ...     &               \\
        &   $c$-C$_3$H$_2$ & J=2(1,2)-1(0,1) &          3.46    $\pm$   0.241         &       10.32   $\pm$   0.56    &       0.20    &               \\
        & H$^{13}$CO$^+$   & J=1-0 &            8.87    $\pm$   0.15    &       13.29         $\pm$   0.24    &       0.63    &               \\
G049.48-00.36   & C$_4$H  & N=9-8   J=19/2-17/2 &               0.60    $\pm$   0.09         &       12.21   $\pm$   2.43    &       0.05    &       green   \\
        & C$_4$H  & N=9-8   J=17/2-15/2 &               0.51    $\pm$   0.09         &       12.77   $\pm$   3.52    &       0.04    &               \\
        &   $c$-C$_3$H$_2$ & J=2(1,2)-1(0,1) &          5.06    $\pm$   0.37         &       7.67    $\pm$   0.85    &       0.62    &               \\
        & H$^{13}$CO$^+$   & J=1-0 &            18.96   $\pm$   0.06    &       7.88         $\pm$   0.03    &       2.26    &               \\
G049.48-00.36   & C$_4$H  & N=9-8   J=19/2-17/2 &               0.37    $\pm$   0.06         &       3.00    $\pm$   0.66    &       0.12    &       cyan    \\
        & C$_4$H  & N=9-8   J=17/2-15/2 &               0.57    $\pm$   0.06         &       5.29    $\pm$   0.62    &       0.10    &               \\
        &   $c$-C$_3$H$_2$ & J=2(1,2)-1(0,1) &          3.96    $\pm$   0.19         &       4.61    $\pm$   0.27    &       0.81    &               \\
        & H$^{13}$CO$^+$   & J=1-0 &            5.39    $\pm$   0.09    &       3.75         $\pm$   0.08    &       1.35    &               \\      
G049.48-00.38   & C$_4$H  & N=9-8   J=19/2-17/2 &               0.72    $\pm$   0.11         &       11.51   $\pm$   1.59    &       0.07    &       green   \\
        & C$_4$H  & N=9-8   J=17/2-15/2 &               0.72    $\pm$   0.04         &       11.03   $\pm$   0.94    &       0.07    &               \\
        &   $c$-C$_3$H$_2$ & J=2(1,2)-1(0,1) &          4.73    $\pm$   0.22         &       11.15   $\pm$   0.52    &       0.40    &               \\
        & H$^{13}$CO$^+$   & J=1-0 &            16.44   $\pm$   0.08    &       7.72         $\pm$   0.04    &       2.00    &               \\
G049.48-00.38   & C$_4$H  & N=9-8   J=19/2-17/2 &               1.25    $\pm$   0.12         &       13.14   $\pm$   1.27    &       0.09    &       cyan    \\
        & C$_4$H  & N=9-8   J=17/2-15/2 &               1.33    $\pm$   0.10         &       13.56   $\pm$   1.11    &       0.09    &               \\
        &   $c$-C$_3$H$_2$ & J=2(1,2)-1(0,1) &          5.10    $\pm$   0.12         &       11.84   $\pm$   0.29    &       0.40    &               \\
        & H$^{13}$CO$^+$   & J=1-0 &            7.99    $\pm$   0.16    &       12.61         $\pm$   0.28    &       0.60    &               \\
G075.76+00.33   & C$_4$H  & N=9-8   J=19/2-17/2 &               0.25    $\pm$   0.05         &       5.17    $\pm$   1.49    &       0.05    &       green   \\
        & C$_4$H  & N=9-8   J=17/2-15/2 &               0.20    $\pm$   0.05         &       3.77    $\pm$   1.25    &       0.05    &               \\
        &   $c$-C$_3$H$_2$ & J=2(1,2)-1(0,1) &          1.11    $\pm$   0.04         &       4.32    $\pm$   0.18    &       0.24    &               \\
        & H$^{13}$CO$^+$   & J=1-0 &            3.42    $\pm$   0.05    &       3.85         $\pm$   0.07    &       0.83    &               \\
G081.75+00.59-1 & C$_4$H  & N=9-8   J=19/2-17/2 &               0.18    $\pm$   0.03         &       1.60    $\pm$   0.25    &       0.11    &       green   \\
        & C$_4$H  & N=9-8   J=17/2-15/2 &               0.10    $\pm$   0.02         &       2.15    $\pm$   0.29    &       0.07    &               \\
        &   $c$-C$_3$H$_2$ & J=2(1,2)-1(0,1) &          2.10    $\pm$   0.03         &       2.51    $\pm$   0.05    &       0.79    &               \\
        & H$^{13}$CO$^+$   & J=1-0 &            6.06    $\pm$   0.03    &       2.19         $\pm$   0.01    &       2.60    &               \\
G081.75+00.59-1 & C$_4$H  & N=9-8   J=19/2-17/2 &               0.27    $\pm$   0.06         &       4.20    $\pm$   1.17    &       0.06    &       cyan    \\
        & C$_4$H  & N=9-8   J=17/2-15/2 &               0.18    $\pm$   0.04         &       4.91    $\pm$   1.05    &       0.03    &               \\
        &   $c$-C$_3$H$_2$ & J=2(1,2)-1(0,1) &          2.00    $\pm$   0.04         &       2.58    $\pm$   0.05    &       0.73    &               \\
        & H$^{13}$CO$^+$   & J=1-0 &            5.11    $\pm$   0.06    &       2.37         $\pm$   0.03    &       2.03    &               \\
G081.75+00.59-2 & C$_4$H  & N=9-8   J=19/2-17/2 &               0.32    $\pm$   0.09         &       3.52    $\pm$   1.12    &       0.09    &       green   \\
        & C$_4$H  & N=9-8   J=17/2-15/2 &               0.46    $\pm$   0.12         &       4.73    $\pm$   1.47    &       0.09    &               \\
        &   $c$-C$_3$H$_2$ & J=2(1,2)-1(0,1) &          3.49    $\pm$   0.09         &       3.73    $\pm$   0.12    &       0.88    &               \\
        & H$^{13}$CO$^+$   & J=1-0 &            7.54    $\pm$   0.10    &       4.14         $\pm$   0.06    &       1.71    &               \\
G081.75+00.59   -2& C$_4$H  & N=9-8   J=19/2-17/2 &             0.51    $\pm$   0.08         &       3.96    $\pm$   0.69    &       0.12    &       cyan    \\
        & C$_4$H  & N=9-8   J=17/2-15/2 &               0.36    $\pm$   0.07         &       3.21    $\pm$   0.70    &       0.10    &               \\
        &   $c$-C$_3$H$_2$ & J=2(1,2)-1(0,1) &          3.46    $\pm$   0.07         &       3.99    $\pm$   0.09    &       0.82    &               \\
        & H$^{13}$CO$^+$   & J=1-0 &            10.37   $\pm$   0.06    &       4.26         $\pm$   0.03    &       2.28    &               \\
G081.87+00.78   & C$_4$H  & N=9-8   J=19/2-17/2 &               0.11    $\pm$   0.03         &       1.59    $\pm$   0.41    &       0.06    &       green   \\
        & C$_4$H  & N=9-8   J=17/2-15/2 &               0.13    $\pm$   0.04         &       1.60    $\pm$   0.32    &       0.06    &               \\
        &   $c$-C$_3$H$_2$ & J=2(1,2)-1(0,1) &          0.99    $\pm$   0.03         &       3.37    $\pm$   0.13    &       0.28    &               \\
        & H$^{13}$CO$^+$   & J=1-0 &            7.34    $\pm$   0.06    &       3.95         $\pm$   0.04    &       1.75    &               \\
G109.87+02.11   & C$_4$H  & N=9-8   J=19/2-17/2 &        $\le$  0.16                     &       ...                     &       ...     &       green   \\
        & C$_4$H  & N=9-8   J=17/2-15/2 &        $\le$  0.19                    &       ...                     &       ...     &               \\
        &   $c$-C$_3$H$_2$ & J=2(1,2)-1(0,1) &          1.44    $\pm$   0.06         &       3.77    $\pm$   0.17    &       0.36    &               \\
        & H$^{13}$CO$^+$   & J=1-0 &            7.17    $\pm$   0.08    &       3.44         $\pm$   0.04    &       1.96    &               \\
G111.54+00.77   & C$_4$H  & N=9-8   J=19/2-17/2 &        $\le$  0.21                    &       ...                     &       ...     &       green   \\
        & C$_4$H  & N=9-8   J=17/2-15/2 &        $\le$  0.20                    &       ...                     &       ...     &               \\
        &   $c$-C$_3$H$_2$ & J=2(1,2)-1(0,1) &          1.37    $\pm$   0.06         &       4.56    $\pm$   0.20    &       0.28    &               \\
        & H$^{13}$CO$^+$   & J=1-0 &            6.09    $\pm$   0.05    &       4.35         $\pm$   0.05    &       1.31    &               \\
G111.54+00.77   & C$_4$H  & N=9-8   J=19/2-17/2 &        $\le$  0.15                    &       ...                     &       ...     &       cyan    \\
        & C$_4$H  & N=9-8   J=17/2-15/2 &        $\le$  0.17                    &       ...                     &       ...     &               \\
        &   $c$-C$_3$H$_2$ & J=2(1,2)-1(0,1) &          1.98    $\pm$   0.09         &       8.70    $\pm$   0.52    &       0.21    &               \\
        & H$^{13}$CO$^+$   & J=1-0 &            4.14    $\pm$   0.07    &       3.77         $\pm$   0.08    &       1.03    &               \\
G121.29+00.65   & C$_4$H  & N=9-8   J=19/2-17/2 &        $\le$  0.12                    &       ...                     &       ...     &       green   \\
        & C$_4$H  & N=9-8   J=17/2-15/2 &        $\le$  0.13                    &       ...                     &       ...     &               \\
        &   $c$-C$_3$H$_2$ & J=2(1,2)-1(0,1) &          2.11    $\pm$   0.06         &       2.54    $\pm$   0.08    &       0.78    &               \\
        & H$^{13}$CO$^+$   & J=1-0 &            5.18    $\pm$   0.04    &       2.46         $\pm$   0.02    &       1.98    &               \\
G133.94+01.06   & C$_4$H  & N=9-8   J=19/2-17/2 &               0.34    $\pm$   0.07         &       4.56    $\pm$   1.03    &       0.07    &       green   \\
        & C$_4$H  & N=9-8   J=17/2-15/2 &               0.25    $\pm$   0.04         &       1.14    $\pm$   0.59    &       0.06    &               \\
        &   $c$-C$_3$H$_2$ & J=2(1,2)-1(0,1) &          1.05    $\pm$   0.09         &       3.62    $\pm$   0.34    &       0.27    &               \\
        & H$^{13}$CO$^+$   & J=1-0 &            4.45    $\pm$   0.08    &       3.40         $\pm$   0.07    &       1.23    &               \\
G188.94+00.88   & C$_4$H  & N=9-8   J=19/2-17/2 &        $\le$  0.16                    &       ...                     &       ...     &       green   \\
        & C$_4$H  & N=9-8   J=17/2-15/2 &        $\le$  0.15                    &       ...                     &       ...     &               \\
        &   $c$-C$_3$H$_2$ & J=2(1,2)-1(0,1) &          0.98    $\pm$   0.12         &       3.67    $\pm$   0.54    &       0.25    &               \\
        & H$^{13}$CO$^+$   & J=1-0 &            3.23    $\pm$   0.08    &       3.03         $\pm$   0.10    &       1.00    &               \\
G192.60-00.04   & C$_4$H  & N=9-8   J=19/2-17/2 &               0.18    $\pm$   0.04         &       2.10    $\pm$   0.61    &       0.07    &       green   \\
        & C$_4$H  & N=9-8   J=17/2-15/2 &               0.37    $\pm$   0.06         &       6.52    $\pm$   0.85    &       0.06    &               \\
        &   $c$-C$_3$H$_2$ & J=2(1,2)-1(0,1) &          1.19    $\pm$   0.04         &       3.01    $\pm$   0.12    &       0.37    &               \\
        & H$^{13}$CO$^+$   & J=1-0 &            2.54    $\pm$   0.05    &       2.81         $\pm$   0.07    &       0.85    &               \\
G192.60-00.04   & C$_4$H  & N=9-8   J=19/2-17/2 &               0.25    $\pm$   0.06         &       5.31    $\pm$   1.69    &       0.04    &       cyan    \\
        & C$_4$H  & N=9-8   J=17/2-15/2 &               0.33    $\pm$   0.07         &       8.60    $\pm$   1.35    &       0.06    &               \\
        &   $c$-C$_3$H$_2$ & J=2(1,2)-1(0,1) &          1.18    $\pm$   0.05         &       3.76    $\pm$   0.19    &       0.30    &               \\
        & H$^{13}$CO$^+$   & J=1-0 &            1.87    $\pm$   0.04    &       2.96         $\pm$   0.08    &       0.59    &               \\

\end{longtable}

\begin{table*}
\centering
\setlength{\tabcolsep}{0.06in}
\centering
\caption{ \label{table_Colunm density}Colunm density \& relative abundance of \cfh,  \cctht and \hcfn.}
\vspace{-0.5mm}
\begin{tabular}{cccccccccl}
  \hline
    \hline
             
  \multirow{2}{*}{Source name}          &       $N$(\cfh)       &    $N$(\cctht)    &        $N$(\hcfn)        & $\dfrac{N_{C_4H}}{N_{H^{13}CO^+}}$  &    $\dfrac{N_{c-C_3H_2}}{N_{H^{13}CO^+}}$   &   $\dfrac{N_{C_4H}}{ N_{c-C_3H_2}}$         & Box colour   & \\
        & $10^{13}$cm$^{-2}$ & $10^{13}$cm$^{-2}$  & $10^{13}$cm$^{-2}$ &     &      &   &      &   \\
\hline
 G005.88-00.39   & $\le$        0.73                    &       15.47    $\pm$   0.24    &       3.53     $\pm$  0.05    & $\le$ 0.13                    &       4.38          $\pm$  0.09    & $\le$ 0.18                    &       green   \\
G011.91-00.61   & $\le$ 0.61                    &       5.86     $\pm$  0.11         &       1.30     $\pm$  0.01    & $\le$ 0.31                    &       4.52          $\pm$  0.09    & $\le$ 0.15                    &       green   \\
G011.91-00.61   & $\le$ 0.79                    &       5.36     $\pm$  0.13         &       1.05     $\pm$  0.02    & $\le$ 0.42                    &       5.11          $\pm$  0.15    & $\le$ 0.16                    &       cyan    \\
G012.80-00.20   & $\le$ 1.16                    &       10.60    $\pm$  1.00         &       1.71     $\pm$  0.03    & $\le$ 0.37                    &       6.20          $\pm$  0.60    & $\le$ 0.23                    &       green   \\
 G015.03-00.67  &       2.22     $\pm$  0.14    &       17.57    $\pm$  0.16    &       2.73     $\pm$   0.02    &       0.60     $\pm$  0.04    &       4.71     $\pm$  0.06         &       0.13     $\pm$  0.01    &       green   \\
G015.03-0067    &       1.64     $\pm$  0.10    &       16.12    $\pm$  0.15    &       1.33     $\pm$   0.01    &       1.23     $\pm$  0.07    &       12.09    $\pm$  0.14    &       0.10     $\pm$   0.01    &       cyan    \\
G023.43-00.18   &       0.77     $\pm$  0.13    &       5.69     $\pm$  0.29    &       1.57     $\pm$   0.02    &       0.49     $\pm$  0.08    &       3.61     $\pm$  0.19         &       0.13     $\pm$  0.02    &       green   \\
G031.28+00.66   & $\le$ 1.15                    &       4.45     $\pm$  0.17         &       1.12     $\pm$  0.01    & $\le$ 0.31                    &       3.96          $\pm$  0.16    & $\le$ 0.13                    &       green   \\
G034.39+00.22   &       0.65     $\pm$  0.19    &       8.67     $\pm$  0.28    &       1.89     $\pm$   0.03    &       0.35     $\pm$  0.10    &       4.60     $\pm$  0.17         &       0.08     $\pm$  0.02    &       green   \\
G035.19-00.74   & $\le$ 0.69                    &       7.87     $\pm$  0.11         &       1.88     $\pm$  0.02    & $\le$ 0.26                    &       4.19          $\pm$  0.07    & $\le$ 0.18                    &       green   \\
G035.19-01.73   & $\le$ 0.88                    &       5.62     $\pm$  0.17         &       1.31     $\pm$  0.01    & $\le$ 0.26                    &       4.29          $\pm$  0.13    & $\le$ 0.13                    &       green   \\
G037.43+01.51   & $\le$ 0.43                    &       2.97     $\pm$  0.10         &       1.06     $\pm$  0.01    & $\le$ 0.41                    &       2.79          $\pm$  0.10    & $\le$ 0.16                    &       green   \\
G043.16+00.01   & $\le$ 1.87                    &       9.43     $\pm$  0.65         &       2.30     $\pm$  0.04    & $\le$ 0.16                    &       4.11          $\pm$  0.29    & $\le$ 0.14                    &       green   \\
G049.48-00.36   &       1.57     $\pm$  0.18    &       13.85    $\pm$  0.10         &       4.92     $\pm$  0.02    &       0.32     $\pm$  0.04    &       2.81          $\pm$  0.21    &       0.11     $\pm$  0.02    &       green   \\
G049.48-00.36   &       1.34     $\pm$  0.12    &       10.84    $\pm$  0.51    &       1.40     $\pm$   0.02    &       0.96     $\pm$  0.08    &       7.75     $\pm$  0.39    &       0.12     $\pm$   0.01    &       cyan    \\
G049.48-00.38   &       2.03     $\pm$  0.16    &       12.93    $\pm$  0.59    &       4.26     $\pm$   0.02    &       0.48     $\pm$  0.04    &       3.03     $\pm$  0.14         &       0.16     $\pm$  0.01    &       green   \\
G049.48-00.38   &       3.67     $\pm$  0.22    &       13.94    $\pm$  0.33    &       2.07     $\pm$   0.04    &       1.77     $\pm$  0.11    &       6.72     $\pm$  0.21    &       0.26     $\pm$   0.02    &       cyan    \\
G075.76+00.33   &       0.62     $\pm$  0.11    &       3.02     $\pm$  0.11    &       0.89     $\pm$   0.01    &       0.71     $\pm$  0.12    &       3.41     $\pm$  0.14         &       0.21     $\pm$  0.04    &       green   \\
G081.75+00.59-1&        0.40     $\pm$  0.05    &       5.75     $\pm$  0.09    &       1.57     $\pm$   0.01    &       0.25     $\pm$  0.03    &       3.66     $\pm$  0.06         &       0.07     $\pm$  0.01    &       green   \\
G081.75+00.59-1 &       0.63     $\pm$  0.10    &       5.46     $\pm$  0.10    &       1.33     $\pm$   0.02    &       0.48     $\pm$  0.08    &       4.12     $\pm$  0.09    &       0.12     $\pm$   0.02    &       cyan    \\
G081.75+00.59-2 &       1.11     $\pm$  0.21    &       9.55     $\pm$  0.24    &       1.96     $\pm$   0.03    &       0.57     $\pm$  0.11    &       4.88     $\pm$  0.14    &       0.12     $\pm$   0.02    &       green   \\              G081.75+00.59-2         &       1.21     $\pm$   0.14    &       9.48     $\pm$  0.19    &       2.69     $\pm$  0.02    &       0.45          $\pm$  0.05    &       3.52     $\pm$  0.07    &       0.13     $\pm$   0.02    &       cyan    \\
G081.87+00.78   &       0.33     $\pm$  0.07    &       2.70     $\pm$  0.09    &       1.90          $\pm$  0.02    &       0.17     $\pm$  0.04    &       1.42     $\pm$   0.05    &       0.12     $\pm$  0.03    &       green   \\
G109.87+02.11   & $\le$ 0.98                    &       3.94     $\pm$  0.16         &       1.86     $\pm$  0.02    & $\le$ 0.23                    &       2.12          $\pm$  0.09    & $\le$ 0.16                    &       green   \\
G111.54+00.77   & $\le$ 0.71                    &       3.75     $\pm$  0.16         &       1.58     $\pm$  0.01    & $\le$ 0.36                    &       2.37          $\pm$  0.10    & $\le$ 0.21                    &       green   \\
G111.54+00.77   & $\le$ 2.66                    &       5.43     $\pm$  0.26         &       1.08     $\pm$  0.02    & $\le$ 0.36                    &       5.05          $\pm$  0.25    & $\le$ 0.15                    &       cyan    \\
G121.29+00.65   & $\le$ 0.64                    &       5.78     $\pm$  0.16         &       1.34     $\pm$  0.01    & $\le$ 0.24                    &       4.30          $\pm$  0.12    & $\le$ 0.12                    &       green   \\
G133.94+01.06   &       0.83     $\pm$  0.12    &       2.86     $\pm$  0.24    &       1.16     $\pm$   0.02    &       0.72     $\pm$  0.10    &       2.48     $\pm$  0.21         &       0.29     $\pm$  0.05    &       green   \\
G188.94+00.88   & $\le$ 1.26                    &       2.68     $\pm$  0.31         &       0.84     $\pm$  0.02    & $\le$ 0.50                    &       3.19          $\pm$  0.39    & $\le$ 0.16                    &       green   \\
G192.60-00.04   &       0.80     $\pm$  0.11    &       3.25     $\pm$  0.10         &       0.66     $\pm$  0.01    &       1.29     $\pm$  0.16    &       4.94          $\pm$  0.19    &       0.25     $\pm$  0.03    &       green   \\
G192.60-00.04   &       0.83     $\pm$  0.14    &       3.24     $\pm$  0.15    &       0.48     $\pm$   0.01    &       1.70     $\pm$  0.29    &       6.69     $\pm$  0.34         &       0.25     $\pm$  0.04    &       cyan    \\                                                                                                                                                                                                                      
\hline 

\end{tabular}
\\

\end{table*}
 \end{appendix}

\end{document}